\documentclass[twocolumn,superscriptaddress,amsmath,amssymb,aps,floatfix]{revtex4-1}

\usepackage{graphicx}
\usepackage{slashed}
\usepackage{bm}
\usepackage{hyperref}
\usepackage{graphicx,color}
\usepackage{multirow}
\usepackage{amsmath,amstext,array,amssymb}
\usepackage{mathtools}
\usepackage[normalem]{ulem}
\usepackage{graphicx}
\usepackage{xfrac}
\usepackage[T1]{fontenc}

\newcolumntype{C}{>{$}c<{$}}

\begin{document}
\title{A Field-Theory Action for the Constructive Standard Model}
\author{Neil Christensen}
\email{nchris3@ilstu.edu}
\affiliation{Department of Physics, Illinois State University, Normal, IL 61790}

\date{\today}

\begin{abstract}
We introduce a field-theory framework in which fields transform under the little group, rather than the Lorentz group, specific to each particle type. By utilizing these fields, along with spinor products and the $x$ factor, we construct a field-theory action that naturally reproduces the vertices of the Constructive Standard Model (CSM). This approach eliminates unphysical components, significantly reduces the degrees of freedom compared to traditional field theory, and offers deeper insights into the power of constructive amplitudes. Our action is momentum-conserving, Lorentz-invariant, Hermitian, and non-local. We also discuss this as a framework for developing new constructive field theories, discussing their essential properties and potential applicability in renormalization theory and non-perturbative calculations.
\end{abstract}

\maketitle

\tableofcontents

\section{Introduction}

Originally, constructive theory was developed from the ground up, based purely on the properties of the $S$ matrix: its analyticity, unitarity, and its behavior under Lorentz transformations and other symmetry transformations. Amplitudes were constructed based on these properties, without reference to fields or field theory, leading to questions about the necessity and importance of the field concept. Two pinnacles of this work were the discovery of an extremely simple expression involving one term for the maximally helicity-violating amplitude for any number of gluons \cite{Parke:1986gb}, and a recursion relation that allowed the construction of any amplitude by gluing together three-point vertices, provided the amplitude could be shown to vanish as the complex parameter, used to complexify the momenta, was taken to infinity \cite{Britto:2005fq}.  Later, \cite{Arkani-Hamed:2017jhn} showed how to extend these methods in a natural way to massive theories with any spin.  Following this, the complete set of three-point and four-point vertices of the Constructive Standard Model (CSM) were described \cite{Christensen:2018zcq,Christensen:2024xzs} and the CSM was shown to be perturbatively unitary, a complete set of four-point amplitudes were calculated and validated against Feynman diagrams \cite{Christensen:2024bdt}, the C++ package SPINAS was created \cite{Christensen:2024xsg}, allowing convenient and efficient phase-space calculations of constructive amplitudes, and many other calculations were performed \cite{Ochirov:2018uyq,Franken:2019wqr,Aoude:2019tzn,Christensen:2019mch,Durieux:2019eor,Baratella:2020lzz,Durieux:2020gip,Balkin:2021dko,Alves:2021rjc,Christensen:2022nja,Liu:2023jbq,Bachu:2023fjn,Lai:2023upa,Ema:2024vww,Ema:2024rss}.

On the other hand, if a momentum shift could not be found, it appeared that an additional ``contact'' term was needed to achieve the correct result. However, in a renormalizable theory such as the Standard Model (SM), additional contact terms beyond those found in Feynman diagrams seem to contradict its renormalizability, where there should only be a finite number of adjustable parameters. Moreover, every four-point amplitude in the CSM has been calculated and found to agree with Feynman diagrams, despite the fact that \textit{neither} contact terms beyond the three CSM four-point vertices \textit{nor} any momentum shifts were needed or used for any of the calculations \cite{Christensen:2024bdt,Christensen:2024xsg}.

In fact, initial suggestions for momentum shifts for $e\bar{e}\bar{\mu}\mu$ did not result in a vanishing amplitude when the complex parameter was taken to infinity \cite{Christensen:2022nja}, which appeared to indicate that a contact term was necessary. However, once all the on-shell identities were discovered and applied, it became clear that a contact term was not needed \cite{Lai:2023upa}. Later, momentum shifts were discovered that did make the amplitudes vanish in the appropriate limit \cite{Ema:2024vww,Ema:2024rss}, thereby clarifying why the constructive amplitude worked and agreed with Feynman diagrams.

Moreover, the discovery of a momentum shift was never required for Feynman diagrams, presumably because they were based on a field theory \cite{Weinberg}, which inherently possessed the desired properties for constructing scattering amplitudes. As long as the traditional field theory satisfies certain properties such as locality, Lorentz invariance, symmetry invariance, and Hermiticity, and the perturbation theory is unitary, its consequences are expected to be consistent, allowing direct comparison with experimental results.

Furthermore, traditional field theory allows for the renormalization of fields, along with other parameters of the theory, which is essential for canceling the infinities that arise in two-point functions. In constructive amplitudes, the formal proof of renormalizability has yet to be established. Nevertheless, the consistent agreement between constructive amplitudes and Feynman diagrams suggests that the theory is renormalizable, meaning that only a finite number of infinities would need to be addressed. However, the absence of fields in constructive theory introduces a challenge in understanding how these infinities are formally canceled.

Another significant advantage of the field-theory approach in particle physics is its ability to facilitate non-perturbative calculations. In particular, the action can be embedded in a functional integral, which can then be discretized on a lattice. Constructive methods, in contrast, have so far been limited to perturbative calculations, and embedding them in a lattice framework has not yet been achieved.

Given this background, it is natural to consider whether a field-theory action can be developed for constructive calculations, thereby harnessing the benefits and insights that field theory provides. Such an approach could offer an alternative criterion for determining whether an amplitude can be derived from the vertices. Moreover, introducing a field theory for constructive amplitudes could provide a more robust theoretical basis for renormalization, particularly if all infinities can be canceled using the fields and parameters of the theory. Additionally, embedding this field theory in a functional integral and placing it on the lattice might offer advantages over traditional lattice field theory, similar to the benefits seen in constructive perturbation theory.

Previous attempts to find a field theory for constructive calculations began with traditional field theory \cite{Kotko:2017nkx,Ananth:2011hu,Buchta:2010qr,Fu:2009gt,Ettle:2008ed,Ettle:2008ey,Feng:2006yy,Ettle:2006bw,Mansfield:2005yd}. However, our goal is a ground-up construction of field theory based on its fundamental properties: momentum conservation, Lorentz invariance, symmetry invariance, Hermiticity, and unitarity of the perturbation theory. Whether these criteria are sufficient for ensuring causality remains to be proven. Nevertheless, if the perturbative amplitudes derived from the CSM continue to agree exactly with Feynman diagrams, we can at least assert that the CSM is causal. This suggests that other field theories constructed with similar principles might also be causal, perhaps with some additional criteria yet to be discovered.

To be clear, locality was never a requirement.  Rather, it satisfied our intuitions that forces should not act through a distance, and it was sufficient, along with the other properties, to achieve causality.  In the past, there wasn't sufficient reason to pursue non-local field theories of the type described here.  However, in the present context with constructive rules, which \textit{are} non-local, agreeing exactly with Feynman diagrams, it is clear that it is time to rethink this expectation.

In this paper, we present a field theory for the CSM in terms of fields that are minimal and \textit{never} contain unphysical components. We achieve this by introducing fields that transform under the little group from the outset, rather than the Lorentz group. This approach aligns with the fact that particles transform under the little group, not the Lorentz group. (The momentum still transforms under the Lorentz group; here, we refer to the degrees of freedom of the particle and field.) We formulate our theory in momentum space, which is natural given its connection with known constructive vertices and because the corresponding spacetime action, which requires a Fourier transform, has yet to be fully developed for constructive theory.

One limitation of our action is that the electroweak symmetry is broken from the beginning. We do not yet have a complete understanding of the Higgs mechanism within this field theory and aim to clarify this in future work. Additionally, we cannot currently explain the relationship between the couplings of the triple-gluon vertex and the gluon-quark vertices. We consider these limitations as deficiencies in the current constructive field theory, but we remain hopeful that these challenges can be addressed in future research.

The importance of this field-theory formulation goes beyond the CSM. If constructive calculations can be obtained directly from a non-local field-theory action, assuming all the required properties are met, then there is no reason we cannot construct new theories beyond the Standard Model directly in a non-local field-theory framework, as described here, without relying on their connection to Feynman diagrams. We can simply write the action, analogous to the approach in traditional field theory, determine its vertices, calculate its amplitudes, and compare these with experimental results. This non-local field theory would stand on its own as a model for particle physics, offering definite predictions at the same level of rigor and consistency as those obtained through local field theory.  Whether this conjecture is correct will require further research, but it presents an exciting avenue to pursue.

The remainder of this paper has the following structure.  In Sec.~\ref{sec:action}, we introduce the non-local field-theory action for the CSM.  We begin by introducing fields for all the particles, their Lorentz transformations and their Hermitian conjugates.  We follow this with the quadratic part of the action, which gives the propagators, in Sec.~\ref{sec:free field}.  In Sec.~\ref{sec:massless 3-point}, we describe the three-point field interactions for all-massless fields.  We include the interactions with massless quarks and leptons, since sometimes taking their massless limit is useful.  Following this, in Sec.~\ref{sec:gluons, photons and gravitons with massive fields}, we give the interactions between photons, gluons and gravitons with massive fields.  The Higgs and the $W$ and $Z$ interactions are described in Sec.~\ref{sec:higgs and weak action}, only after the electroweak symmetry is broken.  We discuss what features of the Higgs mechanism can currently be seen from this non-local field theory in Sec.~\ref{sec:Higgs Mechanism}, and discuss possible ways to make further progress in this important area.  In all these subsections, we also show that the action is Hermitian.  In Sec.~\ref{sec:conclusions}, we conclude.

We have also included several appendices.  In App.~\ref{app:motivation and conventions}, we attempt to motivate our field theory construction.  This discussion will also review the little-group transformations and describe the Lorentz transformation of the fields.  We begin with a very short analogy with Newton's Law in App.~\ref{app:Newton}.  Following this, we review the properties of a single-particle state and its little-group transformations in App.~\ref{app:single particle states}.  Traditional field theory with the embedding of particles in Lorentz-transforming fields is discussed very briefly in App.~\ref{app:traditional Field Theory}.  Our discussion of the little-group transforming fields and their basic properties is discussed in App.~\ref{app:C Field Theory}.  Lorentz transformations are described in greater detail in App.~\ref{app:lorentz transformations}, wherein we discuss the pertinent details of the generators of Lorentz transformations.  We follow this with a description of the helicity spinors and the spin spinors, and their transformation properties under both the Lorentz group and the little-group in App.~\ref{app:spinors}.  Finally, everything is put together to obtain Lorentz invariants and little-group invariants in App.~\ref{app:spinor products}.  It is in this last sub-section that we put everything together and describe the requirements for writing symmetry invariant action terms.  We also introduce $x$ and $\tilde{x}$ here, and we discuss the naive power counting of the interactions.  

In App.~\ref{app:locality}, we discuss locality and why our field theory is non-local.  We begin with an example of a local field theory in App.~\ref{app:local action example}.  Next, we consider a simple field theory with momentum in the denominator and show that it is non-local in App.~\ref{app:non-local denominator momenta}.  Finally, we show that a field theory with spinor products is also non-local in App.~\ref{app:non-local spinor products}.  

We have also included App.~\ref{app:constructive rules}, showing how the vertices and propagators come from the field-theory action in two examples.  We first work out the triple-gluon vertex in App.~\ref{app:ggg vertex helicity}.  This is followed by the $W$-lepton vertex in App.~\ref{app:W-nu vertex chirality}.

\section{\label{sec:action}The Action}

In order to write a field theory action for the particles of the CSM, we will first need to introduce fields for each of the particles.  We have discussed this in detail in App.~\ref{app:motivation and conventions}, which relies significantly on Ref.~\cite{Weinberg}, however, we summarize some relevant aspects here.  Our objective is to introduce fields that do not contain unphysical degrees of freedom and to write free quadratic as well as interaction cubic and quartic action terms for them in a minimal way.  We expect that this field theory will result in the perturbative rules in constructive theory, using the spinors and their products.  

Traditionally, the formulation of a field theory action begins in position space. This approach has historical roots, as fields like the electromagnetic field were originally developed in spatial terms. It also facilitates the construction of a local theory, which is considered desirable both because it aligns with the intuition that interactions should not occur over a distance and because it simplifies the construction of causal theories. However, the action could equally be formulated in momentum space. In many ways, this is more natural for particle physics, where experiments typically involve particles with definite momenta rather than positions. Moreover, if momentum conservation is respected in every term \textit{and} if all explicit momenta appear polynomially, \textit{then} the action in momentum space will be equivalent to a local action in position space. This equivalence holds for traditional field theory, from which Feynman diagrams are derived.

Constructive interactions, however, while conserving momentum, are inherently non-local (see App.~\ref{app:locality}). Despite this, they appear to be causal, as they have consistently agreed with Feynman diagrams for the amplitudes calculated so far, including all tree-level four-point amplitudes in the Standard Model \cite{Christensen:2024bdt}. To clarify, locality was never a strict requirement for causality, but local theories have been both causal and straightforward to construct.  For this reason, non-local particle field theories of the type described here were not extensively explored in the past. However, with the growing importance of constructive calculations and their intrinsic non-locality, it is now time to explore non-local actions in detail. 

This paper aims to formulate a complete non-local action for the CSM. Since the known constructive amplitude rules are formulated in momentum space, it is more straightforward to begin there and, eventually, work back to position space. While the exact form of the constructive field theory in position space remains unknown, this work will lay the foundation for its future discovery.

To determine the properties of the fields in our theory, we must first understand the transformation properties of their corresponding particle excitations. The fields should transform in the same manner as the quantum states of the particles they represent, as reviewed in App.~\ref{app:single particle states}. Generally speaking, the quantum state of a particle transform to a linear combination of the quantum states for that particle at the new momentum, where the coefficients are determined by representations of the little-group. 

For massless particles, this transformation occurs under the helicity little group, where each helicity transforms independently, each within its own invariant subspace. For example, positive and negative helicity photons transform separately and do not mix under Lorentz transformations, a property also shared by massless neutrinos, gluons, and gravitons. Neutrinos, for instance, exist as distinct states -- a negative helicity neutrino and a positive helicity anti-neutrino -- each transforming independently. Similarly, positive and negative helicity gluons and gravitons are distinct states. Therefore, their corresponding fields should reflect these properties, with each helicity represented by a single field degree of freedom and those separate helicity fields should not mix under Lorentz transformations.

In particular, this means that we will not use vector fields for photons and gluons; we will not use second-rank tensor fields for gravitons; and we will not use Dirac or Weyl spinor fields for neutrinos. Not only would such fields introduce unphysical degrees of freedom, but they would necessitate the invention of gauge symmetries for photons and gluons, and diffeomorphism symmetry for gravitons. Moreover, the components of these fields would transform under the full Lorentz group, rather than the little group, and mix their degrees of freedom, thereby having different transformation properties than the particles they are meant to represent. 

Therefore, here, we introduce the following fields, and we give their Lorentz transformations, copied from Eqs.~(\ref{eq:UGU^-1}) through (\ref{eq:UfbU^-1}),
\begin{align}
    U(\Lambda)G^{\pm}(p)U^{-1}(\Lambda) &=
    e^{\pm i2\omega}G^{\pm}(\bar{p})
    \label{eq:UGU^-1}\\
    U(\Lambda)\textsl{g}^{\pm}(p)U^{-1}(\Lambda) &=
    e^{\pm i\omega}\textsl{g}^{\pm}(\bar{p})\\
    U(\Lambda)\gamma^{\pm}(p)U^{-1}(\Lambda) &=
    e^{\pm i\omega}\gamma^{\pm}(\bar{p})\\
    U(\Lambda)f^{\pm}(p)U^{-1}(\Lambda) &=
    e^{\pm i\frac{1}{2}\omega}f^{\pm}(\bar{p})\\
    U(\Lambda)\bar{f}^{\pm}(p)U^{-1}(\Lambda) &=
    e^{\pm i\frac{1}{2}\omega}\bar{f}^{\pm}(\bar{p})
    ,
\end{align}
where $\Lambda$ is a Lorentz transformation, $\bar{p}=\Lambda p$, $U(\Lambda)$ is the quantum Lorentz transformation, $\omega=\omega(\Lambda,p)$ depends on the Lorentz transformation \textit{and} the momentum, $G^{\pm}$ represents a graviton field with helicity $\pm2$, $g^{\pm}$ represents a gluon field with helicity $\pm1$ (with the color-adjoint index suppressed), $\gamma^{\pm}$ represents a photon field with helicity $\pm1$, and $f^{\pm}$ and $\bar{f}^{\pm}$ represent a massless fermion and antifermion, respectively, with helicity $\pm\frac{1}{2}$ (with any other quantum indices suppressed).  Other quantum numbers such as charge and color are not affected by Lorentz transformation.

Massive particles transform under the spin little group, with massive fermions transforming according to the spin-$\frac{1}{2}$ representation. For instance, the electron has two states -- spin up and spin down -- and similarly, the positron also has two states. Notably, there are no separate chirality states for the electron itself; chirality appears only in the context of its interactions with other particles, such as the W and Z bosons. The electron, like all massive leptons and quarks, therefore, has just two states, not four. Consequently, Dirac fields, which possess four degrees of freedom, include unphysical degrees of freedom. This discrepancy arises because Dirac fields transform under representations of the full Lorentz group rather than the spin little group. In this work, we avoid introducing unphysical degrees of freedom for fermions by constructing fields that transform solely under the spin group, not under the chiral group. Specifically, we introduce a fermion field $f^{\mathrm{I}}(p)$ and an antifermion field $\bar{f}^{\mathrm{I}}(p)$, where $\mathrm{I}$ denotes the spin index with two possible values. As we will demonstrate, this choice does not preclude us from writing chiral interactions.

The W and Z bosons transform under spin-1 representations of the spin little group. Although it would be possible to introduce a new index for this purpose, it is more convenient to use a symmetric combination of two spin-$\frac{1}{2}$ indices to form a spin-$1$ field. Therefore, for the $W, \bar{W}$, and $Z$ bosons, we introduce the fields $W^{\mathrm{IK}}, \bar{W}^{\mathrm{IK}}$, and $Z^{\mathrm{IK}}$, where the indices $\mathrm{IK}$ are understood to be symmetrized. In fact, we explicitly symmetrize these indices in the quadratic terms of the action, which leads to symmetrized propagators. This symmetrization suffices for perturbation theory, but in non-perturbative calculations, the interactions involving the $W, \bar{W}$, and $Z$ boson spin indices should also be explicitly symmetrized. We discuss the subtleties of introducing the spin-$1$ fields in this way at the end of Sec.~\ref{app:C Field Theory}.

The Lorentz transformation properties of these fields are governed by the following equations, as derived in Eqs.~(\ref{eq:UhU^-1}) through (\ref{eq:UWbU^-1}):
\begin{align}
    U(\Lambda)h(p)U^{-1}(\Lambda) &=
    h(\bar{p})
    \label{eq:UhU^-1}\\
    U(\Lambda)f^{\mathrm{I}}(p)U^{-1}(\Lambda) &=
    \left(e^{i\vec{\omega}\cdot \vec{J}}\right)^{\ \ \mathrm{I}}_{\mathrm{K}}f^{\mathrm{K}}(\bar{p})\\
    U(\Lambda)\bar{f}^{\mathrm{I}}(p)U^{-1}(\Lambda) &=
    \left(e^{i\vec{\omega}\cdot \vec{J}}\right)^{\ \ \mathrm{I}}_{\mathrm{K}}\bar{f}^{\mathrm{K}}(\bar{p})\\
    U(\Lambda)Z^{\mathrm{IK}}(p)U^{-1}(\Lambda) &=
    \left(e^{i\vec{\omega}\cdot \vec{J}}\right)^{\ \ \mathrm{I}}_{\mathrm{L}}
    \left(e^{i\vec{\omega}\cdot \vec{J}}\right)^{\ \ \mathrm{K}}_{\mathrm{M}}
    Z^{\mathrm{LM}}(\bar{p})\\
    U(\Lambda)W^{\mathrm{IK}}(p)U^{-1}(\Lambda) &=
    \left(e^{i\vec{\omega}\cdot \vec{J}}\right)^{\ \ \mathrm{I}}_{\mathrm{L}}
    \left(e^{i\vec{\omega}\cdot \vec{J}}\right)^{\ \ \mathrm{K}}_{\mathrm{M}}
    W^{\mathrm{LM}}(\bar{p})\\
    U(\Lambda)\bar{W}^{\mathrm{IK}}(p)U^{-1}(\Lambda) &=
    \left(e^{i\vec{\omega}\cdot \vec{J}}\right)^{\ \ \mathrm{I}}_{\mathrm{L}}
    \left(e^{i\vec{\omega}\cdot \vec{J}}\right)^{\ \ \mathrm{K}}_{\mathrm{M}}
    \bar{W}^{\mathrm{LM}}(\bar{p})
    ,
\end{align}
where $\vec{J}$ are the generators of the spin little group, $\vec{\omega}=\vec{\omega}(\Lambda,p)$ depends on the Lorentz transformation \textit{and} on the momentum, and $h(p)$ is the scalar field for the Higgs.

Now that we have defined our fields, the next step is to construct an action that meets several key criteria: it must conserve momentum, be Lorentz invariant, Hermitian, causal, and, of course, consistent with Feynman diagrams. Momentum conservation is straightforward to implement by including a momentum-conserving delta function in every term. Achieving Lorentz invariance, however, requires more consideration. Each field transforms according to an $\omega(\Lambda,p)$ factor, which depends on both the Lorentz transformation and its momentum. This dependency prevents the cancellation of the transformation factors solely among the fields, except in the quadratic case where the momenta are the same. To ensure Lorentz invariance in interaction terms, we must also introduce non-field terms that also transform under the little group representations. These are the spinor products used in constructive calculations. Simplified from Eqs.(\ref{eq:<i|...pl...|j>= transformation}) through (\ref{eq:<i^I|...pl...|j]^K = transformation}), these spinor products transform as follows:
\begin{align}
    \langle ij\rangle &= e^{-\frac{i}{2}\left(\omega_i+\omega_j\right)}
    \langle \bar{i} \bar{j}\rangle
    \\
    \lbrack i j\rbrack &= e^{+\frac{i}{2}\left(\omega_i+\omega_j\right)}
    \lbrack \bar{i} \bar{j}\rbrack
    \\
    \langle i \mathbf{j}\rangle^{\mathrm{K}} &= e^{-\frac{i}{2}\omega_i}
    \left(e^{i\vec{\omega}_j\cdot\vec{J}}\right)_{\mathrm{L}}^{\ \mathrm{K}}
    \langle \bar{i} \bar{\mathbf{j}}\rangle^{\mathrm{L}}
    \\
    \lbrack \mathbf{i}  j\rbrack^{\mathrm{I}} &= e^{+\frac{i}{2}\omega_j}
    \left(e^{i\vec{\omega}_i\cdot\vec{J}}\right)_{\mathrm{L}}^{\ \mathrm{I}}
    \lbrack \bar{\mathbf{i}}  \bar{j}\rbrack^{\mathrm{L}}
    \\
    \langle \mathbf{i} \mathbf{j}\rangle^{\mathrm{IK}} 
    &= 
    \left(e^{i\vec{\omega}_i\cdot\vec{J}}\right)_{\mathrm{L}}^{\ \mathrm{I}}
    \left(e^{i\vec{\omega}_j\cdot\vec{J}}\right)_{\mathrm{M}}^{\ \mathrm{K}}
    \langle \bar{\mathbf{i}} \bar{\mathbf{j}}\rangle^{\mathrm{LM}}
    \\
    \lbrack \mathbf{i} \mathbf{j}\rbrack^{\mathrm{IK}} 
    &= 
    \left(e^{i\vec{\omega}_i\cdot\vec{J}}\right)_{\mathrm{L}}^{\ \mathrm{I}}
    \left(e^{i\vec{\omega}_j\cdot\vec{J}}\right)_{\mathrm{M}}^{\ \mathrm{K}}
    \lbrack \bar{\mathbf{i}} \bar{\mathbf{j}}\rbrack^{\mathrm{LM}} ,
\end{align}
where these spinors are defined in App.~\ref{app:spinors}, their products are defined in App.~\ref{app:spinor products}, and the bars represent the spinors at the transformed momenta.

The importance of these spinor products is that they also transform under the little group at the new momenta, according to whether the momenta are massless (helicity) or massive (spin).
By combining fields with spinor products such that the helicity transformations for each massless momentum cancels and all spin indices are contracted between fields and objects of the same momentum, we ensure that all little group transformations cancel, thereby achieving Lorentz invariance. When an action satisfies these conditions -- namely, that the helicities for each massless momentum sum to zero and all spin indices are contracted between objects of the same momentum -- it is considered manifestly Lorentz invariant in the context of constructive field theory. This approach replaces the contraction of Lorentz indices and closed fermion chains (and the use of covariant derivatives) in traditional field theory.

An important property of the spinors that form the spinor products is their transformation under the chiral components of the Lorentz group (see App.\ref{app:spinors}). Specifically, the angle spinors $\langle i\rvert, \lvert i\rangle, \langle\mathbf{i}\rvert^{\mathrm{I}}$, and $\lvert\mathbf{i}\rangle^{\mathrm{I}}$ transform under the left-chiral group, while the square spinors $\lbrack i\rvert, \lvert i\rbrack, \lbrack\mathbf{i}\rvert^{\mathrm{I}}$, and $\lvert \mathbf{i}\rbrack^{\mathrm{I}}$ transform under the right-chiral group. This chiral behavior also applies to spinors with lower spin indices. The use of these spinor products in constructing interactions determines the chirality of those interactions. Interactions that are symmetric between angle and square spinors are non-chiral and include most interactions within the CSM. In contrast, chiral interactions, such as the W- and Z-boson interactions with fermions, involve different coefficients for angle and square spinors. This distinction will be explored further in Sec.\ref{sec:higgs and weak action}.

The fields and spinor products are sufficient for many interactions. However, when a vertex involves one massless field and two massive fields of the same mass -- such as in the interactions involving photons, gluons, and gravitons -- we require an additional non-field object. This necessity arises because the two helicity spinors for these momenta, $\lvert i\rangle$ and $\lvert i\rbrack$, are not linearly independent in this case. In fact, they are proportional, as shown in Eqs.~(\ref{eq:x_{ij} proportional}) and (\ref{eq:xt_{ij} proportional}):
\begin{align}
    x_{ij}\lvert l\rangle &=
    \frac{(p_j-p_i)}{2m}\lvert l\rbrack
    \\
    \tilde{x}_{ij}\lvert l\rbrack &=
    \frac{(p_j-p_i)}{2m}\lvert l\rangle
    ,
\end{align}
where $m_i=m_j=m$, $m_l=0$ and $p_i+p_j+p_l=0$.  Therefore, in these interactions, we will use $x_{ij}$ and $\tilde{x}_{ij}$ in place of $\lvert l\rangle$ and $\lvert l\rbrack$.  Despite this substitution, we can still cancel the helicity transformations of the massless fields because $x_{ij}$ and $\tilde{x}_{ij}$ transform as [from Eqs.~(\ref{eq:x_ij transform}) and (\ref{eq:xt_ij transform})]:
\begin{align}
    x_{ij} &= e^{i\omega_l}\bar{x}_{ij}
    \\
    \tilde{x}_{ij} &= e^{-i\omega_l}\bar{\tilde{x}}_{ij} ,
\end{align}
where $\omega_l=\omega(\Lambda,p_l)$ and the bar represents that $\bar{x}$ and $\bar{\tilde{x}}$ are at the transformed momenta.  That is, $x$ and $\tilde{x}$ transform as helicity $+1$ and $-1$, respectively, and can be used to cancel the helicity transformations of the photon, gluon and graviton fields.

Now that we have all the pieces we need to achieve Lorentz invariance, we also need to consider the Hermitian conjugates of both the fields and the non-field objects in order to achieve Hermiticity. From Eqs.~(\ref{eq:G^dagger}) through (\ref{eq:h^dagger}), the fields satisfy:
\begin{align}
    \left[G^{\pm}(p)\right]^\dagger &= G^{\mp}(p)
    \\
    \left[\textsl{g}_a^{\pm}(p)\right]^{\dagger} &= \textsl{g}_a^{\mp}(p)\\
    \left[\gamma^{\pm}(p)\right]^{\dagger} &= \gamma^{\mp}(p)\\
    \left[q^{i\pm}(p)\right]^{\dagger} &= \bar{q}_i^{\mp}(p) \\
    \left[l^{\pm}(p)\right]^\dagger &=
    \bar{l}^{\mp}(p)\\
    \left[\nu^-(p)\right]^\dagger &=
    \bar{\nu}^+(p)\\
    \left[Z^{\mathrm{IJ}}(p)\right]^\dagger &= Z_{\mathrm{JI}}(p)\\
    \left[W^{\mathrm{IJ}}(p)\right]^\dagger &= \bar{W}_{\mathrm{JI}}(p)\\
    \left[q^{i\mathrm{I}}(p)\right]^\dagger &=
    \bar{q}_{i\mathrm{I}}(p)\\
    \left[l^{\mathrm{I}}(p)\right]^\dagger &=
    \bar{l}_{\mathrm{I}}(p)\\
    \left[h(p)\right]^\dagger &= h(p),
\end{align}
where $q$ is a quark, $l$ is a lepton, $a$ is the adjoint QCD index and $i$ is the fundamental (or anti-fundamental) QCD index, and lowering a spin index on the left raises it on the right.

The spinor products are also related by hermitian conjugation.  They satisfy, for example, [see Eqs.~(\ref{eq:<ij>^dagger}) through (\ref{eq:<i^Ij_J>^dagger})]:
\begin{align}
    \langle ij\rangle^\dagger &= \lbrack ji \rbrack
    \\
    \langle \mathbf{i}^{\mathrm{I}}\mathbf{j}^{\mathrm{J}}\rangle^\dagger &= \lbrack \mathbf{j}_{\mathrm{J}}\mathbf{i}_{\mathrm{I}} \rbrack
    \\
    \langle \mathbf{i}^{\mathrm{I}}\mathbf{j}_{\mathrm{J}}\rangle^\dagger &= \lbrack \mathbf{j}^{\mathrm{J}}\mathbf{i}_{\mathrm{I}} \rbrack.
\end{align}
In summary, the order is reversed, angle and square brackets are interchanged, and spin indices are lowered and raised.  Finally, $x$ and $\tilde{x}$ satisfy
\begin{align}
    x_{ij}^\dagger &=
    \tilde{x}_{ji},
\end{align}
from Eq.~(\ref{eq:x^dagger = xt}).  With these rules, we can construct action terms that are Hermitian. 

A complete understanding of causality is beyond the scope of this work and will require further study of non-local actions such as the ones presented here. We hope to contribute to this understanding in the future. However, we expect this CSM action to be causal because it should agree with Feynman diagrams for every amplitude. This is an ongoing work, but the results so far are promising in this regard.

\subsection{\label{sec:free field}Quadratic Terms}
In constructive amplitudes, the propagator is always an identity on the helicities or spins and any other quantum numbers of the particles coming from each side divided by the Feynman denominator $(p^2-m^2)$ (at tree-level).  Therefore, the quadratic part of the action is the inverse of this with the field's helicity or spins matched and summed over.  So, for the gravitons, gluons, photons, massless quarks, and massless leptons, we have:
\begin{align}
\mathcal{S}_{2m=0} &=
\int\frac{d^4p_1d^4p_2}{(2\pi)^4} \delta^4(p_1+p_2) p_1^2
    \Big[
    G^+(p_1)G^-(p_2)
    \nonumber\\
    &
    +\textsl{g}_a^+(p_1) \textsl{g}_a^-(p_2)
    +\gamma^+(p_1)\gamma^-(p_2)
    \nonumber\\
    &
    +\bar{q}_i^{+}(p_1) q^{i-}(p_2)
    +\bar{q}_i^{-}(p_1) q^{i+}(p_2)
    \nonumber\\
    &
    +\bar{l}^{+}(p_1) l^{-}(p_2)
    +\bar{l}^{-}(p_1) l^{+}(p_2)
    +\bar{\nu}^{+}(p_1) \nu^{-}(p_2)
    \Big].
    \label{eq:S_2m=0}
\end{align}
We have considered massless quarks and charged leptons for illustrative purposes.  We will give them masses shortly.  At times, it is convenient to consider massless first generation quarks and charged leptons.  

We can easily generalize this to massless fields of any helicity between $0$ and $\pm2$.  We can see that this contribution to the action is Hermitian,  as each term is individually Hermitian.  For example, $\left[\bar{q}^{i+}(p_1) q_i^-(p_2)\right]^\dagger=\bar{q}^{i+}(p_2) q_i^-(p_1)$. However, the momentum integral is symmetric over the momenta, and the dummy momenta can be interchanged, returning us to the original term. We also see that the action is Lorentz invariant since each field transforms oppositely under the helicity little group.

For massive fields, the fields have symmetric spin-$\frac{1}{2}$ indices and the mass squared is subtracted from the momentum squared, but otherwise, the action is analogous:  
\begin{align}
    \mathcal{S}_{2m\neq0} &= 
    \int\frac{d^4p_1d^4p_2}{(2\pi)^4} \delta^4(p_1+p_2)\Big[
    \frac{1}{2}\left(p_1^2-m_h^2\right)h(p_1)h(p_2)
    \nonumber\\
    &
     +\frac{1}{2}\left(p_1^2-M_Z^2\right)
     Z_{\mathrm{IJ}}(p_1)\frac{1}{2}\left(\delta_{\mathrm{K}}^{\mathrm{I}}\delta_{\mathrm{L}}^{\mathrm{J}}+\delta_{\mathrm{L}}^{\mathrm{I}}\delta_{\mathrm{K}}^{\mathrm{J}}\right)Z^{\mathrm{KL}}(p_2)
     \nonumber\\
     &
     +\left(p_1^2-M_W^2\right)
     \bar{W}_{\mathrm{IJ}}(p_1)\frac{1}{2}\left(\delta_{\mathrm{K}}^{\mathrm{I}}\delta_{\mathrm{L}}^{\mathrm{J}}+\delta_{\mathrm{L}}^{\mathrm{I}}\delta_{\mathrm{K}}^{\mathrm{J}}\right)W^{\mathrm{IJ}}(p_2)
     \nonumber\\
     &
    +\left(p_1^2-m_q^2\right)\bar{q}_{i\mathrm{I}}(p_1)q^{i\mathrm{I}}(p_2)
        +\left(p_1^2-m_l^2\right)\bar{l}_{\mathrm{I}}(p_1)l^{\mathrm{I}}(p_2)
    \Big],
    \label{eq:S p^2-m^2}
\end{align}
where we have explicitly symmetrized over the spin indices for the $Z$ and $W$ bosons (the antisymmetric spin-index fields are not propagated).  

Once again, we can see that this can be generalized to any spin by increasing the number of symmetrized spin-\(\frac{1}{2}\) indices. Moreover, this action is Hermitian. For example, $\left[\bar{W}_{\mathrm{IJ}}(p_1)W^{\mathrm{IJ}}(p_2)\right]^\dagger = \bar{W}_{\mathrm{IJ}}(p_2)W^{\mathrm{IJ}}(p_1)$ and $\left[\bar{q}_{i\mathrm{I}}(p_1)q^{i\mathrm{I}}(p_2)\right]^\dagger = \bar{q}_{i\mathrm{I}}(p_2)q^{i\mathrm{I}}(p_1)$. This action is also Lorentz invariant, since the spin indices are contracted and the fields transform oppositely under the spin little group. (The little group transformations that keep $p$ unchanged, also keep $-p$ unchanged.)

We can see that, in constructive field theory, all fields have the same mass dimension, whether bosons or fermions of any helicity or spin.  By inspection, the integration, delta function and momentum squared or mass squared contribute $+6$ to the mass dimension; therefore, the mass dimension of the field is always $-3$.  

Furthermore, we see that the quadratic action is always local as all the momenta are polynomial and in the numerator.  See App.~\ref{app:locality} for a discussion of this property.

Another important feature of the quadratic term, which warrants additional emphasis, is that for both massless and massive fields of any helicity or spin, our kinetic and mass terms directly connect the helicity or spin of the field with a delta function. This direct connection results in a propagator that maintains a delta function between helicities or spins throughout all steps of a constructive amplitude calculation. By the end of the calculation, after all on-shell identities have been applied and the amplitude is allowed to go off shell, the helicity spinors and spin spinors for intermediate states are entirely removed, leaving no internal helicities or spins to connect. Crucially, at no point in the process are the helicities or spins of internal lines not connected by a delta function. This is a key property that contributes to the simpler and more efficient amplitude expressions. 

Moreover, it is important to note that in constructive field theory, the quadratic terms -- and consequently the propagators -- are completely and uniquely determined, leaving no room for ambiguity. This is in stark contrast to traditional field theory, where propagators are uniquely defined on-shell as delta functions on the polarizations, but off-shell, there is significant ambiguity \cite{Christensen:2013aua}, especially for fields with spin greater than $1$. In constructive field theory, however, there is only one possible form for the quadratic term and propagator. This form is uniquely determined to consistently connect helicities and spins, without deviation.

This behavior appears to be fundamentally connected to the non-local nature of the interactions. Either, all the propagator numerators are on-shell and connect helicities and spins \textit{and} the interactions are non-local, as in constructive field theory, \textit{or} all the propagator numerators are (usually) off shell and only connect the polarizations when on shell \textit{and} the interactions are local, as in traditional field theory.

\subsection{\label{sec:massless 3-point}The Massless 3-Point Action}
We will begin with the interactions of massless particles, including the gluon, the photon and the neutrino.  We will also include the quarks and charged leptons for illustration and to support treating the first generation as massless.  There are only 3-point vertices for all-massless particles.  We begin with non-graviton interactions.  The action is
\begin{align}
\mathcal{S}&_{3m=0} =
\sqrt{2}\int\frac{d^4p_1d^4p_2d^4p_3}{(2\pi)^8} \delta^4(p_1+p_2+p_3) \Bigg[
    \nonumber\\
    &
    \hspace{-0.1in}\frac{g_s}{2}\textsl{g}_a^+(p_1) f_{abc}
        \left(\frac{\langle12\rangle^3}{\langle23\rangle\langle31\rangle}\textsl{g}_b^+(p_2)
        -\frac{\lbrack23\rbrack^3}{\lbrack12\rbrack\lbrack31\rbrack}\textsl{g}_b^-(p_2)\right)
    \textsl{g}_c^-(p_3)
    \nonumber\\
    &
    -g_s\bar{q}^{i+}(p_1) T_{a i}^{\ j}\left(
    \frac{\langle12\rangle^2}{\langle31\rangle} \textsl{g}_a^+(p_2)
    +\frac{\lbrack23\rbrack^2}{\lbrack31\rbrack} \textsl{g}_a^-(p_2)
    \right)q_j^-(p_3)
    \nonumber\\
    &
    +g_s\bar{q}^{i-}(p_1) T_{ai}^{\ j} \left(   
    \frac{\langle23\rangle^2}{\langle31\rangle} \textsl{g}_a^+(p_2)
    +\frac{\lbrack12\rbrack^2}{\lbrack31\rbrack} \textsl{g}_a^-(p_2)
    \right)q_j^+(p_3)
    \nonumber\\
    &
    -eQ_f\bar{f}^{+}(p_1)\left(
    \frac{\langle12\rangle^2}{\langle31\rangle} \gamma^+(p_2)
    +\frac{\lbrack23\rbrack^2}{\lbrack31\rbrack} \gamma^-(p_2)
    \right)f^{-}(p_3)
    \nonumber\\
    &
    +eQ_f\bar{f}^{-}(p_1) \left(   
    \frac{\langle23\rangle^2}{\langle31\rangle} \gamma^+(p_2)
    +\frac{\lbrack12\rbrack^2}{\lbrack31\rbrack} \gamma^-(p_2)
    \right)f^+(p_3)
    \Bigg]
    \label{eq:S massless 3-point}
\end{align}
where $g_s$ is the strong coupling, $e$ is the charge of the positron, $eQ_f$ is the charge of $f$, the subscripts $a,b$ and $c$ are QCD adjoint indices and $j$ and $i$ are the fundamental and conjugate-fundamental indices.  If $f$ is a quark, then it also has a fundamental index, which is contracted with the anti-fundamental index of the antiquark.  

A natural, immediate question is what relates the couplings of each of these terms. The terms on a single row are related by Hermiticity, and we will show this below. The third and fourth rows are related by the vectorial nature of quantum chromodynamics (QCD), while the fifth and sixth are related by the vectorial nature of the photon. In fact, when the fermions are massive, the two rows for the gluon or the photon come from the same massive vertex \cite{Christensen:2018zcq}, giving us another way of understanding why these are related. However, we do not yet know how to relate the coupling of the triple-gluon term on the second row with the gluon-quark terms on the third and fourth rows without making reference to gauge field theory, with its accompanying unphysical field components. We also cannot relate the coupling between different quarks without reference to gauge field theory. Since we would like to understand this relationship from the ground up, without reference to gauge field theory, we consider this a current deficiency of constructive amplitudes. However, we do note that this action matches the vertices described in \cite{Christensen:2018zcq}, with the relative sign between the two triple-gluon vertices being a possible exception. Nevertheless, since Hermiticity is a required property, we consider this relative sign to be correct and expect this to be confirmed in more complex amplitude calculations.

We can see that this action is Lorentz invariant.  The little group transformation for each field is cancelled by the little group transformation of the helicity spinor with the same momentum.  A consequence of this is that the spinor products might appear to have the opposite helicities of what we might naively expect.  For example, the coefficient for $\textsl{g}^+\textsl{g}^+\textsl{g}^-$ has $\langle12\rangle^3/(\langle23\rangle\langle31\rangle)$ rather than the $\lbrack12\rbrack^3/(\lbrack23\rbrack\lbrack31\rbrack)$ that we might expect for a vertex with two positive and one negative helicity gluon.  However, as we show in App.~\ref{app:constructive rules}, the three-point amplitude $\mathcal{M}^{++-}$ actually pulls out $-\lbrack12\rbrack^3/(\lbrack23\rbrack\lbrack31\rbrack)$ from the second term with two negative helicity and one positive helicity gluon fields.  So, the helicity structure of these action terms does give the correct constructive vertices.

We can see that the action is dimensionless by considering the mass dimension of the fields and spinor products.  This action has three momentum integrals and a momentum-conserving delta function, contributing +8 to the mass dimension.  Each term has three fields, contributing -9 to the mass dimension and each term has one more spinor product in the numerator than in the denominator, contributing +1 to the mass dimension.  All together, therefore, this action is dimensionless.  The mass dimension of the spinor products is discussed in App.~\ref{app:spinor products}.

We also note that these vertices are non-local.  The presence of spinor products in the vertex leads to non-polynomial momenta, which lead to non-local terms in position space.  Moreover, for these action terms, we can rewrite the vertex in a way that the spinor products only occur in the numerator.  When we do this, we will be left with simple momentum products in the denominator for every term.  For example, if we consider the photon-lepton vertex $\langle12\rangle^2/\langle31\rangle$, we can multiply the numerator and denominator by $\lbrack13\rbrack$, giving $\langle12\rangle^2\lbrack13\rbrack/(2p_1\cdot p_3)$.  The presence of momenta in the denominator is also a sign of non-locality.   See App.~\ref{app:locality} for further discussion.

If we take the Hermitian conjugate of our massless interaction action, we have:
\begin{align}
\mathcal{S}&_{3m=0}^\dagger =
\sqrt{2}\int\frac{d^4p_1d^4p_2d^4p_3}{(2\pi)^8} \delta^4(p_1+p_2+p_3) \Bigg[
    \nonumber\\
    &
    \hspace{-0.1in}\frac{g_s}{2}\textsl{g}_c^+(p_3) f_{abc}
        \left(
        -\frac{\langle32\rangle^3}{\langle21\rangle\langle13\rangle}\textsl{g}_b^+(p_2)
        +\frac{\lbrack21\rbrack^3}{\lbrack32\rbrack\lbrack13\rbrack}\textsl{g}_b^-(p_2)\right)
    \textsl{g}_a^-(p_1)
    \nonumber\\
    &
    -g_s\bar{q}^{j+}(p_3) T_{a j}^{\ i}\left(
    \frac{\langle32\rangle^2}{\langle13\rangle} \textsl{g}_a^+(p_2)
    +\frac{\lbrack21\rbrack^2}{\lbrack13\rbrack} \textsl{g}_a^-(p_2)
    \right)q_i^-(p_1)
    \nonumber\\
    &
    +g_s\bar{q}^{j-}(p_3) T_{aj}^{\ i} \left(  
    \frac{\langle21\rangle^2}{\langle13\rangle} \textsl{g}_a^+(p_2)
    +\frac{\lbrack32\rbrack^2}{\lbrack13\rbrack} \textsl{g}_a^-(p_2)
    \right)q_i^+(p_1)
    \nonumber\\
    &
    -eQ_f\bar{f}^{+}(p_3)\left(
    \frac{\langle32\rangle^2}{\langle13\rangle} \gamma^+(p_2)
    +\frac{\lbrack21\rbrack^2}{\lbrack13\rbrack} \gamma^-(p_2)
    \right)f^{-}(p_1)
    \nonumber\\
    &
    +eQ_f\bar{f}^{-}(p_3) \left(   
    \frac{\langle21\rangle^2}{\langle13\rangle} \gamma^+(p_2)
    +\frac{\lbrack32\rbrack^2}{\lbrack13\rbrack} \gamma^-(p_2)
    \right)f^+(p_1)
    \Bigg],
\end{align}
where we have reversed the order of the fields as part of the Hermitian conjugation.  We have also interchanged the two terms on each row, for convenience.  
We next make the replacement $p_1\longleftrightarrow p_3$ everywhere as well as $a\longleftrightarrow c$ on the second row, and $i\longleftrightarrow j$ on the third and fourth rows, to obtain Eq.~(\ref{eq:S massless 3-point}), 
where we have also used the antisymmetry of $f_{abc}$ to put the color-adjoint indices back in cyclic order.

We will also consider the gravitational action, focusing on the massless fields in this subsection.  All gravitational terms are divided by the Planck mass $M_P$.  We have
\begin{align}
    \mathcal{S}&_{Gm=0} =
    \frac{1}{M_P}\int\frac{d^4p_1d^4p_2d^4p_3}{(2\pi)^8} \delta^4(p_1+p_2+p_3) \Bigg[
    \nonumber\\
    &
    \frac{1}{2}G^+(p_1)\left(\frac{\langle12\rangle^6}{\langle23\rangle^2\langle31\rangle^2}G^+(p_2)
    +\frac{\lbrack23\rbrack^6}{\lbrack12\rbrack^2\lbrack31\rbrack^2}G^-(p_2)\right)
    G^-(p_3)
    \nonumber\\
    &
    +\textsl{g}_a^+(p_1)\left(
    \frac{\langle12\rangle^4}{\langle31\rangle^2} G^+(p_2)
    +\frac{\lbrack23\rbrack^4}{\lbrack31\rbrack^2} G^-(p_2)
    \right)\textsl{g}_a^-(p_3)
    \nonumber\\
    &
    +\gamma^+(p_1)\left(
    \frac{\langle12\rangle^4}{\langle31\rangle^2} G^+(p_2)
    +\frac{\lbrack23\rbrack^4}{\lbrack31\rbrack^2} G^-(p_2)
    \right)\gamma^-(p_3)
    \nonumber\\
    &
    -\bar{f}^{+}(p_1)\left(
    \frac{\langle12\rangle^3\langle23\rangle}{\langle31\rangle^2} G^+(p_2)
    +\frac{\lbrack23\rbrack^3\lbrack12\rbrack}{\lbrack31\rbrack^2} G^-(p_2)
    \right)f^{-}(p_3)
    \nonumber\\
    &
    +\bar{f}^{-}(p_1) \left(   
    \frac{\langle23\rangle^3\langle12\rangle}{\langle31\rangle^2} G^+(p_2)
    +\frac{\lbrack12\rbrack^3\lbrack23\rbrack}{\lbrack31\rbrack^2} G^-(p_2)
    \right)f^+(p_3)
    \nonumber\\
    &
    +\bar{\nu}^{-}(p_1)\left(
    \frac{\langle23\rangle^3\langle12\rangle}{\langle31\rangle^2} G^+(p_2)
    +\frac{\lbrack12\rbrack^3\lbrack23\rbrack}{\lbrack31\rbrack^2} G^-(p_2)
    \right)\nu^{+}(p_3)
    \Bigg].
    \label{eq:S_Gm=0}
\end{align}
This action can be seen to be Hermitian in the same way as $\mathcal{S}_{3m=0}$.  Under Hermitian conjugation, the terms in each row are transformed into each other and each row is Hermitian independent of the others.  Also, similarly to $\mathcal{S}_{3m=0}$, we can understand the relationship of the couplings for the fifth and sixth rows as being due to the ``vectorial'' nature of the graviton, or the fact that they come from massive vertices that contribute to both rows \cite{Christensen:2018zcq}.  However, otherwise, we cannot explain the fact that the couplings on all the rows are the same, without reference to a diffeomorphism-invariant action, with its accompanying unphysical graviton-field components.  Once again, we consider this a current deficiency of constructive amplitudes.

This action is Lorentz invariant for the same reason as the previous one.  The helicity transformation of each field is cancelled by the helicity transformation of the spinors with the same momentum.  Moreover, all the action terms of this section are non-local.  We can see this by the presence of spinor products in the expression.  Finally, this action is dimensionless.  The momentum integrals and delta function contribute +8 to the mass dimension, the fields contribute -9, the spinor products contribute +2, and the entire action is divided by the Planck mass, contributing -1.  All together, the mass dimension is, therefore, zero.

The perturbation theory resulting from the field theory in this subsection will result superficially in all the diagrams (excluding those with four-point vertices) present in Feynman diagrams.  This may seem like a problem for massless amplitudes since their spinor-product structure can come from a single diagram.  Although this is true, each diagram does give important information about the propagator denominators and the associated quantum numbers such as color and electric charge.  This is true even in the massive theory as we saw in four-point amplitudes including $gggg$, $q\bar{q}gg$ and $f\bar{f}\gamma W$ \cite{Christensen:2024bdt}.

\subsection{\label{sec:gluons, photons and gravitons with massive fields}Gluons, Photons and Gravitons with Massive Fields}
When the fields that interact with the gluons, photons and gravitons are massive, we must add spin indices, use massive spinors and introduce the $x$ factor.  For the gluon and photon, we have:
\begin{align}
\mathcal{S}&_{(g/\gamma)ff} =
\sqrt{2}\int\frac{d^4p_1d^4p_2d^4p_3}{(2\pi)^8} \delta^4(p_1+p_2+p_3) \Bigg[
    \nonumber\\
    &
    +g_s\bar{q}^{i}_{\mathrm{I}}(p_1) T_{a i}^{\ j}\Big(
    \tilde{x}_{31} \lbrack\mathbf{31}\rbrack^{\mathrm{JI}} \textsl{g}_a^+(p_2)
    \nonumber\\
    &\hspace{1.5in}
    +x_{31} \langle\mathbf{31}\rangle^{\mathrm{JI}} \textsl{g}_a^-(p_2)
    \Big)q_{j\mathrm{J}}(p_3)
    \nonumber\\
    &
    +eQ_f\bar{f}_{\mathrm{I}}(p_1)\Big(
    \tilde{x}_{31} \lbrack\mathbf{31}\rbrack^{\mathrm{JI}} \gamma^+(p_2)
    \nonumber\\
    &\hspace{1.5in}
    +x_{31} \langle\mathbf{31}\rangle^{\mathrm{JI}} \gamma^-(p_2)
    \Big)f_{\mathrm{J}}(p_3)
    \nonumber\\
    &
    +\frac{e}{M_W}\bar{W}_{\mathrm{IJ}}(p_1)\Big(
    \tilde{x}_{31} \lbrack\mathbf{31}\rbrack^{\mathrm{KI}} \lbrack\mathbf{31}\rbrack^{\mathrm{LJ}}
    \gamma^+(p_2)
    \nonumber\\
    &\hspace{1in}
    +x_{31} \langle\mathbf{31}\rangle^{\mathrm{KI}} \langle\mathbf{31}\rangle^{\mathrm{LJ}} \gamma^-(p_2)
    \Big)W_{\mathrm{KL}}(p_3)
    \Bigg],
    \label{eq:S_gAff}
\end{align}
where the spin indices on the spin-1 boson fields are assumed symmetrized, as always.  When doing perturbation theory, it is not necessary to explicitly symmetrize them, since the symmetrized propagator effectively symmetrizes the vertices.  However, on the lattice, the spin indices must be explicitly symmetrized, in order to get correct results.  This is necessary to remove the antisymmetric, spin-0 part of the fields from contributing.  The high-energy limit of the fermion terms in $\mathcal{S}_{(g/\gamma)ff}$ gives the fermion terms in the action in Eq.~(\ref{eq:S massless 3-point}) \cite{Christensen:2018zcq}.

We can see that this action is Hermitian by taking the Hermitian conjugate, giving:
\begin{align}
\mathcal{S}&_{(g/\gamma)ff}^{\dagger} =
\sqrt{2}\int\frac{d^4p_1d^4p_2d^4p_3}{(2\pi)^8} \delta^4(p_1+p_2+p_3) \Bigg[
    \nonumber\\
    &
    +g_s \bar{q}^{j\mathrm{J}}(p_3) T_{a j}^{\ i}\Big(
    \tilde{x}_{13} \lbrack\mathbf{13}\rbrack_{\mathrm{IJ}} \textsl{g}_a^+(p_2)
    \nonumber\\
    &\hspace{1.5in}
    +x_{13} \langle\mathbf{13}\rangle_{\mathrm{IJ}} \textsl{g}_a^-(p_2)
    \Big)q_{i}^{\mathrm{I}}(p_1)
    \nonumber\\
    &
    +e Q_f \bar{f}^{\mathrm{J}}(p_3) \Big(
    \tilde{x}_{13} \lbrack\mathbf{13}\rbrack_{\mathrm{IJ}} \gamma^+(p_2)
    \nonumber\\
    &\hspace{1.5in}
    +x_{13} \langle\mathbf{13}\rangle_{\mathrm{IJ}} \gamma^-(p_2)
    \Big)f^{\mathrm{I}}(p_1)
    \nonumber\\
    &
    +\frac{e}{M_W}\bar{W}^{\mathrm{LK}}(p_3)\Big(
    \tilde{x}_{13} \lbrack\mathbf{13}\rbrack_{\mathrm{IK}} \lbrack\mathbf{13}\rbrack_{\mathrm{JL}} \gamma^+(p_2)
    \nonumber\\
    &\hspace{1in}
    +x_{13} \langle\mathbf{13}\rangle_{\mathrm{IK}} \langle\mathbf{13}\rangle_{\mathrm{JL}}
    \gamma^-(p_2)
    \Big)W^{\mathrm{JI}}(p_1)
    \Bigg],
\end{align}
where we have interchanged the order of the terms on each row.  We now make the label changes $1\longleftrightarrow3$ and $i\longrightarrow j$.  On the second and third rows, we make the change $\mathrm{I}\longleftrightarrow\mathrm{J}$, while on the fourth row, we make the change $\mathrm{I}\longleftrightarrow\mathrm{L}$ and $\mathrm{J}\longleftrightarrow\mathrm{K}$.  We also use the fact that raising and lowering spin indices results in a sign change ($e.g. A^{\mathrm{I}}B_{\mathrm{I}}=-A_{\mathrm{I}}B^{\mathrm{I}}$), but we have an even number of contracted spin indices in each row.  Therefore, we can raise and lower all the spin indices with no sign change.  After all this, we have returned back to the original action in Eq.~(\ref{eq:S_gAff}). 

This action is Lorentz invariant.  The helicity transformation of the gluons and photons is cancelled by the helicity transformation of $x$ and $\tilde{x}$, while the spin transformations of the massive fields is cancelled by the spin transformation of the massive spin spinors of the same momenta.  This action is not local.  We can see this by noting the presence of $x$, $\tilde{x}$ and the spinor products.  This action is dimensionless.  The mass dimension of the momentum integrals and the delta function is +8, the fields contribute -9, the $x$ contributes 0, and the spinor products and mass give +1, for a total of zero.  The couplings for each Hermitian pair are related by Hermiticity, of course, but they are otherwise not related to other couplings, within the current constructive framework.

The contribution to the action for the graviton interacting with massive fields is
\begin{align}
    \mathcal{S}&_{Gm\neq0} =
    \frac{1}{M_P}\int\frac{d^4p_1d^4p_2d^4p_3}{(2\pi)^8} \delta^4(p_1+p_2+p_3) \Bigg[
    \nonumber\\
    &
    +m_h^2 h(p_1) \Big(   
    \tilde{x}_{31}^2 G^+(p_2)
    +x_{31}^2 G^-(p_2)
    \Big) h(p_3)
    \nonumber\\
    &
    +m_f\bar{f}_{\mathrm{I}}(p_1) \Big(   
    \tilde{x}_{31}^2 \lbrack\mathbf{31}\rbrack^{\mathrm{JI}} G^+(p_2)
    \nonumber\\
    &\hspace{1in}
    +x_{31}^2 \langle\mathbf{31}\rangle^{\mathrm{JI}} G^-(p_2)
    \Big)f_{\mathrm{J}}(p_3)
    \nonumber\\
    &
    +\bar{W}_{\mathrm{IJ}}(p_1)\Big(
    \tilde{x}_{31}^2 \lbrack\mathbf{31}\rbrack^{\mathrm{KI}} \lbrack\mathbf{31}\rbrack^{\mathrm{LJ}}
    G^+(p_2)
    \nonumber\\
    &\hspace{1in}
    +x_{31}^2 \langle\mathbf{31}\rangle^{\mathrm{KI}} \langle\mathbf{31}\rangle^{\mathrm{LJ}} G^-(p_2)
    \Big)W_{\mathrm{KL}}(p_3)
    \nonumber\\
    &
    +\frac{1}{2}Z_{\mathrm{IJ}}(p_1)\Big(
    \tilde{x}_{31}^2 \lbrack\mathbf{31}\rbrack^{\mathrm{KI}} \lbrack\mathbf{31}\rbrack^{\mathrm{LJ}}
    G^+(p_2)
    \nonumber\\
    &\hspace{1in}
    +x_{31}^2 \langle\mathbf{31}\rangle^{\mathrm{KI}} \langle\mathbf{31}\rangle^{\mathrm{LJ}} G^-(p_2)
    \Big)Z_{\mathrm{KL}}(p_3)
    \Bigg].
    \label{eq:S_Gm neq0}
\end{align}
The Hermitian conjugate of this action is
\begin{align}
    \mathcal{S}&_{Gm\neq0}^{\dagger} =
    \frac{1}{M_P}\int\frac{d^4p_1d^4p_2d^4p_3}{(2\pi)^8} \delta^4(p_1+p_2+p_3) \Bigg[
    \nonumber\\
    &
    +m_h^2 h(p_3) \Big(   
    \tilde{x}_{13}^2 G^+(p_2)
    +x_{13}^2 G^-(p_2)
    \Big) h(p_1)
    \nonumber\\
    &
    +m_f\bar{f}^{\mathrm{J}}(p_3) \Big(   
    \tilde{x}_{13}^2 \lbrack\mathbf{13}\rbrack_{\mathrm{IJ}} G^+(p_2)
    \nonumber\\
    &\hspace{1in}
    +x_{13}^2 \langle\mathbf{13}\rangle^{\mathrm{IJ}} G^-(p_2)
    \Big)f^{\mathrm{I}}(p_1)
    \nonumber\\
    &
    +\bar{W}^{\mathrm{LK}}(p_3)\Big(
    \tilde{x}_{13}^2 \lbrack\mathbf{13}\rbrack_{\mathrm{IK}} \lbrack\mathbf{13}\rbrack_{\mathrm{JL}}
    G^+(p_2)
    \nonumber\\
    &\hspace{1in}
    +x_{13}^2 \langle\mathbf{13}\rangle_{\mathrm{IK}} \langle\mathbf{13}\rangle_{\mathrm{JL}} G^-(p_2)
    \Big)W^{\mathrm{JI}}(p_1)
    \nonumber\\
    &
    +\frac{1}{2}Z^{\mathrm{LK}}(p_3)\Big(
    \tilde{x}_{13}^2 \lbrack\mathbf{13}\rbrack_{\mathrm{IK}} \lbrack\mathbf{13}\rbrack_{\mathrm{JL}}
    G^+(p_2)
    \nonumber\\
    &\hspace{1in}
    +x_{13}^2 \langle\mathbf{13}\rangle_{\mathrm{IK}} \langle\mathbf{13}\rangle_{\mathrm{JL}} G^-(p_2)
    \Big)Z^{\mathrm{JI}}(p_1)
    \Bigg].
\end{align}
After switching $1\longleftrightarrow3$ throughout, 
$\mathrm{I}\longleftrightarrow\mathrm{J}$ on the third and fourth rows, 
$\mathrm{I}\longleftrightarrow\mathrm{L}$ and 
$\mathrm{J}\longleftrightarrow\mathrm{K}$ 
on the fifth through eighth rows, and raising and lowering the spin indices, we return to Eq.~(\ref{eq:S_Gm neq0}), showing this action is Hermitian.

We also see this action contribution is Lorentz invariant.  All the spin indices are contracted appropriately and the helicity transformation of the graviton is cancelled by the helicity transformation of $x$ and $\tilde{x}$.  It is non-local due to the presence of spinor products and $x$ and $\tilde{x}$.  It is dimensionless.  The momentum integrals and delta function contribute +8, the fields contribute -9, $x$ and $\tilde{x}$ contribute 0 and the Planck scale contributes -1, for a total of -2 before including the masses and spinor products.  The Higgs line contains a mass squared making its line dimensionless.  The fermion line has a linear mass and one spinor product resulting in dimensionless terms.  The $W$ and $Z$ lines have two spinor products, leaving their lines dimensionless as well.  We also see that the high-energy limit of the terms results in the action of Eq.~(\ref{eq:S_Gm=0}) \cite{Christensen:2018zcq}.  The couplings are only related within Hermitian pairs.  Otherwise, constructive theory does not yet relate them to each other.

\subsection{\label{sec:higgs and weak action}The Higgs and Weak Action}

We begin with the Higgs self interactions.
\begin{align}
    \mathcal{S}_{h} =& 
    -\frac{e m_h^2}{2M_W s_W}\int\frac{d^4p_1d^4p_2d^4p_3}{(2\pi)^{8}} 
    \nonumber\\
    &\hspace{0.5in}h(p_1)h(p_2)h(p_3)\delta^4(p_1+p_2+p_3)
    \nonumber\\
    &-\frac{e^2m_h^2}{16M_W^2s_W^2}\int\frac{d^4p_1d^4p_2d^4p_3d^4p_4}{(2\pi)^{12}} 
    \nonumber\\
    &\hspace{0.5in} h(p_1)h(p_2)h(p_3)h(p_4) \delta^4(p_1+p_2+p_3+p_4).
    \label{eq:S_h}
\end{align}
This action is trivially Hermitian since the Higgs field is real.  It is Lorentz invariant since the Higgs field is a singlet under the spin little group.  It is also local since there are no momenta in the vertex in any form.  This is the only local interaction in the CSM.  It is dimensionless.  On the first integral, the momentum integrals and delta function contribute +8, the fields -9 and the masses +1.  On the second integral, the contributions are +12 for the momentum integrals and delta function and -12 for the fields.  In fact, this action is exactly equal to the Higgs self-interaction action in traditional field theory.  As a consequence, this is the only action terms where the couplings could currently be related to each other, by writing the Higgs field as an electroweak doublet with a symmetry-breaking potential that leads to these terms.  

The interaction between the Higgs and the other massive fields is given by
\begin{align}
    \mathcal{S}_{hf/b} =& 
    -\frac{e}{M_Ws_W} \int\frac{d^4p_1d^4p_2d^4p_3}{(2\pi)^{8}} \delta^4(p_1+p_2+p_3) \Bigg[
    \nonumber\\
    &\frac{m_f}{2}h(p_1) 
    \bar{f}_{\mathrm{I}}(p_2)\left(\langle\mathbf{32}\rangle^{\mathrm{JI}}  + \lbrack\mathbf{32}\rbrack^{\mathrm{JI}}\right)f_{\mathrm{J}}(p_3)
    \nonumber\\
    &+h(p_1)
    \langle\mathbf{23}\rangle^{\mathrm{IK}} \lbrack\mathbf{23}\rbrack^{\mathrm{JL}}
    \Big(\bar{W}_{\mathrm{IJ}}(p_2)W_{\mathrm{KL}}(p_3)
    \nonumber\\
    &\hspace{1in}
    +\frac{1}{2} Z_{\mathrm{IJ}}(p_2)Z_{\mathrm{KL}}(p_3)\Big)\Bigg]
    \nonumber\\
    &\hspace{-0.3in}+\frac{e^2}{2M_W^2s_W^2} \int\frac{d^4p_1d^4p_2d^4p_3d^4p_4}{(2\pi)^{12}} \delta^4(p_1+p_2+p_3+p_4)
    \nonumber\\
    &\frac{1}{2}h(p_1)h(p_2)  \langle\mathbf{34}\rangle^{\mathrm{IK}} \lbrack\mathbf{34}\rbrack^{\mathrm{JL}}
    \Big(\bar{W}_{\mathrm{IJ}}(p_3)W_{\mathrm{KL}}(p_4)
    \nonumber\\
    &\hspace{1in}
    +\frac{1}{2} Z_{\mathrm{IJ}}(p_3)Z_{\mathrm{KL}}(p_4)\Big).
    \label{eq:S_hbf}
\end{align}
The Hermitian conjugate of this action is
\begin{align}
    \mathcal{S}_{hf/b}^{\dagger} =& 
    -\frac{e}{M_Ws_W} \int\frac{d^4p_1d^4p_2d^4p_3}{(2\pi)^{8}} \delta^4(p_1+p_2+p_3) \Bigg[
    \nonumber\\
    &\frac{m_f}{2}h(p_1) 
    \bar{f}^{\mathrm{J}}(p_3) \left(\langle\mathbf{23}\rangle_{\mathrm{IJ}}  + \lbrack\mathbf{23}\rbrack_{\mathrm{IJ}}\right) f^{\mathrm{I}}(p_2)
    \nonumber\\
    &+h(p_1)
    \langle\mathbf{32}\rangle_{\mathrm{LJ}} \lbrack\mathbf{32}\rbrack_{\mathrm{KI}} 
    \Big( \bar{W}^{\mathrm{LK}}(p_3) W^{\mathrm{JI}}(p_2)
    \nonumber\\
    &\hspace{1in}
    +\frac{1}{2} Z_{\mathrm{LK}}(p_3) Z_{\mathrm{JI}}(p_2) \Big)\Bigg]
    \nonumber\\
    &\hspace{-0.3in}+\frac{e^2}{2M_W^2s_W^2} \int\frac{d^4p_1d^4p_2d^4p_3d^4p_4}{(2\pi)^{12}} \delta^4(p_1+p_2+p_3+p_4)
    \nonumber\\
    &\frac{1}{2}h(p_1)h(p_2)  \langle\mathbf{43}\rangle_{\mathrm{LJ}} \lbrack\mathbf{43}\rbrack_{\mathrm{KI}} 
    \Big( \bar{W}^{\mathrm{LK}}(p_4) W^{\mathrm{JI}}(p_3)
    \nonumber\\
    &\hspace{1in}
    +\frac{1}{2} Z^{\mathrm{LK}}(p_4) Z^{\mathrm{JI}}(p_3) \Big).
\end{align}
For the first integral, we make the replacement $2\longleftrightarrow3$, while we make the replacement $3\longleftrightarrow4$ on the second integral.  On the fermion terms, we interchange $\mathrm{I}\longleftrightarrow\mathrm{J}$, while on the weak boson terms, we interchange $\mathrm{I}\longleftrightarrow\mathrm{L}$ and $\mathrm{J}\longleftrightarrow\mathrm{K}$.  We also lower and raise all the contracted spin indices.  Since there are an even number of them per term, there is no sign change.  This brings us back to Eq.~(\ref{eq:S_hbf}), so this action is Hermitian.  

Since the spin transformation of the fields is cancelled by the spin transformation of the spinors, this action is Lorentz invariant.  This action is generally non-local due to the presence of spinor products.  The total mass dimension is zero.  For the first integral, the momentum integrals and delta function contribute +8, the fields -9 and $M_W$ -1.  Addtionally, the fermion line has a mass and a spinor product, cancelling the other contributions.  The weak boson lines contain two spinor products, bringing the total to zero.  On the second integral, the momentum integrals and delta function contribute +12, the fields contribute -12, $M_W$ contributes -2 and the two spinor products contribute +2, for a total of zero.  

We do not explicitly symmetrize the spin indices on the $W$ and $Z$ bosons here because in perturbation theory, the symmetrization of the indices on the propagator are sufficient to induce a symmetrization of the interactions.  However, when non-perturbative calculations are performed, on the lattice for example, then these interaction terms must be explicitly symmetrized.  

The interaction between the weak bosons and the fermions is given by
\begin{align}
    \mathcal{S}_{(W/Z)f} &=
    \frac{e}{M_W s_W}
    \int\frac{d^4p_1d^4p_2d^4p_3}{(2\pi)^{8}} \delta^4(p_1+p_2+p_3)\Bigg[
    \nonumber\\
    &
    \bar{\nu}^+(p_1)l_{\mathrm{I}}(p_2)W_{\mathrm{KL}}(p_3)
    \lbrack\mathbf{32}\rbrack^{\mathrm{KI}} \langle1\mathbf{3}\rangle^{\mathrm{L}}
    \nonumber\\
    &\hspace{0.5in}
    -
    \bar{l}_{\mathrm{I}}(p_1)\nu^-(p_2)\bar{W}_{\mathrm{KL}}(p_3)
    \langle\mathbf{13}\rangle^{\mathrm{IL}} \lbrack\mathbf{3}2\rbrack^{\mathrm{K}}
    \nonumber\\
    &+
    \bar{u}_{\mathrm{I}}(p_1)d_{\mathrm{J}}(p_2) W_{\mathrm{KL}}(p_3)
    \langle\mathbf{23}\rangle^{\mathrm{JL}} \lbrack\mathbf{31}\rbrack^{\mathrm{KI}}
    \nonumber\\
    &\hspace{0.5in}
    +\bar{d}_{\mathrm{I}}(p_1)u_{\mathrm{J}}(p_2)\bar{W}_{\mathrm{KL}}(p_3) 
    \lbrack\mathbf{31}\rbrack^{\mathrm{KI}} \langle\mathbf{23}\rangle^{\mathrm{JL}}
    \nonumber\\
    &
    -\frac{1}{\sqrt{2}}\bar{f}_{\mathrm{I}}(p_1) f_{\mathrm{J}}(p_2) Z_{\mathrm{KL}}(p_3) 
    \nonumber\\
    &\hspace{0.5in}
    \big( g_{Lf}\langle\mathbf{23}\rangle^{\mathrm{JK}} \lbrack\mathbf{31}\rbrack^{\mathrm{LI}}
    +g_{Rf}\lbrack\mathbf{23}\rbrack ^{\mathrm{JK}}\langle\mathbf{31}\rangle^{\mathrm{LI}}
    \big)
    \Bigg].
    \label{eq:S_W/Zf}
\end{align}
This action contribution is our first to be chiral.  We can see this because it is not symmetric between angle and square spinors.  As discussed in the introduction to this section and in App.~\ref{app:spinors}, the angle spinors transform under the left-chiral part of the Lorentz group, while the square spinors transform under the right-chiral part.  (The Lorentz transformations are cancelled in the spinor products between the spinors, but the chirality of the interactions is still a consequence.)  At first, naively, this appears to give the opposite chirality for the interactions.  However, as we show in detail in App.~\ref{app:W-nu vertex chirality}, this does give the correct chirality for vertices.  For example, for the negative-helicity neutrino, anti-lepton, $\bar{W}$ boson amplitude it gives $\lbrack\mathbf{32}\rbrack^{\mathrm{JI}}\langle1\mathbf{3}\rangle^{\mathrm{K}}$, where the negative-helicity neutrino is particle $1$ and represented by $\langle1\rvert$, which transforms under the left-chiral group and the charged anti-lepton is represented by $\lvert\mathbf{2}\rbrack^{\mathrm{I}}$, which transforms under the right-chiral group.  The chirality of the vertex and the amplitude are determined by the interaction with the opposite-helicity fields.  We also saw this in the case of the triple-gluon vertex, as discussed in detail in App.~\ref{app:ggg vertex helicity}.  This is a general property of the constructive action.  The quantum numbers of the fields are the opposite of those in the vertex and amplitudes.  This is necessary to cancel the symmetry transformations and obtain a symmetry-invariant action.  However, due to the way the vertices come from the action, they are correct.

The relative sign on the neutrino term is required to satisfy Hermiticity.  Moreover, the massless limit of the up-quark interaction would produce similarly signed terms, as can be seen in Ref.~\cite{Christensen:2018zcq}.
The Hermitian conjugate is
\begin{align}
    \mathcal{S}_{(W/Z)f}^{\dagger} &=
    \frac{e}{M_W s_W}
    \int\frac{d^4p_1d^4p_2d^4p_3}{(2\pi)^{8}} \delta^4(p_1+p_2+p_3)\Bigg[
    \nonumber\\
    &
    -
    \bar{\nu}^+(p_2) l^{\mathrm{I}}(p_1) W^{\mathrm{LK}}(p_3)
    \lbrack\mathbf{31}\rbrack_{\mathrm{LI}} \langle2\mathbf{3}\rangle_{\mathrm{K}}
    \nonumber\\
    &\hspace{0.5in}
    +\bar{l}^{\mathrm{I}}(p_2) \nu^-(p_1) \bar{W}^{\mathrm{LK}}(p_3)
    \langle\mathbf{23}\rangle_{\mathrm{IK}} \lbrack\mathbf{3}1\rbrack_{\mathrm{L}}
    \nonumber\\
    &
    +\bar{u}^{\mathrm{J}}(p_2) d^{\mathrm{I}}(p_1) W^{\mathrm{LK}}(p_3) 
    \langle\mathbf{13}\rangle_{\mathrm{IK}} \lbrack\mathbf{32}\rbrack_{\mathrm{LJ}}
    \nonumber\\
    &\hspace{0.5in}+
    \bar{d}^{\mathrm{J}}(p_2) u^{\mathrm{I}}(p_1) \bar{W}^{\mathrm{LK}}(p_3)
    \lbrack\mathbf{32}\rbrack_{\mathrm{LJ}} \langle\mathbf{13}\rangle_{\mathrm{IK}}
    \nonumber\\
    &
    -\frac{1}{\sqrt{2}} \bar{f}^{\mathrm{J}}(p_2) f^{\mathrm{I}}(p_1)  Z^{\mathrm{LK}}(p_3) 
    \nonumber\\
    &\hspace{0.5in}
    \big( g_{Lf}\lbrack\mathbf{32}\rbrack_{\mathrm{KJ}} \langle\mathbf{13}\rangle_{\mathrm{IL}}
    +g_{Rf}\langle\mathbf{32}\rangle_{\mathrm{KJ}}\lbrack\mathbf{13}\rbrack_{\mathrm{IL}}
    \big)
    \Bigg],
\end{align}
where we have interchanged the second and third rows and the fourth and fifth rows.
We next switch $1\longleftrightarrow2$, $\mathrm{I}\longleftrightarrow\mathrm{J}$ and $\mathrm{K}\longleftrightarrow\mathrm{L}$.  After this, we raise and lower the spin indices.  This will result in a minus sign on the $W$-lepton terms since there are an odd number of spin indices, but there won't be a sign change on the other terms since there are an even number of spin indices in those cases.  With these changes, the Hermitian conjugate comes into agreement with Eq.~(\ref{eq:S_W/Zf}), so this action is Hermitian.

This action is Lorentz invariant as the helicity of the neutrino fields cancels with the helicity of the helicity spinors of the same momentum and all the spin indices are contracted among objects with the same momentum.  This action is non-local because of the spinor products.  The mass dimension is zero, as usual because the momentum integrals and delta function contribute +8, the fields -9, the spinor products +2 and $M_W$ -1, for a total of zero.

The action for the triple-vector-boson interaction is given by
\begin{align}
    \mathcal{S}_{ZWW} &=
    \frac{-e}{\sqrt{2}M_Z s_W}
    \int\frac{d^4p_1d^4p_2d^4p_3}{(2\pi)^{8}} \delta^4(p_1+p_2+p_3)
    \nonumber\\
    &
    W_{\mathrm{IJ}}(p_1)\bar{W}_{\mathrm{KL}}(p_2)Z_{\mathrm{MN}}(p_3)\Bigg[
    \nonumber\\
    &
    \frac{
    \langle\mathbf{12}\rangle^{\mathrm{IK}} \langle\mathbf{23}\rangle^{\mathrm{LM}} \lbrack\mathbf{31}\rbrack^{\mathrm{NJ}} +\lbrack\mathbf{12}\rbrack^{\mathrm{IK}} \lbrack\mathbf{23}\rbrack^{\mathrm{LM}} \langle\mathbf{31}\rangle^{\mathrm{NJ}}
    }{M_Z}
    \nonumber\\
    &+\frac{
    \langle\mathbf{12}\rangle^{\mathrm{IK}} \lbrack\mathbf{23}\rbrack^{\mathrm{LM}} \langle\mathbf{31}\rangle^{\mathrm{NJ}} +\lbrack\mathbf{12}\rbrack^{\mathrm{IK}} \langle\mathbf{23}\rangle^{\mathrm{LM}} \lbrack\mathbf{31}\rbrack^{\mathrm{NJ}}
    }{M_Z}
    \nonumber\\
    &+\frac{
    \lbrack\mathbf{12}\rbrack^{\mathrm{IK}} \langle\mathbf{23}\rangle^{\mathrm{LM}} \langle\mathbf{31}\rangle^{\mathrm{NJ}} +\langle\mathbf{12}\rangle^{\mathrm{IK}} \lbrack\mathbf{23}\rbrack^{\mathrm{LM}} \lbrack\mathbf{31}\rbrack^{\mathrm{NJ}}
    }{M_W}
    \Bigg].
    \label{eq:S_ZWW}
\end{align}
The Hermitian conjugate of this action is
\begin{align}
    \mathcal{S}_{ZWW}^{\dagger} &=
    \frac{-e}{\sqrt{2}M_Z s_W}
    \int\frac{d^4p_1d^4p_2d^4p_3}{(2\pi)^{8}} \delta^4(p_1+p_2+p_3)
    \nonumber\\
    &
    W^{\mathrm{LK}}(p_2)\bar{W}^{\mathrm{JI}}(p_1)Z^{\mathrm{NM}}(p_3)\Bigg[
    \nonumber\\
    &\frac{
    \langle\mathbf{21}\rangle_{\mathrm{KI}} \lbrack\mathbf{32}\rbrack_{\mathrm{ML}} \langle\mathbf{13}\rangle_{\mathrm{JN}} +\lbrack\mathbf{21}\rbrack_{\mathrm{KI}} \langle\mathbf{32}\rangle_{\mathrm{ML}} \lbrack\mathbf{13}\rbrack_{\mathrm{JN}}
    }{M_Z}
    \nonumber\\
    &+
    \frac{
    \langle\mathbf{21}\rangle_{\mathrm{KI}} \langle\mathbf{32}\rangle_{\mathrm{ML}} \lbrack\mathbf{13}\rbrack_{\mathrm{JN}} +\lbrack\mathbf{21}\rbrack_{\mathrm{KI}} \lbrack\mathbf{32}\rbrack_{\mathrm{ML}} \langle\mathbf{13}\rangle_{\mathrm{JN}}
    }{M_Z}
    \nonumber\\
    &+\frac{
    \lbrack\mathbf{21}\rbrack_{\mathrm{KI}} \langle\mathbf{32}\rangle_{\mathrm{ML}} \langle\mathbf{13}\rangle_{\mathrm{JN}} +\langle\mathbf{21}\rangle_{\mathrm{KI}} \lbrack\mathbf{32}\rbrack_{\mathrm{ML}} \lbrack\mathbf{13}\rbrack_{\mathrm{JN}}
    }{M_W}
    \Bigg],
\end{align}
where we have switched the first and second terms on each row and we have also switched the third and fourth rows.  We next interchange $1\longleftrightarrow2$, $\mathrm{I}\longleftrightarrow\mathrm{K}$, $\mathrm{J}\longleftrightarrow\mathrm{L}$ and $\mathrm{M}\longleftrightarrow\mathrm{N}$.  We also raise and lower the spin indices and use the symmetry of the spin indices on the $W$ and $\bar{W}$ to bring it back to the form of Eq.~(\ref{eq:S_ZWW}).

This action is Lorentz invariant as all the spin indices are contracted with objects of the same momenta.  It is non-local due to the spinor products.  It is dimensionless because the momentum integrals and delta function contribute +8, the fields -9, the masses -2 and the spinor products +3.

\subsection{\label{sec:Higgs Mechanism}An Incomplete Higgs-Like Mechanism for the CSM}

In this subsection, we outline our current understanding of the Higgs mechanism in the CSM. As we will demonstrate, there are clear indications of the Higgs mechanism at work (see also Ref.~\cite{Bachu:2023fjn}). However, the mechanism remains incomplete, particularly if we insist on avoiding the introduction of unphysical degrees of freedom. While it might be possible to begin with the Higgs field, Dirac fields, and vector gauge fields -- along with their unphysical components -- and then deconstruct them to extract the relevant pieces for the action described in the previous subsection, this approach contradicts the principles of constructive theory and the goals of this paper. Therefore, we will focus solely on what can be achieved without unphysical degrees of freedom, and highlight where the current understanding falls short. A significant challenge we face is the inability to explain how to decouple the Goldstone bosons from interactions with fields that have non-zero spin. Despite this limitation, we will proceed by provisionally excluding the Goldstone bosons to make progress, considering only the vacuum expectation value (vev) and the remaining Higgs field in the following discussion.

Since the Higgs action is local and identical with the traditional Higgs field-theory action, it has the same potential and the same vev.  We will not review it from the position-space point of view, which can be found in textbooks \cite{Cheng:1984vwu}.  We will simply take for granted that, after developing a vev, and after removing the Goldstone bosons, the Higgs doublet can be written, in momentum space (after Fourier transformation), 
\begin{align}
    \Phi(p) &= 
    \frac{1}{\sqrt{2}}
    \left(\begin{array}{c}
    0 \\ v(2\pi)^4\delta^4(p)  + h(p)
    \end{array}\right),
    \label{eq:Phi(p)}
\end{align}
where $\int\frac{d^4p}{(2\pi)^4} v(2\pi)^4\delta^4(p)e^{ip\cdot x} = v$, our vev in position space.  We also note that each term has the same mass dimension of $-3$.

We can now insert this into the action that includes $\Phi$.  We begin with the pure-Higgs-doublet action.
\begin{align}
    \mathcal{S}_{\Phi} =& 
    \int\frac{d^4p_1d^4p_2}{(2\pi)^4}\Phi^\dagger(p_1)(p_1^2+\mu^2)\Phi(p_2)\delta^4(p_1+p_2)
    \nonumber\\
    &-\lambda\int\frac{dp_1dp_2dp_3dp_4}{(2\pi)^{12}}
    \delta^4(p_1+p_2+p_3+p_4)
    \nonumber\\
    &\hspace{0.5in}
    \Phi^\dagger(p_1)\Phi(p_2)\Phi^\dagger(p_3)\Phi(p_4).
    \label{eq:S_Phi}
\end{align}
We can see that this action is manifestly Hermitian and Lorentz invariant, as usual.  Inserting Eq.~(\ref{eq:Phi(p)}), we have
\begin{align}
    \mathcal{S}_{\Phi} =& 
    \int\frac{d^4p_1d^4p_2}{(2\pi)^4}\frac{1}{2}h(p_1)(p_1^2-\mu^2)h(p_2)\delta^4(p_1+p_2)
    \nonumber\\
    &-\lambda v\int\frac{d^4p_1d^4p_2d^4p_3}{(2\pi)^{8}}\delta^4(p_1+p_2+p_3)
    \nonumber\\
    &
     \hspace{0.5in}h(p_1)h(p_2)h(p_3)
    \nonumber\\
    &-\frac{\lambda}{4}\int\frac{d^4p_1d^4p_2d^4p_3d^4p_4}{(2\pi)^{12}}\delta^4(p_1+p_2+p_3+p_4)
    \nonumber\\
    &
     \hspace{0.5in}h(p_1)h(p_2)h(p_3)h(p_4),
\end{align}
where we have thrown away non-dynamical constant terms.  The linear terms cancel and the quadratic terms combine to give the result we have shown.  Plugging in the well-known values,
\begin{align}
    \mu &= m_h \\
    v &= \frac{2 M_W s_W}{e}\\
    \lambda &= \frac{e^2m_h^2}{4M_W^2 s_W^2},
\end{align}
we obtain Eq.~(\ref{eq:S_h}).  We note that, here, the Goldstone bosons cancel in Eq.~(\ref{eq:S_Phi}) and there is no need for a gauge transformation for this result.

We now turn to the Higgs-fermion action terms. The case of the leptons presents a significant challenge due to the differing transformation properties of the massive charged leptons and the massless neutrinos under the little group. Specifically, the charged lepton, being massive, carries a spin index, while the neutrino, being massless, does not. As a result, combining them into a single electroweak doublet would produce an object that does not transform uniformly under the little group. This discrepancy is a fundamental issue in the context of the constructive Higgs mechanism, where we avoid introducing unphysical degrees of freedom. Currently, this presents a notable deficiency in our understanding, and overcoming this obstacle is an area we hope to address in future work. In contrast, the quarks do not face this issue, and they can be combined into doublets. We present their potential action below:
\begin{align}
    \mathcal{S}_{\Phi q} &= 
    -
    \sqrt{2}\int\frac{d^4p_1d^4p_2d^4p_3}{(2\pi)^8}
    \delta^4(p_1+p_2+p_3)
    \Bigg[
    \nonumber\\
    &
    +\frac{m_d}{v}
    \left(\begin{array}{c}
        \bar{u}_{\mathrm{I}}(p_2) \\
        \bar{d}_{\mathrm{I}}(p_2)
    \end{array}\right)^{T}
    \Phi(p_1)
    d_{\mathrm{J}}(p_3)
    \langle\mathbf{32}\rangle^{\mathrm{JI}}  
    \nonumber\\
    &+\frac{m_d}{v}
    \bar{d}_{\mathrm{I}}(p_2)
    \Phi^{\dagger}(p_1)
    \left(\begin{array}{c}
        u_{\mathrm{J}}(p_3) \\
        d_{\mathrm{J}}(p_3)
    \end{array}\right)
    \lbrack\mathbf{32}\rbrack^{\mathrm{JI}}
    \nonumber\\
    &
    +\frac{m_u}{v}
    \left(\begin{array}{c}
        \bar{u}_{\mathrm{I}}(p_2) \\
        \bar{d}_{\mathrm{I}}(p_2)
    \end{array}\right)^{T}
    \tilde{\Phi}(p_1)
    u_{\mathrm{J}}(p_3)
    \langle\mathbf{32}\rangle^{\mathrm{JI}}  
    \nonumber\\
    &+\frac{m_u}{v}
    \bar{u}_{\mathrm{I}}(p_2)
    \tilde{\Phi}^{\dagger}(p_1)
    \left(\begin{array}{c}
        u_{\mathrm{J}}(p_3) \\
        d_{\mathrm{J}}(p_3)
    \end{array}\right)
    \lbrack\mathbf{32}\rbrack^{\mathrm{JI}}
    \Bigg],
    \label{eq:S_Phi q}
\end{align}
where, as usual,
\begin{align}
    \tilde{\phi}(p) &= i\sigma_2\Phi^*(p) 
    = \frac{1}{\sqrt{2}}\left(\begin{array}{c}
    v(2\pi)^4\delta^4(p)  + h(p) \\ 0
    \end{array}\right),
    \label{eq:PhiTilde(p)}
\end{align}
and we have again removed the Goldstone bosons.  Unlike the pure Higgs action in Eq.~(\ref{eq:S_Phi}), we cannot explain the absence of the Goldstone bosons here, without reference to a gauge transformation, another current deficiency.  

The Hermitian conjugate of this action is
\begin{align}
    \mathcal{S}^{\dagger}_{\Phi q} &= 
    -
    \sqrt{2}\int\frac{d^4p_1d^4p_2d^4p_3}{(2\pi)^8}
    \delta^4(p_1+p_2+p_3)
    \Bigg[
    \nonumber\\
    &+\frac{m_d}{v}
    \left(\begin{array}{c}
        \bar{u}^{\mathrm{J}}(p_3) \\
        \bar{d}^{\mathrm{J}}(p_3)
    \end{array}\right)^T
    \Phi(p_1)
    d^{\mathrm{I}}(p_2)
    \langle\mathbf{23}\rangle_{\mathrm{IJ}}
    \nonumber\\
    &
    +\frac{m_d}{v}
    \bar{d}^{\mathrm{J}}(p_3)
    \Phi^{\dagger}(p_1)
    \left(\begin{array}{c}
        u^{\mathrm{I}}(p_2) \\
        d^{\mathrm{I}}(p_2)
    \end{array}\right)
    \lbrack\mathbf{23}\rbrack_{\mathrm{IJ}}   
    \nonumber\\
    &+\frac{m_u}{v}
    \left(\begin{array}{c}
        \bar{u}^{\mathrm{J}}(p_3) \\
        \bar{d}^{\mathrm{J}}(p_3)
    \end{array}\right)^T
    \tilde{\Phi}(p_1)
    u^{\mathrm{I}}(p_2)
    \langle\mathbf{23}\rangle_{\mathrm{IJ}}
    \nonumber\\
    &
    +\frac{m_u}{v}
    \bar{u}^{\mathrm{J}}(p_3)
    \tilde{\Phi}^{\dagger}(p_1)
    \left(\begin{array}{c}
        u^{\mathrm{I}}(p_2) \\
        d^{\mathrm{I}}(p_2)
    \end{array}\right)
    \lbrack\mathbf{23}\rbrack_{\mathrm{IJ}} 
    \Bigg],
\end{align}
where we have interchanged the second and third rows and the fourth and fifth rows.  Next, we interchange $2\longleftrightarrow3$ and $\mathrm{I}\longleftrightarrow\mathrm{J}$ and raise and lower the spin indices.  This brings us back to Eq.~(\ref{eq:S_Phi q}), so this action is Hermitian. 

If we insert the Higgs doublet from Eqs.(\ref{eq:Phi(p)}) and (\ref{eq:PhiTilde(p)}), we can reproduce the interaction term from Eq.(\ref{eq:S_hbf}). In fact, switching the angle and square brackets in this action would also yield the correct interaction term. However, this approach fails to produce the correct mass term for the fermions, highlighting two critical issues. First, the mass term appears only to the first power here, whereas it should be squared in the quadratic part of the action [see Eq.~(\ref{eq:S p^2-m^2})]. Second, the quadratic term for the fermions includes the spinor product \((\langle\mathbf{32}\rangle^{\mathrm{JI}} + \lbrack\mathbf{32}\rbrack^{\mathrm{JI}})\), which is absent in the correct quadratic action for the fermions. This discrepancy marks another significant deficiency in the Higgs action for the CSM.

One possible workaround would be to remove the spinor products, which would correct the mass term but at the expense of producing an incorrect interaction term. Moreover, considering any interaction term, the spinor products are necessary to cancel the little-group transformations, as the momenta of the fields are not identical before the Higgs attains a vev. Thus, removing the spinor products seems poorly motivated. This deficiency in the Higgs mechanism as applied to the CSM requires further investigation and clarification.

Our final action involves the Higgs doublet and the weak bosons. However, we cannot include the photon because it transforms under the helicity little group, unlike the  $Z$  boson, which transforms under the spin little group. This difference in transformation properties means that we cannot construct a unified field that incorporates both the  $Z$  boson and the photon (representing the neutral bosons of  $SU(2) \times U(1)$ ) while preserving a consistent transformation under the little group, without introducing unphysical degrees of freedom. Since the goal of this paper is to explore what can be achieved without adding such unphysical components, this limitation prevents us from combining these fields.

Additionally, it is important to note that there is no interaction between the Higgs and the photon in this framework \cite{Christensen:2018zcq,Christensen:2024xzs}, contrary to what is observed in Feynman diagrams derived from standard gauge field theory. As a result, we cannot construct an action that would generate such a vertex. Therefore, we will attempt to build our action term with the  $Z$  boson alone in the diagonal entry of the weak matrix, even though this approach is not correct. Nonetheless, this exercise will provide further insights into this interaction. With this, we have:
\begin{align}
    \mathcal{S}_{\Phi (W/Z)} =& 
    -\frac{e^2}{2M_W^2s_W^2}\int\frac{d^4p_1d^4p_2d^4p_3d^4p_4}{(2\pi)^{12}}
    \nonumber\\
    &
    \delta^4(p_1+p_2+p_3+p_4)
    \langle\mathbf{23}\rangle_{IK}\lbrack\mathbf{23}\rbrack_{JL}
    \nonumber\\
    &
    \Phi^\dagger(p_1)
    \left(\begin{array}{cc} \frac{1}{\sqrt{2}}Z_{\mathrm{IJ}}(p_2) & W_{\mathrm{IJ}}(p_2) \\ \bar{W}_{\mathrm{IJ}}(p_2) & \frac{1}{\sqrt{2}}Z_{\mathrm{IJ}}(p_2) \end{array}\right)
    \nonumber\\
    &
    \left(\begin{array}{cc} \frac{1}{\sqrt{2}}Z_{\mathrm{KL}}(p_3) & W_{\mathrm{KL}}(p_3) \\ \bar{W}_{\mathrm{KL}}(p_3) & \frac{1}{\sqrt{2}}Z_{\mathrm{KL}}(p_3)\end{array}\right)
    \Phi(p_4).
\end{align}
After an initial expansion, we have
\begin{align}
    \mathcal{S}_{\Phi (W/Z)} =& 
    -\frac{1}{2}\frac{e^2}{2M_W^2s_W^2}\int\frac{d^4p_1d^4p_2d^4p_3d^4p_4}{(2\pi)^{12}}
    \nonumber\\
    &
    \delta^4(p_1+p_2+p_3+p_4)
    \langle\mathbf{23}\rangle_{IK}\lbrack\mathbf{23}\rbrack_{JL}
    \nonumber\\
    &
    \left(v(2\pi)^4\delta^4(p_1)  + h(p_1)\right)
    \left(v(2\pi)^4\delta^4(p_4)  + h(p_4)\right)
    \nonumber\\
    &
    \left(
        \bar{W}_{\mathrm{IJ}}(p_2)W_{\mathrm{KL}}(p_3)
        +\frac{1}{2}Z_{\mathrm{IJ}}(p_2)Z_{\mathrm{KL}}(p_3)
    \right).
\end{align}
Completing the expansion and integrating over the delta functions, we have
\begin{align}
    \mathcal{S}_{\Phi (W/Z)} =& 
    -\int\frac{d^4p_2d^4p_3}{(2\pi)^4}
    \delta^4(p_2+p_3)
    \langle\mathbf{23}\rangle_{IK}\lbrack\mathbf{23}\rbrack_{JL}
    \nonumber\\
    &
    \left(
        \bar{W}_{\mathrm{IJ}}(p_2)W_{\mathrm{KL}}(p_3)
        +\frac{1}{2}Z_{\mathrm{IJ}}(p_2)Z_{\mathrm{KL}}(p_3)
    \right)
    \nonumber\\
    & 
    -\frac{e}{M_W s_W}\int\frac{d^4p_1d^4p_2d^4p_3}{(2\pi)^8}
    \delta^4(p_1+p_2+p_3)
    \nonumber\\
    &
    \langle\mathbf{23}\rangle_{IK}\lbrack\mathbf{23}\rbrack_{JL}
    \nonumber\\
    &
    h(p_1)
    \left(
        \bar{W}_{\mathrm{IJ}}(p_2)W_{\mathrm{KL}}(p_3)
        +\frac{1}{2}Z_{\mathrm{IJ}}(p_2)Z_{\mathrm{KL}}(p_3)
    \right)
    \nonumber\\
    & 
    -\frac{e^2}{4M_W^2s_W^2}\int\frac{d^4p_1d^4p_2d^4p_3d^4p_4}{(2\pi)^{12}}
    \nonumber\\
    &
    \delta^4(p_1+p_2+p_3+p_4)
    \langle\mathbf{23}\rangle_{IK}\lbrack\mathbf{23}\rbrack_{JL}
    h(p_1) h(p_4)
    \nonumber\\
    &
    \left(
        \bar{W}_{\mathrm{IJ}}(p_2)W_{\mathrm{KL}}(p_3)
        +\frac{1}{2}Z_{\mathrm{IJ}}(p_2)Z_{\mathrm{KL}}(p_3)
    \right).
\end{align}
The mass term is clearly incorrect. It does not include the masses and, instead, contains spinor products. Even if we replaced the spinor products with a mass squared, it would still be wrong because it would incorrectly equate the masses of the $W$ and $Z$ bosons. This outcome should not be surprising given the way we constructed this term. Since we did not use the Higgs mechanism to break $SU(2) \times U(1) \to U(1)_{\text{EM}}$, there is no reason to expect it to differentiate between the $W$ and $Z$ bosons.

On the other hand, this construction did yield the correct form for the vertices with the Higgs, including the correct coupling constant. This is particularly impressive considering that the three-point vertices were derived purely from symmetry arguments \cite{Christensen:2018zcq}, while the four-point vertices were designed to achieve perturbative unitarity \cite{Christensen:2024xzs}. The fact that the correct interaction structure and coupling constants emerge from this flawed post-vev action suggests that it could offer valuable insights into the development of the correct action terms in future work.

A further potential issue we observe is that our current construction seems to predict that the signs of the three-point and four-point Higgs vertices with the massive vector bosons are the same. However, in \cite{Christensen:2024xzs}, these signs were found to be opposite. This discrepancy could be viewed from two perspectives, that could be related.

On the one hand, it may reflect an incomplete understanding of the signs involved in the constructive diagram procedure. Specifically, we used a positive sign for all our propagators, $+i/(p^2-m^2)$. If the propagator sign should have been negative, then the sign of the four-point vertex would flip, aligning with the result of our Higgs-like toy calculation here. We think this is possible considering the signs of the propagators for Feynman diagrams are not all the same. We maintained a consistent sign for our propagators because, at the time, we lacked a guiding principle to choose one sign over the other, and there was no distinct propagator structure -- other than quantum-number-preserving delta functions -- to differentiate them. Moreover, we did not have a field-theory action from which to derive these signs. Our approach was constructive, built from the ground up, relying solely on symmetries at the level of individual propagators and vertices. This approach was not sufficient to definitively determine these signs.

From a perturbation theory perspective, the way to resolve this issue is to work through a larger number of diagrams using a fixed set of rules and identify the rule set that yields the correct scattering amplitudes. We plan to undertake this in future calculations. However, achieving a complete understanding of constructive field theory that can fix these signs based on an underlying principle would not only resolve this issue but also provide a more fundamental understanding of these signs.

Another perspective on this issue might stem from our incomplete understanding of the Higgs mechanism within the CSM. With a fully developed Higgs mechanism, we might indeed find that the two vertices have opposite signs. Additionally, it's important to note that, in traditional field theory, the quadratic terms for the $W, Z$ , and Higgs bosons -- which give rise to the propagators -- along with their interaction terms, all originate from the same two action terms [schematically,  $-\frac{1}{4}F_{\mu\nu}F^{\mu\nu}  and  \left(D_{\mu}\Phi\right)^\dagger D^{\mu}\Phi$ ]. This suggests that, with a complete understanding of the weak symmetry and its breaking within the constructive action, all these signs and their relationships would naturally emerge, eliminating any ambiguity. Therefore, it is crucial to pursue a comprehensive Higgs mechanism within the CSM action in future research.

As we have stated, the Higgs mechanism in constructive field theory remains incompletely understood, particularly when adhering to the constraint of excluding any unphysical degrees of freedom. We have attempted to highlight the specific challenges this approach encounters. The primary difficulty lies in the non-unified transformation under the little group of fields that should ideally be combined into multiplets under the electroweak group. To put it another way, we lack separate $SU(2)$- and $U(1)$-boson fields that transform under a single representation of the little group, which could then combine into the massive $W$- and $Z$-boson fields (transforming under spin) and the photon field (transforming under helicity). This misalignment complicates the differentiation between the W and Z bosons.

Moreover, we are currently unable to explain how to remove the Goldstone bosons, as this would seemingly require a gauge symmetry -- something we aim to avoid in this framework. Finally, we do not yet have a unified action that simultaneously provides the kinetic terms, masses, \textit{and} interaction terms. The action either includes spinor products that are inappropriate for mass terms or correctly generates mass terms but lacks the spinor products necessary for interaction terms.

Before concluding this section, we will briefly discuss two potential approaches that might help overcome these limitations. One approach, which aligns with our preference, involves adhering strictly to our goal of not introducing any unphysical degrees of freedom. To achieve this, we would likely need to start with massless fields for all the electroweak bosons, with all fields transforming under helicity. Then, after the Higgs acquires a vev, we could combine the massless helicity-$\pm1$ $W$ and $Z$ bosons with the Goldstone bosons to form the final massive $W$ and $Z$ spin-$1$ bosons. A crucial part of this approach would be to demonstrate that these final combined fields transform correctly under spin. Importantly, in this scenario, the photon would never possess unphysical degrees of freedom.

Alternatively, we could follow a more traditional approach, where the massless fields are embedded within larger spin-1 fields that transform under spin (with two symmetrized spin indices). This approach would presumably necessitate adding a gauge symmetry to eliminate the negative effects of unphysical degrees of freedom. In this framework, the photon would initially be embedded in a spin-$1$ field, alongside the $Z$ boson. However, to ensure that the final theory contains only physical degrees of freedom -- after electroweak symmetry breaking and the absorption of the Goldstone bosons into the $W$ and $Z$ bosons -- the spin-$1$ photon field would need to be decomposed into its helicity components, with the unphysical spin-$0$ field removed. While this route might be easier to implement, it comes at the cost of introducing unphysical degrees of freedom, at least temporarily.

However, it is important to note that even if we introduce a gauged symmetry in the weak sector, its role in constructive field theory may differ from that in traditional gauge theory. In traditional gauge theory, gauged symmetries are \textit{required} to cancel the effects of unphysical degrees of freedom and to ensure the consistency of the theory. In constructive field theory, while a gauge symmetry could potentially provide deeper insights into the relationships between couplings, its necessity may not be as clear-cut since the fields are constructed to transform correctly from the outset. Nonetheless, if the photon is embedded in a spin-$1$ field, gauge symmetry might still play a crucial role in maintaining the consistency of the theory.

With either approach, a crucial aspect will be understanding how the mass terms in the action can avoid spinor products, while the interaction terms include the necessary spinor products. Additionally, to gain a better understanding of the Higgs mechanism, it seems increasingly likely that we will need to Fourier transform the constructive action into position space and perform the relevant operations there. This presents its own set of challenges, which we discuss in App.~\ref{app:non-local spinor products}. Finally, achieving a full understanding of the electroweak sector and its symmetry breaking will likely provide valuable insights into the relationships between the couplings of different vertices in constructive theories. This will be particularly relevant for the weakly coupled fields, but it may also shed light on the coupling constants governing quark-gluon interactions and their consistency with the triple-gluon vertex.

\section{\label{sec:conclusions}Conclusions}

We have developed a complete field-theory action for the CSM, ensuring that the fields introduced for all particles transform under the same little-group representation as the particles themselves. A key outcome of this approach is that no unphysical degrees of freedom are introduced for any of the particles, whether they are massless or massive. In the CSM, this applies to massless helicity-$\pm1$ fields (the photons and gluons), massless helicity-$\pm2$ fields (gravitons), and massive spin-$1$ fields (the $W$ and $Z$ bosons). Additionally, it includes both massless and massive fermions (neutrinos, charged leptons, and quarks). Unlike in traditional field theory, where the fields for these particles carry unphysical degrees of freedom, the constructive field-theory framework eliminates them entirely. This results in a significant reduction in the number of field degrees of freedom compared to traditional local field theory.

In order to achieve Lorentz invariance in our action, we must introduce non-field objects that also transform under the little group, thereby canceling the little-group transformations of the fields. We demonstrate that these objects are the spinor products and $x$ from constructive-amplitude theory, and we work out their full transformations. Furthermore, we show that for a contribution to the action to be Lorentz invariant, the helicities of all objects with the same momentum (both fields and non-field objects) must sum to zero, and all spin indices must be contracted among objects with the same momentum (both fields and spinor products). Using these criteria, we construct fully Lorentz-invariant action terms for the CSM, a property we refer to as manifest Lorentz invariance in a constructive action.

We formulated the action in momentum space and explicitly conserved momentum in every interaction by including a momentum-conserving delta function. This approach is natural given that particle-physics experiments typically focus on definite-momentum states rather than definite-position states, and because the constructive vertices are already defined in momentum space, not in position space. However, we believe that a complete understanding of the theory will eventually require a position-space form of the action, and we plan to explore this further in the future.

We detailed the Hermitian conjugates of both the fields and non-field objects, demonstrating that our action is Hermitian. Furthermore, we showed that Hermiticity imposes a relationship between the couplings of interactions that are Hermitian conjugates of each other. However, we also noted that the relationship between the couplings in the triple-gluon interaction and the gluon-quark interactions is not currently explained by constructive theory. We consider this a significant deficiency and believe that further investigation is needed to bring clarity to this important topic.

The Higgs and weak interaction terms were successfully introduced \textit{after} electroweak symmetry breaking. However, a complete description of the Higgs mechanism within constructive field theory remains lacking, representing another significant deficiency that needs to be addressed in the future. We proposed several approaches to tackle this problem and suggested that it might be necessary to formulate the action in position space to fully resolve the issues related to this symmetry breaking.

We also outlined general features of constructive field theories, including a basic power counting of the energy growth of minimal interactions. Our focus was on ensuring that the theory satisfies the essential requirements of momentum conservation, Lorentz invariance, and Hermiticity -- properties that are necessary for any consistent theory. While these are crucial, it is possible that additional requirements are needed to ensure causality, and we hope this can be further clarified in the future. However, if the CSM produces results that exactly match those of the traditional SM, as we believe it does, then the CSM must be causal, given that the SM is causal. This suggests that the set of properties we have identified is nearly complete, if not entirely so, for causality.  

On the other hand, ensuring renormalizability and perturbative unitarity will certainly require an additional property. At the very least, this will involve a mechanism that relates the couplings of all vertices involving fields transforming under the same symmetry. In traditional field theory, this role is played by gauge symmetry. However, gauge symmetry also typically leads to the presence of many four-point vertices, such as the four-gluon vertex, which are absent in constructive theory. Therefore, whatever this additional property in constructive field theory turns out to be, it must also prevent the emergence of these four-point vertices.

A natural implication of this work is that, once the remaining deficiencies are addressed, any new particle model could be constructed directly as a constructive field theory from the outset, without the need to begin with a traditional local field theory. Theorists could start with a set of symmetries and fields that transform under them, and then write the constructive field-theory action directly, incorporating all interactions that are invariant under these symmetries, as well as those that conserve momentum, are Lorentz invariant, Hermitian, and satisfy any other necessary criteria. If the couplings are weak and perturbative unitarity is satisfied, a perturbation series will naturally follow. These fields will also be important for understanding the renormalization of the amplitudes as it allows the fields to be renormalized as well as the other parameters of the theory. Furthermore, this non-local field theory could, in principle, be placed on the lattice to obtain non-perturbative results directly, potentially leading to new insights from this lattice approach.

There is, of course, a trade-off between working with a local field theory and the non-local field theory described here. In perturbative calculations using constructive vertices, expressions in terms of spinor products tend to be simpler and more efficient to compute. However, this efficiency comes with the requirement that all on-shell identities must be faithfully and completely applied. While this leads to significant simplification of the final results -- a major advantage -- missing even a single on-shell identity can render the result incorrect. In contrast, Feynman diagrams do not require or permit the use of on-shell identities, resulting in more complex final expressions. However, this approach ensures that the amplitude is correct at every intermediate step, regardless of simplifications. The downside is that a typical Feynman diagram result involves numerous interconnected components, such as momenta and gamma matrices, leading to a more complicated and less efficient phase-space calculation.  Furthermore, individual diagrams are not physically meaningful; only gauge-invariant sets of diagrams, often with large cancellations, are physically meaningful.

When performing lattice calculations, the primary advantage of using constructive field theory is the significantly smaller number of field degrees of freedom to integrate over, along with the absence of a local symmetry that would otherwise need to be managed. However, the downside is that the field theory action is no longer local, meaning that the integration must account for fields at different spacetime points. Naturally, other benefits and challenges -- beyond those we have considered -- may emerge as lattice calculations are carried out.

\section{Acknowledgements}
We would like to thank Zhen Liu, Mahbub Ishmam, Sekhar Chivukula, Kirtimaan Mohan, Xing Wang, Stephen Martin, and Jaroslov Trnka for helpful discussions.  
We also acknowledge the profound influence of Steven Weinberg's work, The Quantum Theory of Fields (Volume 1) \cite{Weinberg}, which has significantly shaped our understanding of particles and fields, making this research possible. 
We are deeply grateful to Nima Arkani-Hamed for his deeply insightful presentations on constructive amplitudes for any mass and spin at multiple Pheno Symposia, and for the extensive and thought-provoking discussions that followed at the University of Pittsburgh. These interactions were a significant source of inspiration and directly led to our interest in this research direction. 
This work was supported, in part, by the National Science Foundation under Grant No. PHY-2411482.

\appendix

\section{\label{app:motivation and conventions}Motivation and Conventions}
In this appendix, we aim to motivate the constructive calculations and the action presented in this paper. We begin by noting that the fields in traditional field theory inherently include extra unphysical degrees of freedom, which necessitates the introduction of a gauge symmetry -- a localized form of a physical global symmetry -- to cancel out these unphysical components. This also requires multiplying the Feynman amplitude by polarization vectors to extract the physical degrees of freedom from the amplitude. To better understand this, we examine how single-particle quantum states transform under Lorentz transformations, following the approach outlined by Weinberg in Ref.~\cite{Weinberg}, where further details can be found.

We then describe our fields and review the spinor products used in this article, which, unlike traditional fields, do not possess any unphysical degrees of freedom and therefore do not require a gauge symmetry to cancel them. This presents a more economical approach to constructing a particle field theory. We also discuss the spinors and their transformation properties under Lorentz transformations, emphasizing that spinor products transform only under the Little Group. This characteristic makes them particularly suitable for constructing an action that exclusively contains physical degrees of freedom in the fields.

\subsection{\label{app:Newton}A Metaphor with Newton's Law}
We begin with a quick metaphor involving Newton's law. Suppose we are trying to determine the position of a particle under a constant force after some time $t$. Now, imagine that we had not yet developed the concept of a vector. However, we had already established the concept of a second-rank tensor -- an object with nine degrees of freedom that transforms in a specific way under rotations. Although this second-rank tensor has more components than we need, we might discover that we could embed the position, velocity, and force into second-rank tensors in the following way:
\begin{align}
    r_{ij}(t) &= 
    \left(\begin{array}{ccc}
    a(t) & z(t)+\omega(t) & -y(t)+\tau(t)\\
    -z(t)+\omega(t) & b(t) & x(t)+\eta(t)\\
    y(t)+\tau(t) & -x(t)+\eta(t) & c(t)
    \end{array}\right)\\
    v_{ij}(t) &=
    \left(\begin{array}{ccc}
    \alpha(t) & v_z(t)+\phi(t) & -v_y(t)+\psi(t) \\
    -v_z(t)+\phi(t) & \beta(t) & v_x(t)+\mu(t) \\
    v_y(t)+\psi(t) & -v_x(t)+\mu(t) & \gamma(t)
    \end{array}\right)\\
    F_{ij} &= 
    \left(\begin{array}{ccc}
    \theta & F_z+\delta & -F_y+\epsilon \\
    -F_z+\delta & \kappa & F_x+\zeta \\
    F_y+\epsilon & -F_x+\zeta & \lambda
    \end{array}\right).
\end{align}

To achieve this, we would need to add eighteen extra unphysical degrees of freedom -- six to each tensor. Despite this, we would find that the position, velocity, and force all transform correctly under rotations. Since the final position must transform properly, it must be expressible as a linear combination of the tensors at the initial time. The coefficients could be determined, up to a numerical constant, through dimensional analysis. The exact numerical constants could then be determined either experimentally or by repeating the process for the acceleration and integrating. The result would be:
\begin{equation}
    r_{ij}(t) = r_{ij}(0) + v_{ij}(0)t + \frac{1}{2m}F_{ij}t^2.
\end{equation}
While the final result would contain many extraneous values that we would discard, the correct values for $x(t), y(t)$, and $z(t)$ would emerge.

This approach seems unnecessarily wasteful. We might wonder if there is a more efficient method -- one that avoids introducing unphysical degrees of freedom from the outset and works exclusively with the physical degrees of freedom.

\subsection{\label{app:single particle states}Single-Particle States}
The single-particle states differ for massive and massless particles. We begin with massless particles. It can be shown that massless particles possess a well-defined helicity that remains unchanged under Lorentz transformations. Even when a particle has two possible helicities, these correspond to distinct states that do not mix under Lorentz transformations. The components of the particle quantum states transform under the little group, and under a Lorentz transformation, these transformations can be block-diagonalized so that each helicity transforms independently within its own block. This contrasts with massive particles, where all spin states transform together within the same block.

Now, consider a massless particle with helicity $h$. If we begin with a quantum single-particle state at a reference momentum $k$,
\begin{align}
    \Psi^h(k),
\end{align}
where  $k$  can be any light-like momentum [though a standard convention is to use  $k = (1,0,0,1)$], we can then \textit{define} the quantum state for any other momentum by applying a standard boost that takes  $k$  directly to the new momentum. Specifically, we have
\begin{align}
    \sqrt{p^0}\Psi^h(p) = U\left[L(p)\right]\Psi^h(k),
\end{align}
where $p^0$ is the energy, $U$ is the quantum Lorentz transformation operator and
\begin{align}
    p = L(p)k,
\end{align}
where $L(p)$ boosts $k$ directly to $p$.  

Now that we have defined the single-particle states for any momentum, we can consider what happens to a general single-particle state $\Psi^h(p)$ under a Lorentz transformation $\bar{p} = \Lambda p$. We expect that the state should transform to $\Psi^h(\bar{p})$, maintaining the same helicity, up to some phase. Indeed, we find that
\begin{align}
    \sqrt{\frac{\bar{p}^0}{p^0}}
    e^{ih\omega(\Lambda,p)}\Psi^h(\bar{p}) = 
    U\left[\Lambda\right]\Psi^h(p).
\end{align}
That is, the single-particle states transform under the helicity little group, with the phase determined by the Lorentz transformation, the momentum, and the helicity of the particle. This phase satisfies
\begin{align}
    e^{ih\omega(\Lambda,p)}\Psi^h(\bar{p}) = U\left[\Lambda L(p)L^{-1}(\bar{p})\right]\Psi^h(\bar{p}),
    \label{eq:U[Lambda L(p1)L-1(p2)] Psi^h}
\end{align}
where $L^{-1}(\bar{p})$ takes $\bar{p}\to k$, $L(p)$ takes $k\to p$, and $\Lambda$ takes $p\to \bar{p}$.  In other words, $U\left[\Lambda L(p)L^{-1}(\bar{p})\right]$ is the transformation that leaves the momentum of the quantum particle state unchanged -- it represents the quantum little-group transformation.

Massive particles, unlike massless particles, have multiple spin states that transform together. While we can align the spin along any direction, including the direction of motion (often referred to as helicity), these spin states will still mix with each other under a Lorentz transformation. The spin states of a massive particle are all components of the same particle, and they generally mix under Lorentz transformations when observed from different frames. In contrast, massless particles do not have a spin degree of freedom that can be projected. Instead, they possess a fixed helicity -- an intrinsic property -- that remains invariant under Lorentz transformations and is independent of any chosen axis or reference frame. Massless particles with different helicities do not mix under Lorentz transformations.  The helicities of massive and massless particle states are only superficially related.

For a massive particle state, we label the spin component with $\mathrm{I}$. If we begin with a quantum single-particle state at a reference momentum $k$,
\begin{align}
    \Psi^{\mathrm{I}}(k),
\end{align}
where $k$ can be any time-like momentum corresponding to the correct mass, though a convenient choice is the particle's rest frame, $k = (m,0,0,0)$, then, as before, we can \textit{define} the quantum state for any other momentum by applying a standard boost that takes $k$ directly to the new momentum. Specifically, we \textit{define}
\begin{align}
    \sqrt{p^0}\Psi^{\mathrm{I}}(p) = 
    U\left[L(p)\right]\Psi^{\mathrm{I}}(k),
\end{align}
where $p=L(p)k$, with this $L(p)$ being different than the $L(p)$ in the massless case.

Now that we have defined the single-particle states, we can once again consider the effect of a Lorentz transformation acting directly on the state. This transformation will generally result in a linear combination of states at the new momentum. This combination is governed by a spin transformation:
\begin{align}
    \sqrt{\frac{\bar{p}^0}{\bar{p}^0}}
    \left(e^{i\omega J}\right)^{\ \mathrm{I}}_{\mathrm{K}}\Psi^{\mathrm{K}}(\bar{p}) = 
    U\left[\Lambda\right]\Psi^{\mathrm{I}}(p),
    \label{eq:e^iomega J Psi^K = U[Lambda]Psi^I}
\end{align}
where $J^{\ \mathrm{I}}_{\mathrm{K}}$ is the generator of spin transformations, $\omega=\omega(\Lambda,p)$ depends on the Lorentz transformation and momentum, and $\bar{p}=\Lambda p$.  Although $J$ could be the generator for any spin representation, we focus here on the spin-$\frac{1}{2}$ representation and we take higher spin as a totally symmetric product representation.  For example, spin-$1$ states would have two symmetrized spin indices and would transform under the product representation. The spin transformation is determined by the Lorentz transformation, the momentum, and the spin of the original state. It satisfies:
\begin{align}
    \left(e^{i\omega J}\right)^{\ \mathrm{I}}_{\mathrm{K}}\Psi^{\mathrm{K}}(\bar{p}) = 
    U\left[\Lambda L(p)L^{-1}(\bar{p})\right]\Psi^{\mathrm{I}}(\bar{p}).
    \label{eq:e^iomega J Psi^K = U[Lambda L(p1)L-1(p2)]Psi^I}
\end{align}
Once again, $U\left[\Lambda L(p)L^{-1}(\bar{p})\right]$ represents the quantum little-group transformation, which leaves the momentum of the state unchanged. In this case, the little group corresponds to spin.

With the single-particle states for both massless and massive particles defined, we can now construct the amplitude as an inner product of in-states and out-states. Each of these states can be treated as a direct product of single-particle states, and because the particles begin and end far apart, we can consider them as non-interacting. Consequently, the amplitude will transform as the direct product of single-particle states:
\begin{align}
    \sqrt{\frac{\cdots \bar{p}_i^0 \cdots \bar{p}_j^0 \cdots}{\cdots p_i^0 \cdots p_j^0 \cdots}}
    \cdots \left(e^{i\omega_i J}\right)^{\ \mathrm{I}_i}_{\mathrm{K}_i} 
    \cdots e^{ih_j\omega_j} 
    \cdots
    \nonumber\\
    \mathcal{M}^{\cdots \mathrm{K}_i\cdots h_j\cdots}
    (\cdots, \bar{p}_i, \cdots, \bar{p}_j, \cdots)  =
    \nonumber\\
    U\left[\Lambda\right]
    \mathcal{M}^{\cdots \mathrm{I}_i\cdots h_j\cdots}
    (\cdots, p_i, \cdots, p_j, \cdots),
\end{align}
where $\bar{p}=\Lambda p$.  Notably, under a Lorentz transformation, only the little-group transformations remain after shifting to the new momenta. The amplitude transforms under Lorentz transformations as a direct product of little-group representations at the new momenta.

Our goal is to identify a set of objects that can be used to construct this amplitude. These objects must transform in a way that ensures the amplitude exhibits the correct transformation properties under Lorentz transformations.

\subsection{\label{app:traditional Field Theory}Traditional Field Theory}
If we aren't familiar with modern spinors, our initial approach when looking for objects to construct amplitudes would likely involve those that transform under representations of the Lorentz group. Specifically, we might begin with scalars, vectors, and higher-order tensors. While it is not the purpose of this appendix to review these details in depth -- these can be found in Ref. \cite{Weinberg} -- we will comment on the relevant features as they pertain to our discussion.

Let's focus on the photon again. As we mentioned earlier, the photon has two distinct helicities that transform independently within their own invariant subspaces. Under a Lorentz transformation, each helicity picks up a phase. However, when we examine the available objects, none of them seem to fit this requirement. A scalar, for instance, does not undergo a phase change under a Lorentz transformation, so it cannot be used. The smallest object that can accommodate the helicity transformation is a vector. However, a vector has four degrees of freedom, rather than the two that we need for the photon. As a result, we are forced to include two additional unphysical degrees of freedom.

This approach seems wasteful and uneconomical. Moreover, under Lorentz transformations, the vector field mixes its various components and does not keep them distinct. In fact, it turns out that the photon ``vector'' field does not even transform as a vector. Under a Lorentz transformation, the photon vector field transforms as follows \cite{Weinberg}:
\begin{align}
    U(\Lambda)A_{\mu}(x)U^{-1}(\Lambda) = 
\Lambda^{\nu}_{\ \mu}A_{\nu}(\Lambda x) + \partial_{\mu}\Omega(x,\Lambda).
\end{align}
We see that, in addition to the usual Lorentz transformation, there is an additional term that is a total derivative of a scalar function. As a result, we can only use this embedding of the photon \textit{if} we include it in the action in such a way that this extra total derivative piece cancels out. This requirement leads to the introduction of a new symmetry, which, although not physical, is necessary to avoid the problematic consequences of this embedding. This newly invented symmetry is a local version of the physical global symmetry, known as a gauge symmetry. To obtain correct physical results, we must strictly maintain this gauge symmetry at every step of our calculations. The same holds true for the gluon.

Massive spin-$1$ particles, such as the $W$ and $Z$ bosons, possess multiple spin states that are related under Lorentz transformations. However, these particles still have only three physical degrees of freedom. Traditionally, we also embed them in a vector field, which necessitates the addition of one unphysical degree of freedom. Although the situation is slightly different from that of the photon and gluon, a gauge symmetry, along with a spontaneous symmetry breaking mechanism, is still required to obtain correct physical results.

Massive fermions, such as the electron, have two spin states (spin $+\frac{1}{2}$  and spin $-\frac{1}{2}$), and their antiparticles also have two spin states. As we have discussed, the spins mix with each other under Lorentz transformations. Unlike gauge bosons, we do not require a new gauge symmetry for each fermion, so it might seem that we have not introduced any unphysical degrees of freedom. However, we argue that unphysical degrees of freedom are indeed present for fermions. Fermions are traditionally embedded in Dirac spinors, which are a direct sum of a left-chiral and a right-chiral spinor, and have four degrees of freedom. (Even if we separate them into Weyl spinors, the same degrees of freedom remain.)

It's important to note that the chiral Lorentz group $SU(2)_L \times SU(2)_R$ is locally isomorphic to the more familiar $SO(3,1)$ (the generators of the groups are linearly related). It is a covering group. The electron, however, does not have chirality. As mentioned earlier, it transforms under the little-group spin, which is the diagonal subgroup of the chiral Lorentz group [$SU(2)_{spin} \subset SU(2)_L \times SU(2)_R$]. While the interactions of the electron possess chirality, the electron itself does not. As a result, we are effectively adding two unphysical degrees of freedom to the electron and two to the anti-electron. Despite this, the Dirac field's Lorentz transformations are consistent, meaning that, unlike the photon and gluon, no additional gauge symmetries are required for each Dirac fermion field. This observation applies to all massive spin-$\frac{1}{2}$ fermions: while traditional field theory adds unphysical degrees of freedom for them, no additional gauge symmetries are necessary to accommodate them.

Massless fermions, such as neutrinos, are also typically embedded in Dirac spinors or Weyl spinors. In the case of Dirac spinors, a projection operator is used for every interaction to eliminate the contribution from the opposite chirality. However, regardless of this, massless fermions inherently possess only one helicity, while massless anti-fermions possess the opposite helicity. This means that even when using Weyl spinors and their conjugates, we are still introducing an extra unphysical degree of freedom for each.

Once our traditional field theory is constructed, we can generate Feynman rules and compute scattering amplitudes. However, since Feynman amplitudes are constructed using Lorentz representations, they inherently transform under direct products of Lorentz representations. As a result, we must contract the Feynman amplitude with polarization vectors that extract the correct helicities and spins from the amplitude. In other words, the Feynman amplitude still contains many unphysical degrees of freedom, and we cannot directly square it and sum over the Lorentz and Dirac indices. We must first isolate the physical degrees of freedom -- namely, the helicities and spins of the external particles -- before squaring and summing over them. For example, the scattering amplitude can be schematically written as:
\begin{align}
    \mathcal{M}^{\cdots \mathrm{I}_i\cdots h_j\cdots} = 
    \cdots \epsilon^{\mathrm{I}_i}_{\mu_i}(p_i)
    \cdots \epsilon^{h_j}_{\mu_j}(p_j)
    \tilde{\mathcal{M}}^{\cdots \mu_i \cdots \mu_j \cdots},
\end{align}
where $\tilde{\mathcal{M}}$ is the Feynman amplitude that transforms under a direct product of Lorentz representations. In this example, for clarity, we have used the case of a spin-1 massive boson, such as a $W$ or $Z$ boson, and a helicity-$\pm1$ boson, such as a photon or gluon. However, the same procedure applies to fermions. The Feynman amplitude would involve a product of gamma matrices that transform under the $SU(2)_L \times SU(2)_R$ Lorentz transformation. To extract the physical spin components, we multiply on the left and right of every fermion chain by the ``polarization'' vectors $\bar{u}^{\mathrm{I}}(p)$ and $u^{\mathrm{I}}(p)$ for particles, and $\bar{v}^{\mathrm{I}}(p)$ and $v^{\mathrm{I}}(p)$ for antiparticles. See Ref.~\cite{Christensen:2013aua} for a detailed discussion of all the polarization vectors for spin up to $2$.

\subsection{\label{app:C Field Theory}Constructive Fields}
In constructive theory, we do not use fields that transform under Lorentz representations. Instead, we employ fields that transform under the little group -- spin for massive particles and helicity for massless particles. Since we do not introduce any extra unphysical degrees of freedom, we expect no subtleties in their transformations. These fields transform exactly as their particle excitations do. Accordingly, we introduce the following fields and their corresponding Lorentz transformations. For massless particles, we have:
\begin{align}
    U(\Lambda)G^{\pm}(p)U^{-1}(\Lambda) &=
    e^{\pm i2\omega}G^{\pm}(\bar{p})
    \label{eq:UGU^-1}\\
    U(\Lambda)\textsl{g}^{\pm}(p)U^{-1}(\Lambda) &=
    e^{\pm i\omega}\textsl{g}^{\pm}(\bar{p})\\
    U(\Lambda)\gamma^{\pm}(p)U^{-1}(\Lambda) &=
    e^{\pm i\omega}\gamma^{\pm}(\bar{p})\\
    U(\Lambda)f^{\pm}(p)U^{-1}(\Lambda) &=
    e^{\pm i\frac{1}{2}\omega}f^{\pm}(\bar{p})\\
    U(\Lambda)\bar{f}^{\pm}(p)U^{-1}(\Lambda) &=
    e^{\pm i\frac{1}{2}\omega}\bar{f}^{\pm}(\bar{p})
    \label{eq:UfbU^-1},
\end{align}
where $f$ is a massless fermion, and the transformation phase $\omega$ depends on the Lorentz transformation and the initial momentum $\omega(\Lambda,p)$. Any other quantum numbers, such as those related to QCD or electric charge, remain unchanged under a Lorentz transformation.

We can see that if the helicities of the fields with the same momentum in a product sum to zero, the product will be invariant under helicity transformations. For a simple concrete example, consider a term arising from the quadratic part of the action for photons:
\begin{align}
    U(\Lambda)\gamma^+(p_1)\gamma^-(p_2)U^{-1}(\Lambda) 
    &= \gamma^+(\bar{p}_1)e^{+i\omega}e^{-i\omega}\gamma^-(\bar{p}_2)
    \nonumber\\
    &= \gamma^+(\bar{p}_1)\gamma^-(\bar{p}_2),
\end{align}
as expected, where we have used the fact that there is a momentum conserving delta function that equates the two phase angles, $\omega_2=\omega_1=\omega$.  More complex products will behave similarly, as long as the helicity contributions of the transformations for each momentum sum to zero.

For massive fields, the transformations are given by:
\begin{align}
    U(\Lambda)h(p)U^{-1}(\Lambda) &=
    h(\bar{p})
    \label{eq:UhU^-1}\\
    U(\Lambda)f^{\mathrm{I}}(p)U^{-1}(\Lambda) &=
    \left(e^{i\vec{\omega}\cdot \vec{J}}\right)^{\ \ \mathrm{I}}_{\mathrm{K}}f^{\mathrm{K}}(\bar{p})\\
    U(\Lambda)\bar{f}^{\mathrm{I}}(p)U^{-1}(\Lambda) &=
    \left(e^{i\vec{\omega}\cdot \vec{J}}\right)^{\ \ \mathrm{I}}_{\mathrm{K}}\bar{f}^{\mathrm{K}}(\bar{p})\\
    U(\Lambda)Z^{\mathrm{IK}}(p)U^{-1}(\Lambda) &=
    \left(e^{i\vec{\omega}\cdot \vec{J}}\right)^{\ \ \mathrm{I}}_{\mathrm{L}}
    \left(e^{i\vec{\omega}\cdot \vec{J}}\right)^{\ \ \mathrm{K}}_{\mathrm{M}}
    Z^{\mathrm{LM}}(\bar{p})\\
    U(\Lambda)W^{\mathrm{IK}}(p)U^{-1}(\Lambda) &=
    \left(e^{i\vec{\omega}\cdot \vec{J}}\right)^{\ \ \mathrm{I}}_{\mathrm{L}}
    \left(e^{i\vec{\omega}\cdot \vec{J}}\right)^{\ \ \mathrm{K}}_{\mathrm{M}}
    W^{\mathrm{LM}}(\bar{p})\\
    U(\Lambda)\bar{W}^{\mathrm{IK}}(p)U^{-1}(\Lambda) &=
    \left(e^{i\vec{\omega}\cdot \vec{J}}\right)^{\ \ \mathrm{I}}_{\mathrm{L}}
    \left(e^{i\vec{\omega}\cdot \vec{J}}\right)^{\ \ \mathrm{K}}_{\mathrm{M}}
    \bar{W}^{\mathrm{LM}}(\bar{p})
    \label{eq:UWbU^-1},
\end{align}
where $\vec{\omega}$ depends on the Lorentz transformation and the momentum, $\vec{\omega}(\Lambda,p)$, and $\vec{J}$ are the generators of the little-group spin. Once again, all other quantum numbers remain unaffected by the Lorentz transformation.

We will describe these transformations in greater detail shortly. To understand the invariance of products with contracted spin indices, we need to introduce the generators for spin, exponentiate them, and then consider products with contracted spin indices. It is important to emphasize that in our framework, fermion fields transform under spin and possess only two degrees of freedom each, in contrast to traditional field theory. The $W$ and $Z$ bosons have two symmetrized spin indices, leaving only the three physical spin-1 degrees of freedom. To avoid introducing extra unphysical degrees of freedom, it is necessary to symmetrize their indices in all action terms. We have done this explicitly in our quadratic terms, which should be sufficient for perturbative calculations. However, if this action is placed on a lattice, it is necessary to explicitly symmetrize the interactions as well. We will come back to these spin-1 fields after discussing the spin transformations.

It may seem that the conjugate fields should transform under the conjugate spin representation. This is correct, but it is important to note that conjugation lowers the spin index. In the transformations provided so far, we have given the transformation laws for fields and conjugated fields with upper spin indices. The same principle applies to massless fields: conjugation flips the helicity, but the Lorentz transformation we provided corresponds to the helicity of the conjugated field. To clarify this point, and because it is necessary for understanding the Hermiticity of the action, we now specify the conjugation of the fields. The Hermitian conjugates of the fields are given by:
\begin{align}
    \left[G^{\pm}(p)\right]^\dagger &= G^{\mp}(p)
    \label{eq:G^dagger}\\
    \left[\textsl{g}_a^{\pm}(p)\right]^{\dagger} &= \textsl{g}_a^{\mp}(p)\\
    \left[\gamma^{\pm}(p)\right]^{\dagger} &= \gamma^{\mp}(p)\\
    \left[q^{i\pm}(p)\right]^{\dagger} &= \bar{q}_i^{\mp}(p) \\
    \left[l^{\pm}(p)\right]^\dagger &=
    \bar{l}^{\mp}(p)\\
    \left[\nu^-(p)\right]^\dagger &=
    \bar{\nu}^+(p)\\
    \left[Z^{\mathrm{IJ}}(p)\right]^\dagger &= Z_{\mathrm{JI}}(p)\\
    \left[W^{\mathrm{IJ}}(p)\right]^\dagger &= \bar{W}_{\mathrm{JI}}(p)\\
    \left[q^{i\mathrm{I}}(p)\right]^\dagger &=
    \bar{q}_{i\mathrm{I}}(p)\\
    \left[l^{\mathrm{I}}(p)\right]^\dagger &=
    \bar{l}_{\mathrm{I}}(p)\\
    \left[h(p)\right]^\dagger &= h(p),
    \label{eq:h^dagger}
\end{align}
where we have interchanged the order of the spin indices on the $W$ and $Z$, although they are symmetric, so this has no effect.  If the spin indices are lowered on the left, then they are raised on the right.

When we examine the list of Hermitian conjugates, we notice that for all charged fields, Hermitian conjugation transforms the field into its antifield, with the opposite helicity or lowered (or raised) spin. This behavior is essential for maintaining both Hermiticity and symmetry invariance in the action, starting from the quadratic terms. Interestingly, this property extends to massless neutral fields as well. Hermitian conjugation transforms these fields into their opposite-helicity counterparts. The parallel between massless neutral fields and charged fields is particularly striking in constructive field theory.

Additionally, massless fields of opposite helicity exhibit distinct interactions with other particles, dictated by their helicities. These interactions are Hermitian conjugates of each other, analogous to the behavior of charged particles, where the field and antifield (with opposite helicity) are interchanged. Furthermore, the propagators of massless particles feature a particle of one helicity propagating in one direction, while its counterpart, with the opposite helicity, propagates in the opposite direction -- again, similarly to the behavior of charged particle-antiparticle pairs on propagator lines.

This stands in contrast to traditional field theory, where opposite-helicity neutral particles are embedded together within a single Lorentz-representation field, that is self conjugate. In particular, $A^{\dagger}_{\mu}=A_{\mu}$, $g^{\dagger}_{a\mu}=g_{a\mu}$, and $G^{\dagger}_{\mu\nu}=G_{\mu\nu}$. As a result, photons, gluons, and gravitons interact with other particles through a single unified interaction, and their Feynman propagators do not differentiate between opposite directions of propagation.

When considering the relationship between helicity states, it is common terminology to refer to massless neutrinos of opposite helicity, $\nu^-$ and $\bar{\nu}^+$, as antiparticles of each other. Extending this terminology, photons, gluons, and gravitons with opposite helicity could similarly be called antiparticles of one another, as they exhibit the same behavior under Hermitian conjugation, and similar behavior under $CP$, in their interactions, and in their propagators. However, if we instead define antiparticles as those obtained by reversing the charge while preserving the \textit{same} helicity or spin -- as occurs under the $C$ operator -- we see that the photon is its own antiparticle, with $C\gamma^{\pm}C^{-1}=-\gamma^{\pm}$ \cite{Weinberg} aligning with common terminology for the photon, and that massless neutrinos do not have antiparticles. Similar considerations apply to gluons and gravitons. Regardless of the definition used, constructive field theory treats these fields in a manner more closely aligned with the physical particles they represent than traditional field theory does.

In Appendix A3 of \cite{Christensen:2024xsg}, we detailed the generators for spin transformations as part of our validation of the SPINAS package. In this subsection, we review those generators, exponentiate them for the full transformations, and discuss invariance when spin indices are contracted.

The generators of spin are given by:
\begin{align}
J^{(3)\ \mathrm{J}}_{\quad \mathrm{K}} &=
    \frac{-1}{2}\left(\begin{array}{cc}
        1 & 0 \\
        0 & -1
    \end{array}\right)
    \label{eq:J^3_K^J on spin}
\\
J^{(+)\ \mathrm{J}}_{\quad \mathrm{K}} &=
    -e^{i \phi}\left(\begin{array}{cc}
        0 & 0\\
        1 & 0
    \end{array}\right)
\\
J^{(-)\ \mathrm{J}}_{\quad \mathrm{K}} &=
    -e^{-i \phi}\left(\begin{array}{cc}
    0 & 1\\
    0 & 0
    \end{array}\right),
    \label{eq:J^-_K^J on spin}
\end{align}
where these operators act on upper spin indices (from the right in matrix notation).  For lower spin indices, the generators are:
\begin{align}
J^{(3)\mathrm{K}}_{\quad \ \ \mathrm{J}} &=
    \frac{1}{2}\left(\begin{array}{cc}
        1 & 0 \\
        0 & -1
    \end{array}\right)
\\
J^{(+)\mathrm{K}}_{\quad \ \ \mathrm{J}} &=
    e^{i \phi}\left(\begin{array}{cc}
        0 & 1\\
        0 & 0
    \end{array}\right)
\\
J^{(-)\mathrm{K}}_{\quad \ \ \mathrm{J}} &=
    e^{-i \phi}\left(\begin{array}{cc}
    0 & 0\\
    1 & 0
    \end{array}\right) .
\end{align}

It is important to note that the spin generators acting on lower indices form the conjugate representation when compared with those acting on upper indices. This relationship can be seen by recalling that the conjugate representation is given by $\bar{J} = -J^*$, which satisfies:
\begin{align}
    \left[\bar{J}^i,\bar{J}^j\right] &=
    \left[J^{i*},J^{j*}\right]
    \nonumber\\
    &= \left[J^i,J^j\right]^*
    \nonumber\\
    &= \left(i\epsilon^{ijk}J^k\right)^*
    \nonumber\\
    &= i\epsilon^{ijk}\bar{J}^k.
\end{align}

In terms of the raising and lowering generators, this becomes:
\begin{align}
    \bar{J}^{3} &= - J^{3*}\\
    \bar{J}^{\pm} &= 
    \bar{J}^{1} \pm i \bar{J}^{2}
    \nonumber\\
    &= -\left(J^{1*} \pm i J^{2*}\right)
    \nonumber\\
    &= -\left(J^{1} \mp i J^{2}\right)^*
    \nonumber\\
    &= -J^{\mp*}.
\end{align}

Thus, the generators acting on lower indices indeed correspond to the conjugate representation of those acting on upper indices. Moreover, the generators acting on lower spin indices can be obtained from those acting on upper spin indices by raising and lowering the indices using the $\epsilon_{\mathrm{IJ}}$ and $\epsilon^{\mathrm{IJ}}$ tensors. This relationship arises because $SU(2)$ is pseudoreal, meaning that the conjugate representation is linearly related to the original representation.

To explicitly demonstrate the exponentiation of our spin generators without delving too deeply into the detailed exponentiation of each generator, we will use the fact that the choice of generator axis can be aligned with the rotation axis. For convenience, we will call this axis the $z$-axis. Although we perform the calculation with this specific choice of axes, the result is more general.

Exponentiating $J^{3}$, we have:
\begin{align}
    e^{i\omega J^3}
    &= \sum_{n=0}\frac{\left(i\omega\right)^n}{n!}\left(J^3\right)^n
    \nonumber\\
    &\hspace{-0.25in}= 
    \sum_{n=0}\frac{(-1)^n}{(2n)!}\left(\frac{\omega}{2}\right)^{2n}
    + i 2 J^3\sum_{n=0}\frac{(-1)^n}{(2n+1)!}\left(\frac{\omega}{2}\right)^{2n+1}
    \nonumber\\
    &= \cos\left(\frac{\omega}{2}\right)
    + i 2 J^3\sin\left(\frac{\omega}{2}\right)
    \nonumber\\    
    &= \left(\begin{array}{cc}
    \cos\left(\frac{\omega}{2}\right) \mp i \sin\left(\frac{\omega}{2}\right) & 0\\
    0 & \cos\left(\frac{\omega}{2}\right) \pm i \sin\left(\frac{\omega}{2}\right)
     \end{array}\right),
\end{align}
where we have split the sum into its even and odd parts on the second line and used the fact that $(J^3)^2=\left(\frac{1}{2}\right)^2$.  The extra factor of $2$ in front of the $J^3$ is because we have an extra division of $2$ inside the sum.  The sign on the last row depends on whether this transformation is acting on upper spin indices (upper sign) or lower spin indices (lower sign).  Therefore, the full rotation in spin space is given by:
\begin{align}
    \left(e^{i\omega J^3}\right)^{\ \ \mathrm{I}}_{\mathrm{K}} 
    &= \left(\begin{array}{cc}
    e^{-i\frac{\omega}{2}} & 0\\
    0 & e^{+i\frac{\omega}{2}}
     \end{array}\right)^{\ \ \mathrm{I}}_{\mathrm{K}} 
     \\
     \left(e^{i\omega J^3}\right)^{\mathrm{I}}_{\ \mathrm{K}} 
    &= \left(\begin{array}{cc}
    e^{+i\frac{\omega}{2}} & 0\\
    0 & e^{-i\frac{\omega}{2}}
     \end{array}\right)^{\mathrm{I}}_{\ \mathrm{K}}
     \nonumber\\
     &= \left[\left(e^{i\omega J^3}\right)^{\ \ \mathrm{I}}_{\mathrm{K}} \right]^\dagger.
\end{align}

We can now observe the impact of a spin transformation on a product with a contracted spin index, where the momentum is the same so that the phase $\omega$ is the same. As a concrete example, let's consider a quadratic term, as it appears in the quadratic part of the action. For simplicity, we will focus on a fermion field with one spin index. The transformation is given by:
\begin{align}
    U(\Lambda)\bar{f}_{\mathrm{I}}(p_1)f^{\mathrm{I}}(p_2) U^{-1}(\Lambda)
    \nonumber\\
    &\hspace{-1in}= \bar{f}_{\mathrm{K}}(\bar{p}_1)
    \left(e^{i\omega J^3}\right)^{\mathrm{K}}_{\ \mathrm{I}} 
    \left(e^{i\omega J^3}\right)^{\ \ \mathrm{I}}_{\mathrm{L}}
    f^{\mathrm{L}}(\bar{p}_2)
    \nonumber\\
    &= \bar{f}_{\mathrm{K}}(\bar{p}_1)
    f^{\mathrm{L}}(\bar{p}_2),
\end{align}
as expected, where we have used the fact that there is a momentum-conserving delta function that equates the two phase angles, $\omega_2=\omega_1=\omega$. Since we chose the axes without loss of generality, this result holds for any rotation.

Furthermore, the same principle applies to more complex action terms involving multiple fields and spinor products. As long as all the spin indices are contracted \textit{between objects with the same momentum}, the spin transformations cancel out in pairs, ensuring invariance under the transformation.

We have emphasized multiple times that our field theory does not introduce any extra unphysical degrees of freedom. However, there is an important caveat that must be discussed, regarding the spin structure of the $W$ and $Z$ bosons.  Although it is true that symmetrizing their spin-$\frac{1}{2}$ indices leaves only the spin-1 degrees of freedom, and symmetrizing the propagators is sufficient to remove the antisymmetric spin-0 piece from perturbation theory, we must be more careful when considering non-perturbative calculations.  We must also symmetrize the interaction terms.  Moreover, integrating over all possible field configurations for the $W$ and $Z$ bosons should also only consider symmetric field configurations for efficiency, even if the antisymmetric field configurations don't contribute.

On the other hand, strictly speaking, the functional integral formulation of quantum mechanics integrates over all possible field configurations, including the antisymmetric spin-0 configurations for these fields, which may raise concerns. However, it is important to note that our formulation of these fields using two spin-$\frac{1}{2}$ indices is merely a shorthand for the spin-$1$ fields, which inherently have only one spin-$1$ index.

In order to clarify this, we introduce a spin-1 index $\mathfrak{N}$, which runs over the values $-1, 0$ and $1$. Since the interaction of this field must be Lorentz invariant and, as we will see, requires the contraction of this spin-$1$ index with the spin-$\frac{1}{2}$ indices of spinor products, we need a tensor to connect these spins. Therefore, we define:
\begin{align}
     \mathrm{P}_{\mathfrak{N}}^{\mathrm{IJ}} &= 
     \begin{cases}
        \mathfrak{N},\mathrm{I},\mathrm{J} = \hspace{0.08in}1, \hspace{0.1in}\frac{1}{2}, \hspace{0.1in}\frac{1}{2} & 1\\
        \mathfrak{N},\mathrm{I},\mathrm{J} =\hspace{0.08in}0, \hspace{0.1in}\frac{1}{2}, -\frac{1}{2} & \frac{1}{\sqrt{2}} \\
        \mathfrak{N},\mathrm{I},\mathrm{J} =\hspace{0.08in}0, -\frac{1}{2}, \hspace{0.1in}\frac{1}{2} & \frac{1}{\sqrt{2}} \\
        \mathfrak{N},\mathrm{I},\mathrm{J} =-1, -\frac{1}{2}, -\frac{1}{2} & 1\\
        \mathrm{else} & 0 ,
     \end{cases}
\end{align}
and its conjugate with its spin indices raised and lowered, which contain the Clebsch-Gordon coefficients.  Using this tensor, we \textit{define} our fields with two spin-$\frac{1}{2}$ indices as follows:
\begin{align}
    Z^{\mathrm{IJ}} &= \mathrm{P}_{\mathfrak{N}}^{\mathrm{IJ}} Z^{\mathfrak{N}}
    \label{eq:app:Z^IJ=P Z^N}\\
    W^{\mathrm{IJ}} &= \mathrm{P}_{\mathfrak{N}}^{\mathrm{IJ}} W^{\mathfrak{N}}\\
    \bar{W}^{\mathrm{IJ}} &= \mathrm{P}_{\mathfrak{N}}^{\mathrm{IJ}} \bar{W}^{\mathfrak{N}}
    \label{eq:app:W^IJ=P W^N}.
\end{align}
The fields with a single spin-1 index are the quantum fields we actually introduce, and they truly do not possess any unphysical degrees of freedom. However, whenever these fields are used in an interaction, they must be contracted with a symmetric combination of spin-$\frac{1}{2}$ indices from the spinor products. The tensor $\mathrm{P}_{\mathfrak{N}}^{\mathrm{IJ}}$ is necessary to achieve this contraction. On the other hand, since $\mathrm{P}_{\mathfrak{N}}^{\mathrm{IJ}}$ and the field always appear together, we use the shorthand notation of writing the field with two spin-$\frac{1}{2}$ indices.

As an example of how this works, we consider the quadratic term.  We note that a product of two of these tensors gives:
\begin{align}
    \mathrm{P}^{\mathfrak{M}}_{\mathrm{IJ}}
    \frac{1}{2}\left(
        \delta_{\mathrm{K}}^{\mathrm{I}}\delta_{\mathrm{L}}^{\mathrm{J}}
        +\delta_{\mathrm{L}}^{\mathrm{I}}\delta_{\mathrm{K}}^{\mathrm{J}}
    \right)
    \mathrm{P}_{\mathfrak{N}}^{\mathrm{KL}}
    &= \frac{1}{2}\left(
        \mathrm{P}^{\mathfrak{M}}_{\mathrm{IJ}}
        \mathrm{P}_{\mathfrak{N}}^{\mathrm{IJ}}
        +
        \mathrm{P}^{\mathfrak{M}}_{\mathrm{IJ}}
        \mathrm{P}_{\mathfrak{N}}^{\mathrm{JI}}
    \right)
    \nonumber\\
    &= 
    \delta^{\mathfrak{M}}_{\mathfrak{N}}
    .
\end{align}
Therefore, the quadratic terms for both $Z^{\mathrm{IJ}}$ bosons and $W^{\mathrm{IJ}}$ bosons give the correct quadratic terms for $Z^{\mathfrak{N}}$ and $W^{\mathfrak{N}}$, since using this result in Eq.~(\ref{eq:S p^2-m^2}) results in:
\begin{align}
    \frac{1}{2}\left(p_1^2-M_Z^2\right)
     Z_{\mathrm{IJ}}(p_1)\frac{1}{2}\left(\delta_{\mathrm{K}}^{\mathrm{I}}\delta_{\mathrm{L}}^{\mathrm{J}}+\delta_{\mathrm{L}}^{\mathrm{I}}\delta_{\mathrm{K}}^{\mathrm{J}}\right)Z^{\mathrm{KL}}(p_2)
     \nonumber\\
     \hspace{0.75in}
     =\frac{1}{2}\left(p_1^2-M_Z^2\right)
     Z_{\mathfrak{N}}(p_1)Z^{\mathfrak{N}}(p_2)
     \\
     \left(p_1^2-M_W^2\right)
     \bar{W}_{\mathrm{IJ}}(p_1)\frac{1}{2}\left(\delta_{\mathrm{K}}^{\mathrm{I}}\delta_{\mathrm{L}}^{\mathrm{J}}+\delta_{\mathrm{L}}^{\mathrm{I}}\delta_{\mathrm{K}}^{\mathrm{J}}\right)W^{\mathrm{IJ}}(p_2)
     \nonumber\\
     =\left(p_1^2-M_W^2\right)
     \bar{W}_{\mathfrak{N}}(p_1)W^{\mathfrak{N}}(p_2).
\end{align}

In summary, although we use a symmetric combination of spin-$\frac{1}{2}$ indices for our $Z$- and $W$-boson fields, this is merely a shorthand for the fields $Z^{\mathfrak{N}}$ and $W^{\mathfrak{N}}$. Therefore, we maintain our assertion that no unphysical degrees of freedom are introduced for any of the fields, including the $Z$ and $W$ bosons. The use of two spin-$\frac{1}{2}$ indices is purely a notational convenience.  Moreover, this procedure can be generalized to any spin.

On a practical level, as long as we explicitly symmetrize the propagator, as we have done, and symmetrize the interactions in non-perturbative calculations, the antisymmetric component will not contribute. However, in the functional integral, strictly speaking, we should integrate over $Z^{\mathfrak{N}}, W^{\mathfrak{N}}$, and $\bar{W}^{\mathfrak{N}}$ to ensure that no unphysical degrees of freedom are introduced. Nevertheless, if preferred, \textit{all} $Z^{\mathrm{IJ}}$, $W^{\mathrm{IJ}}$ and $\bar{W}^{\mathrm{IJ}}$ can be replaced, using Eqs.~(\ref{eq:app:Z^IJ=P Z^N}) through (\ref{eq:app:W^IJ=P W^N}), completely removing even the hint of unphysical degrees of freedom.

Before leaving this subsection, we note that we have simply defined all our fields with the properties they should have.  We do not think there are any issues with our definitions.  However, in principle, we should follow \cite{Weinberg} and build up the field properties based on the creation and annihilation operators for single-particle states and the requirements of causality.  Since Weinberg assumes locality from an early point in his discussion, and he works in position space, we would need to derive the analogous properties without the assumption of locality and also in momentum space.  We think this would make an interesting and instructive exercise and would like to pursue it in the future.  However, it is beyond the scope of the present article.

\subsection{\label{app:lorentz transformations}Lorentz Transformations}
To construct interaction terms in our action, we need to consider how to combine fields that transform under the little group. For quadratic terms, which give rise to propagators, the process is straightforward: we can pair fields of opposite helicity or combine a massive field with its conjugate, contracting the spin indices. We have already established that these terms are Lorentz invariant. However, when we move on to interaction terms, additional elements are required for several reasons.

First and foremost, the momenta of the fields in an interaction are generally not identical, resulting in different phase angles $\omega_i = \omega(\Lambda,p_i)$ for each field. These phase angles cannot cancel each other out unless they act on objects with the same momentum, which already presents a fundamental and sufficient challenge. However, there is an secondary issue: in some interactions, the spin indices and helicity transformations of the fields cannot be made to cancel each other out, even in principle. Consider, for example, the cubic interaction involving a $W$ boson, a charged lepton, and a neutrino. The $W$ boson field has two spin indices, the charged lepton field has one spin index, and the neutrino field has no spin indices but instead transforms under helicity. In this case, it is not even possible to contract the indices of these fields, let alone cancel their transformations, without introducing additional components. Therefore, to construct Lorentz-invariant interaction terms, we must introduce non-field objects that transform under the little group to effectively cancel the transformations of the fields.

In traditional field theory, fields transform under Lorentz representations rather than little-group representations, necessitating the use of non-field objects that also transform under Lorentz representations to construct invariant terms. For this purpose, we employed the covariant derivative $D^{\mu}$ and the Dirac gamma matrices $\gamma^\mu$ and $\gamma_5$, which allowed us to form products that are Lorentz invariant. However, in the context of constructive field theory, we now need to identify analogous objects that transform under little-group representations to maintain invariance while working with our fields.

One way to approach the challenge of forming Lorentz-invariant interaction terms in constructive field theory is by considering how we represent momentum. Traditionally, momentum is represented by the four-vector $p^\mu$, which naturally incorporates the four degrees of freedom of the momentum, with one-to-one correspondence. However, this is not the only possible representation. Another way to represent momentum is through a Hermitian $2 \times 2$ matrix, which also contains four degrees of freedom (two along the diagonal and two on the off-diagonals). There are two convenient embeddings of momentum into these matrices:
\begin{align}
    p_{\alpha \dot{\beta}} &=
    \left(\begin{array}{cc}
    p^0+p^3 & p^1-i p^2\\
    p^1 + i p^2 & p^0 - p^3
    \end{array}\right)
    \nonumber\\
    &= \left(\begin{array}{cc}
    \mathcal{E}+p c_{\theta} & p s_{\theta}e^{-i\phi}\\
    p s_{\theta}e^{i\phi} & \mathcal{E} - p c_{\theta}
    \end{array}\right)\\
    p^{\dot{\alpha} \beta} &=
    \left(\begin{array}{cc}
    p^0-p^3 & -p^1+i p^2\\
    -p^1 - i p^2 & p^0 + p^3
    \end{array}\right)
    \nonumber\\
    &= \left(\begin{array}{cc}
    \mathcal{E} -p c_{\theta} & -p s_{\theta}e^{-i\phi}\\
    -p s_{\theta}e^{i\phi} & \mathcal{E} + p c_{\theta}
    \end{array}\right),
\end{align}
where the signs of the spatial momenta are flipped with upper Lorentz indices; $\mathcal{E}$ is the energy, $p=\lvert\vec{p}\rvert$ is the magnitude of the vector momentum, $c_{\theta}=\cos(\theta), s_{\theta}=\sin(\theta)$, $\theta$ is the polar angle, and $\phi$ is the azimuthal angle; and we will explain the need for the different indices (undotted and dotted) shortly.  In this notation, the Lorentz indices $\alpha$ and $\dot{\beta}$ take two values each.  The convenience of this definition is that the determinant is equal to the mass squared
\begin{align}
    \det(p)=m^2,
\end{align}
and that they are related to the well-known Pauli sigma matrices.  

Our next step is to understand how Lorentz transformations act on this matrix form of momentum. We anticipate that the transformation will involve two $2 \times 2$ matrices, each acting on one of the Lorentz indices. To do this in general, we need to find the set of transformations that preserve the determinant, which corresponds to the mass. We find that the momentum matrix transforms as follows:
\begin{align}
    p^{\prime}_{\alpha\dot{\beta}} = \Lambda_{\frac{1}{2}\alpha}^{\ \ \ \lambda}
    \Lambda_{\frac{1}{2}\dot{\beta}}^{*\ \ \dot{\omega}}
    p_{\lambda\dot{\omega}}
    \\
    p^{\prime\dot{\beta}\alpha}
    = 
    \Lambda_{\frac{1}{2}\ \dot{\omega}}^{*\dot{\beta}}
    \Lambda_{\frac{1}{2}\ \lambda}^{\ \alpha}
    p^{\dot{\omega}\lambda}
    ,
\end{align}
where $\Lambda_{\frac{1}{2}}$ is the defining representation of the Lorentz group $SL(2,\mathbb{C})$, and $\Lambda_{\frac{1}{2}}^*$ is the conjugate representation. The group $SL(2,\mathbb{C})$ consists of complex $2 \times 2$ matrices with unit determinant, possessing six degrees of freedom -- corresponding to three rotations and three boosts. This group is a covering group for $SO(3,1)$ and shares the same Lie algebra.

Given that this representation is complex, $\Lambda_{\frac{1}{2}}$ and $\Lambda_{\frac{1}{2}}^*$ are not related by a similarity transformation, necessitating the use of different Lorentz indices -- hence the undotted and dotted indices. It is important to note that $SO(3,1)$ itself does not have a two-dimensional representation. This is because a $2\pi$ rotation in $SO(3,1)$ is always equivalent to the identity, whereas all two-dimensional representations of $SL(2,\mathbb{C})$ require a $4\pi$ rotation to return to the identity. This distinction highlights the nature of $SL(2,\mathbb{C})$ as a double cover of $SO(3,1)$, where each element of $SO(3,1)$ corresponds to two elements in $SL(2,\mathbb{C})$.

A special set of $SL(2,\mathbb{C})$ generators, acting on lower Lorentz indices, are given and validated in \cite{Christensen:2024xsg}.  For convenience, we repeat them here:
\begin{align}
    \mathcal{J}^{(3)\ \beta}_{\quad \alpha} &=
    \frac{-1}{2}\left(
\begin{array}{cc}
 c_\theta &
   s_\theta e^{-i \phi } \\
   s_\theta e^{i \phi }  & -c_\theta \\
\end{array}
\right)
\label{eq:J^3 around p undotted}
\\
\mathcal{J}^{(+)\ \beta}_{\quad \alpha} &=
    \frac{-e^{-\eta}}{2}\left(
\begin{array}{cc}
 -s_\theta & e^{-i \phi
   } (c_\theta-1) \\
 e^{i \phi } (c_\theta+1) & s_\theta \\
\end{array}
\right)
\\
\mathcal{J}^{(-)\ \beta}_{\quad \alpha} &=
    \frac{-e^{\eta}}{2}\left(
\begin{array}{cc}
 -s_\theta & e^{-i \phi
   } (c_\theta+1) \\
 e^{i \phi } (c_\theta-1) & s_\theta \\
\end{array}
\right)
\\
\mathcal{J}^{(3)\ \dot{\beta}}_{\quad \dot{\alpha}} &=
    \frac{1}{2}\left(
\begin{array}{cc}
 c_\theta & e^{i \phi }
   s_\theta \\
 e^{-i \phi } s_\theta &
   -c_\theta \\
\end{array}
\right)
\label{eq:J^3 around p dotted}
\\
\mathcal{J}^{(+)\ \dot{\beta}}_{\quad \dot{\alpha}} &=
    \frac{e^{\eta}}{2}\left(
\begin{array}{cc}
 -s_\theta & e^{i \phi }
   (c_\theta+1) \\
 e^{-i \phi } (c_\theta-1) & s_\theta \\
\end{array}
\right)
\\
\mathcal{J}^{(-)\ \dot{\beta}}_{\quad \dot{\alpha}} &=
    \frac{e^{-\eta}}{2}\left(
\begin{array}{cc}
 -s_\theta & e^{i \phi }
   (c_\theta-1) \\
 e^{-i \phi } (c_\theta+1) & s_\theta \\
\end{array}
\right)
\label{eq:J^- around p dotted}
\end{align}
where 
\begin{equation}
    \eta = \frac{1}{2}\ln\left(\frac{E+p}{E-p}\right),
\end{equation}
in the massive case.  This representation of generators is aligned with the particle's momentum (its orientation and magnitude).  We can see that these generators satisfy the $SU(2)$ algebra, $\left[\mathcal{J}^{(+)},\mathcal{J}^{(-)}\right] = 2\mathcal{J}^{(3)}$ and $\left[\mathcal{J}^{(3)},\mathcal{J}^{(\pm)}\right] = \pm\mathcal{J}^{(\pm)}$.  In fact, as we will soon see, this is because these are the generators of the little-group spin for time-like momenta, that act on the Lorentz space. 

In the massless case $\mathcal{J}^{(3)}$ is unchanged.  However, we will not need $\mathcal{J}^{(\pm)}$.  This is because $\mathcal{J}^{(3)}$ is a representation of helicity, and not a member of spin, in this case.  Indeed, for massless fields, we will only need $\mathcal{J}^{(3)}$ in this form, and will not use $\mathcal{J}^{(\pm)}$ at all.  While normalized versions of these operators would still belong to $SL(2,\mathbb{C})$, they would not be part of the little group for light-like momenta and are therefore not of interest in this context.

We note that the generators with dotted indices are related to those with undotted indices by $\mathcal{J}^{(3)\ \dot{\beta}}_{\quad \dot{\alpha}} = - \left[\mathcal{J}^{(3)\ \beta}_{\quad \alpha}\right]^*$ and $\mathcal{J}^{(\pm)\ \dot{\beta}}_{\quad \dot{\alpha}} = -\left[\mathcal{J}^{(\mp)\ \beta}_{\quad \alpha}\right]^*$, as expected for a conjugate representation.  The generators acting on upper Lorentz indices are given by raising and lowering these Lorentz indices with the antisymmetric epsilon tensor, which is numerically the same as the tensor used to raise and lower spin indices.  These generators are given by:
\begin{align}
    \mathcal{J}^{(3)\alpha}_{\quad \ \ \beta} &= 
    \frac{1}{2}\left(
\begin{array}{cc}
 c_\theta & e^{i \phi }
   s_\theta \\
 e^{-i \phi } s_\theta &
   -c_\theta \\
\end{array}
\right)
\\
    \mathcal{J}^{(+)\alpha}_{\quad \ \ \beta} &= \frac{e^{-\eta}}{2}\left(
\begin{array}{cc}
 -s_\theta & e^{i \phi }
   (c_\theta+1) \\
 e^{-i \phi } (c_\theta-1) & s_\theta \\
\end{array}
\right)
\\
    \mathcal{J}^{(-)\alpha}_{\quad \ \ \beta} &= \frac{e^{\eta}}{2}\left(
\begin{array}{cc}
 -s_\theta & e^{i \phi }
   (c_\theta-1) \\
 e^{-i \phi } (c_\theta+1) & s_\theta \\
\end{array}
\right) 
\\
    \mathcal{J}^{(3)\dot{\alpha}}_{\quad \ \ \dot{\beta}} &= 
    \frac{-1}{2}\left(
\begin{array}{cc}
 c_\theta & e^{-i \phi }
   s_\theta \\
 e^{i \phi } s_\theta &
   -c_\theta \\
\end{array}
\right)
\\
    \mathcal{J}^{(+)\dot{\alpha}}_{\quad \ \ \dot{\beta}} &= \frac{-e^{\eta}}{2}\left(
\begin{array}{cc}
 -s_\theta & e^{-i \phi
   } (c_\theta-1) \\
 e^{i \phi } (c_\theta+1) & s_\theta \\
\end{array}
\right)
\\
    \mathcal{J}^{(-)\dot{\alpha}}_{\quad \ \ \dot{\beta}} &= \frac{-e^{-\eta}}{2}\left(
\begin{array}{cc}
 -s_\theta & e^{-i \phi
   } (c_\theta+1) \\
 e^{i \phi } (c_\theta-1) & s_\theta \\
\end{array}
\right) .
\end{align}
Once again, we will only need $\mathcal{J}^{(3)}$ in the massless case.

To determine the action of these generators on the momentum, we need to exponentiate them. In order to keep this section simple, we will only consider a rotation around the axis of the momentum, meaning we will only exponentiate $\mathcal{J}^{(3)}$ here.  However, the result we find will be more general, as we will discuss below. As with the spin generators, we find that $\left(\mathcal{J}^{(3)}\right)^2 = \left(\frac{1}{2}\right)^2$, leading to the expression $e^{i\omega\mathcal{J}^3} = \cos\left(\frac{\omega}{2}\right) + i2\mathcal{J}^{3}\sin\left(\frac{\omega}{2}\right)$. This results in:
\begin{align}
    \left(e^{i\omega\mathcal{J}^3}\right)_{\alpha}^{\ \beta}  
   = &
   \\
   &\hspace{-0.65in}
   \left(\begin{array}{cc}
        \cos\left(\frac{\omega}{2}\right) - i c_{\theta} \sin\left(\frac{\omega}{2}\right)  & 
        -i s_{\theta} e^{-i\phi} \sin\left(\frac{\omega}{2}\right)\\
        -i s_{\theta} e^{i\phi} \sin\left(\frac{\omega}{2}\right) & 
        \cos\left(\frac{\omega}{2}\right) + i c_{\theta} \sin\left(\frac{\omega}{2}\right)
   \end{array}\right)
   \nonumber
   \\
   \left(e^{i\omega\mathcal{J}^3}\right)_{\dot{\alpha}}^{\ \dot{\beta}}  
   = &
   \\
   &\hspace{-0.65in}
   \left(\begin{array}{cc}
        \cos\left(\frac{\omega}{2}\right) + i c_{\theta} \sin\left(\frac{\omega}{2}\right)  & 
        i s_{\theta} e^{i\phi} \sin\left(\frac{\omega}{2}\right)\\
        i s_{\theta} e^{-i\phi} \sin\left(\frac{\omega}{2}\right) & 
        \cos\left(\frac{\omega}{2}\right) - i c_{\theta} \sin\left(\frac{\omega}{2}\right)
   \end{array}\right)
   \nonumber
   \\
   \left(e^{i\omega\mathcal{J}^3}\right)^{\alpha}_{\ \beta}  
   = &
   \\
   &\hspace{-0.65in}
   \left(\begin{array}{cc}
        \cos\left(\frac{\omega}{2}\right) + i c_{\theta} \sin\left(\frac{\omega}{2}\right)  & 
        i s_{\theta} e^{i\phi} \sin\left(\frac{\omega}{2}\right)\\
        i s_{\theta} e^{-i\phi} \sin\left(\frac{\omega}{2}\right) & 
        \cos\left(\frac{\omega}{2}\right) - i c_{\theta} \sin\left(\frac{\omega}{2}\right)
   \end{array}\right)
   \nonumber
   \\
   \left(e^{i\omega\mathcal{J}^3}\right)^{\dot{\alpha}}_{\ \dot{\beta}}  
   = &
   \\
   &\hspace{-0.65in}
   \left(\begin{array}{cc}
        \cos\left(\frac{\omega}{2}\right) - i c_{\theta} \sin\left(\frac{\omega}{2}\right)  & 
        -i s_{\theta} e^{-i\phi} \sin\left(\frac{\omega}{2}\right)\\
        -i s_{\theta} e^{i\phi} \sin\left(\frac{\omega}{2}\right) & 
        \cos\left(\frac{\omega}{2}\right) + i c_{\theta} \sin\left(\frac{\omega}{2}\right)
   \end{array}\right)
   \nonumber  .
\end{align}

Consequently, we find that:
\begin{align}
    \left(e^{i\omega\mathcal{J}^3}\right)_{\alpha}^{\ \zeta} 
    \left(e^{i\omega\mathcal{J}^3}\right)_{\dot{\beta}}^{\ \dot{\eta}} 
    p_{\zeta\dot{\eta}}
    &=
    \left(\begin{array}{cc}
    \mathcal{E}+p c_{\theta} & p s_{\theta}e^{-i\phi}\\
    p s_{\theta}e^{i\phi} & \mathcal{E}-p c_{\theta}
    \end{array}\right)
    \nonumber\\
    &= p_{\alpha\dot{\beta}}.
\end{align}
That is, these generators possess the special property of leaving the momentum unchanged.  The generator $\mathcal{J}^{(3)}$ is responsible for rotations around the momentum's axis, which will leave the momentum unchanged, whether massive or massless.  However, $\mathcal{J}^{(1)}$ and $\mathcal{J}^{(2)}$ only correspond with spatial rotations in the rest frame of the particle.  In other frames, they are more complicated, but still have the property that they do not change the momentum.  Generally speaking, these operators should not be thought of as generators of spatial rotation.  Rather, they are the generators of spin, which act on Lorentz indices, whereas, the generators of spin introduced in the last subsection act on spin indices.  Although they act on different ``spaces'' (Lorentz versus spin), they are generators of the same group.

As we have seen, the Lorentz spin generators we have discussed so far are specific to each particle, and are aligned with the particle's momentum. These spin generators will be used in the next subsection. However, in order to understand the Lorentz invariance of the action and handle general Lorentz transformations, it is also essential to consider the generators of spatial rotations that are fixed and independent of any particle's momentum.

For this purpose, we introduce the following fixed representation of the Lorentz rotation generators:
\begin{align}
    \mathfrak{J}^{(3)\ \beta}_{\quad \alpha} &=
    \left(
\begin{array}{cc}
 -\frac{1}{2} & 0 \\
   0  & \frac{1}{2} \\
\end{array}
\right)
& 
\mathfrak{J}^{(3)\alpha}_{\quad \ \ \beta} &= 
    \left(
\begin{array}{cc}
 \frac{1}{2} & 0 \\
 0 & -\frac{1}{2} \\
\end{array}
\right)
\\
\mathfrak{J}^{(1)\ \beta}_{\quad \alpha} &=
    \left(
\begin{array}{cc}
 0 & -\frac{1}{2} \\
 -\frac{1}{2} & 0 \\
\end{array}
\right)
&
\mathfrak{J}^{(1)\alpha}_{\quad \ \ \beta} &= \left(
\begin{array}{cc}
 0 & \frac{1}{2} \\
 \frac{1}{2} & 0 \\
\end{array}
\right)
\\
\mathfrak{J}^{(2)\ \beta}_{\quad \alpha} &=
    \left(
\begin{array}{cc}
 0 & -\frac{i}{2} \\
 \frac{i}{2} & 0 \\
\end{array}
\right)
&
\mathfrak{J}^{(2)\alpha}_{\quad \ \ \beta} &= \left(
\begin{array}{cc}
 0 & -\frac{i}{2} \\
 \frac{i}{2} & 0 \\
\end{array}
\right)
\\
\mathfrak{J}^{(3)\ \dot{\beta}}_{\quad \dot{\alpha}} &=
    \left(
\begin{array}{cc}
 \frac{1}{2} & 0 \\
 0 & -\frac{1}{2} \\
\end{array}
\right)
&
\mathfrak{J}^{(3)\dot{\alpha}}_{\quad \ \ \dot{\beta}} &= 
    \left(
\begin{array}{cc}
 -\frac{1}{2} & 0 \\
 0 & \frac{1}{2} \\
\end{array}
\right)
\\
\mathfrak{J}^{(1)\ \dot{\beta}}_{\quad \dot{\alpha}} &=
    \left(
\begin{array}{cc}
 0 & \frac{1}{2} \\
 \frac{1}{2} & 0 \\
\end{array}
\right)
&
\mathfrak{J}^{(1)\dot{\alpha}}_{\quad \ \ \dot{\beta}} &= \left(
\begin{array}{cc}
 0 & -\frac{1}{2} \\
 -\frac{1}{2} & 0 \\
\end{array}
\right)
\\
\mathfrak{J}^{(2)\ \dot{\beta}}_{\quad \dot{\alpha}} &=
    \left(
\begin{array}{cc}
 0 & -\frac{i}{2} \\
 \frac{i}{2} & 0 \\
\end{array}
\right)
&
\mathfrak{J}^{(2)\dot{\alpha}}_{\quad \ \ \dot{\beta}} &= \left(
\begin{array}{cc}
 0 & -\frac{i}{2} \\
 \frac{i}{2} & 0 \\
\end{array}
\right).
\end{align}
These generators are valid for both massive and massless momenta.  They are also generators of $SU(2)$ however, strictly speaking, they are not generators of the little group, except in special cases.

To fully understand Lorentz invariance, we also need to consider the group elements generated by exponentiation of the rotation generators. Without loss of generality, we will choose the $z$-axis as the rotation axis. The group element corresponding to a rotation by an angle $\bar{\omega}$ around the $z$-axis is given by: 
\begin{align}
    e^{i\bar{\omega}\mathfrak{J}^3}
    &= \cos\left(\frac{\bar{\omega}}{2}\right)
    + i 2 \mathfrak{J}^3\sin\left(\frac{\bar{\omega}}{2}\right).
\end{align}
Therefore,
\begin{align}
   \left(e^{i\bar{\omega}\mathfrak{J}^3}\right)_{\alpha}^{\ \beta}  
   &= 
   \left(\begin{array}{cc}
        e^{-i\frac{\bar{\omega}}{2}}  & 0\\
        0 & e^{+i\frac{\bar{\omega}}{2}}
   \end{array}\right)
   \\
   \left(e^{i\bar{\omega}\mathfrak{J}^3}\right)_{\dot{\alpha}}^{\ \dot{\beta}}  
   &= 
   \left(\begin{array}{cc}
        e^{+i\frac{\bar{\omega}}{2}} & 0\\
        0 & e^{-i\frac{\bar{\omega}}{2}}
   \end{array}\right) 
   \\
   \left(e^{i\bar{\omega}\mathfrak{J}^3}\right)^{\alpha}_{\ \beta}  
   &= 
   \left(\begin{array}{cc}
        e^{+i\frac{\bar{\omega}}{2}} & 0\\
        0 & e^{-i\frac{\bar{\omega}}{2}}
   \end{array}\right)
   \\
   \left(e^{i\bar{\omega}\mathfrak{J}^3}\right)^{\dot{\alpha}}_{\ \dot{\beta}}  
   &= 
   \left(\begin{array}{cc}
        e^{-i\frac{\bar{\omega}}{2}}  & 0\\
        0 & e^{+i\frac{\bar{\omega}}{2}}
   \end{array}\right)  .
\end{align}

From this, we observe that if Lorentz indices are contracted, the product remains Lorentz invariant under rotations. For instance, consider two objects $A_{\alpha}$ and $B^{\alpha}$ that transform under $\Lambda_{\frac{1}{2}}$. The contracted product of these objects transforms as:
\begin{align}
    A'_{\alpha}B'^{\alpha} 
    &= A_{\beta}\left(e^{i\bar{\omega}\mathfrak{J}^3}\right)_{\alpha}^{\ \beta}  
    \left(e^{i\bar{\omega}\mathfrak{J}^3}\right)^{\alpha}_{\ \lambda} 
    B^{\lambda}
    = A_{\alpha}B^{\alpha}.
    \label{eq:A'_alpha B'^alpha = A_alpha B^alpha}
\end{align}

We can confirm that this representation yields the correct rotation of the momentum around the $z$-axis, without loss of generality. For a rotation by an angle $\bar{\omega}$, we find:
\begin{align}
    \left(e^{i\bar{\omega}\mathfrak{J}^3}\right)_{\alpha}^{\ \zeta} 
    \left(e^{i\bar{\omega}\mathfrak{J}^3}\right)_{\dot{\beta}}^{\ \dot{\eta}} 
    p_{\zeta\dot{\eta}}
    =
    \left(\begin{array}{cc}
    \mathcal{E}+p c_{\theta} & p s_{\theta}e^{-i(\phi+\bar{\omega})}\\
    p s_{\theta}e^{i(\phi+\bar{\omega})} & \mathcal{E}-p c_{\theta}
    \end{array}\right).
\end{align}

We now turn to the Lorentz boosts, which, like the rotations, can be derived from the condition that $det(p) = m^2$ is unchanged. Interestingly, despite this general derivation, the boosts exhibit a straightforward relationship with the generators of rotations. Specifically, we find:
\begin{align}
    \mathcal{K} = \pm i \mathfrak{J},
    \label{eq:K=-iJ}
\end{align}
where the upper sign corresponds to undotted indices, and the lower sign corresponds to dotted indices. This result naturally satisfies the Lorentz algebra:
\begin{align}
    \left[ \mathfrak{J}_j , \mathfrak{J}_k \right] &= i \epsilon_{jkl}\mathfrak{J}_l
    \\
    \left[ \mathfrak{J}_j , \mathcal{K}_k \right] &= i \epsilon_{jkl}\mathcal{K}_l
    \\
    \left[ \mathcal{K}_j , \mathcal{K}_k \right] &= -i \epsilon_{jkl}\mathfrak{J}_l.
\end{align}
As we can see, in contrast to the rotation generators, the boost generators do not form a closed subalgebra on their own.

Exponentiating the generator gives the corresponding group element. As with rotations, we align our $z$-axis along the direction of the boost, without loss of generality. However, unlike rotations, the square of our boost generator flips signs, yielding $-\left(\frac{1}{2}\right)^2$. Otherwise, following a similar logic, we obtain:
\begin{align}
    e^{i\eta\mathcal{K}^3}
    &= \cosh\left(\frac{\eta}{2}\right)
    \mp 2 \mathfrak{J}^3\sinh\left(\frac{\eta}{2}\right).
\end{align}
Consequently, we have:
\begin{align}
     \left(e^{i\eta\mathcal{K}^3}\right)_{\alpha}^{\ \beta}  
   &=  
   \left(\begin{array}{cc}
        e^{\frac{\eta}{2}} & 0\\
        0 & e^{-\frac{\eta}{2}}
   \end{array}\right)
   \\
   \left(e^{i\eta\mathcal{K}^3}\right)_{\dot{\alpha}}^{\ \dot{\beta}}  
   &=  
   \left(\begin{array}{cc}
        e^{\frac{\eta}{2}} & 0\\
        0 & e^{-\frac{\eta}{2}}
   \end{array}\right)
   \\
   \left(e^{i\eta\mathcal{K}^3}\right)^{\alpha}_{\ \beta}  
   &=  
   \left(\begin{array}{cc}
        e^{\frac{-\eta}{2}} & 0\\
        0 & e^{\frac{\eta}{2}}
   \end{array}\right)
   \\
   \left(e^{i\eta\mathcal{K}^3}\right)^{\dot{\alpha}}_{\ \dot{\beta}}  
   &=  
   \left(\begin{array}{cc}
        e^{-\frac{\eta}{2}} & 0\\
        0 & e^{\frac{\eta}{2}}
   \end{array}\right).
\end{align}
Therefore, products with contracted Lorentz indices are also invariant under boosts.  For example,
\begin{align}
    A'_{\alpha}B'^{\alpha} 
    &= A_{\beta}\left(e^{i\eta\mathcal{K}^3}\right)_{\alpha}^{\ \beta}  
    \left(e^{i\eta\mathcal{K}^3}\right)^{\alpha}_{\ \lambda} 
    B^{\lambda}
    = A_{\alpha}B^{\alpha}.
    \label{eq:A'_alpha B'^alpha = A_alpha B^alpha Boosts}
\end{align}

We will also confirm that we get the expected boosts in a simple example, where we choose the boost direction to be the $z-$direction, without loss of generality.  We have
\begin{align}
    \left(e^{i\eta\mathcal{K}^3}\right)_{\alpha}^{\ \zeta} 
    \left(e^{i\eta\mathcal{K}^3}\right)_{\dot{\beta}}^{\ \dot{\lambda}} 
    p_{\zeta\dot{\lambda}}
    &= \left(\begin{array}{cc}
    \bar{\mathcal{E}}+\bar{p} c_{\bar{\theta}} & \bar{p} s_{\bar{\theta}}e^{-i\phi}\\
    \bar{p} s_{\bar{\theta}}e^{i\phi} & \bar{\mathcal{E}}-\bar{p} c_{\bar{\theta}}
    \end{array}\right),
\end{align}
where
\begin{align}
    \bar{\mathcal{E}}\pm\bar{p}c_{\bar{\theta}} 
    &= (\mathcal{E}\pm pc_{\theta})e^{\pm\eta}
    \\
    \bar{p}s_{\bar{\theta}} 
    &= ps_{\theta},
\end{align}
leading to 
\begin{align}
    \bar{\mathcal{E}} &= 
    \mathcal{E}\cosh(\eta) 
    + p c_{\theta}\sinh(\eta)\\
    \bar{p}c_{\bar{\theta}} &= 
    \mathcal{E}\sinh(\eta)
     + p c_{\theta}\cosh(\eta), 
\end{align}
as expected for a boost.  

This is a good point to talk about the chiral group $SU(2)_L\times SU(2)_R$.  Noticing that the generators for boosts and rotations are related in Eq.~(\ref{eq:K=-iJ}), we see that we can combine them to form
\begin{align}
    \mathcal{A} &= 
    \frac{1}{2}\left( \mathfrak{J} - i \mathcal{K} \right) \\
    \mathcal{B} &= 
    \frac{1}{2}\left( \mathfrak{J} + i \mathcal{K} \right).
\end{align}
These generators satisfy two separate closed $SU(2)$ algebras,
\begin{align}
    \left[\mathcal{A}_j , \mathcal{A}_k\right] 
    &= i\epsilon_{jkl}\mathcal{A}_l
    \\
    \left[\mathcal{B}_j , \mathcal{B}_k\right] 
    &= i\epsilon_{jkl}\mathcal{B}_l
    \\
    \left[\mathcal{A}_j , \mathcal{B}_k\right] 
    &= 0.
\end{align}
In fact, for undotted and dotted indices, we find
\begin{align}
    \mathcal{A}_{\alpha}^{\ \beta}
    &= \mathfrak{J}_{\alpha}^{\ \beta}
    \\
    \mathcal{B}_{\alpha}^{\ \beta}
    &= 0
    \\
    \mathcal{A}_{\dot{\alpha}}^{\ \dot{\beta}}
    &= 0
    \\
    \mathcal{B}_{\dot{\alpha}}^{\ \dot{\beta}}
    &= \mathfrak{J}_{\dot{\alpha}}^{\ \dot{\beta}}
\end{align}
This is the chiral form of the Lorentz group.  Representations of $\mathcal{A}$ are said to be left chiral, while representations of $\mathcal{B}$ are right chiral.  We see that the undotted indices transform under the left-chiral group, while dotted indices transform under the right-chiral group.

Even so, it should be remembered that a rotation on its own will transform under both $\mathcal{A}$ and $\mathcal{B}$ at the same time.  The same is true for boosts.  Only a special combination of rotations and boosts, $\vec{\eta} = \pm i\vec{\theta}$, would transform purely under one or the other.  However, physical boosts and rotations are real.  Nevertheless, this group will be crucial for understanding chiral interactions in constructive field theory.

Before concluding this subsection, it is important to emphasize again that these Lorentz operators do \textit{not} act directly on our fields. Instead, our fields transform under the little group, which corresponds purely to helicity and spin representations, rather than under the full Lorentz group representations. In particular, they have spin indices (or no index in the massless case).  They do not have Lorentz indices. This is a significant distinction from traditional field theory, where fields do transform under Lorentz group representations, and the Lorentz operators act directly on them and their Lorentz indices. Despite this difference, we will still require the Lorentz representations to construct Lorentz-invariant terms in the action.

\subsection{\label{app:spinors}Helicity and Spin Spinors}
As we discussed at the beginning of the previous subsection, in order to construct field theory action terms, we need non-field objects that transform under the little group that can be combined with the fields to create Lorentz invariants.  Since the fields transform with different phase angles $\omega_i=\omega(\Lambda,p_i)$, the little-group transformations can not cancel directly between the fields.  Moreover, as a further example, we considered the interaction of a $W$ boson, a charged lepton and a neutrino.  Between the three of these, we have three spin indices and one helicity transformation, with no way to even naively pair them up.  Since these transformations cannot be cancelled purely among these fields, we need to multiply these with non-field objects that also transform under spin and helicity and cancel their transformations.  

In order to build this up naturally, we will begin with a massless momentum, which can be written:
\begin{align}
    p_{\alpha \dot{\beta}} &=
    2\mathcal{E}\left(\begin{array}{cc}
    c^2 & c s^*\\
    c s & s s^*
    \end{array}\right)\\
    p^{\dot{\alpha} \beta} &=
    2\mathcal{E}\left(\begin{array}{cc}
    s s^* & -c s^*\\
    -c s & c^2
    \end{array}\right),
\end{align}
where $c=\cos\left(\frac{\theta}{2}\right)$ and $s=\sin\left(\frac{\theta}{2}\right)e^{i\phi}$.  As expected for a light-like momentum, the determinant of these are zero.  This means that these matrices are rank one and, therefore, their rows (or columns) are linearly dependent.  Consequently, these matrices can be factorized as a product of a $2 \times 1$ column and a $1 \times 2$ row.

Focusing first on $p_{i\alpha \dot{\beta}}$, we call the column $\lvert i \rangle_{\alpha}$ and the row $\lbrack i \rvert_{\dot{\beta}}$, where $i$ represents the $i^{\text{th}}$ momentum. This notation is shorthand for $\lvert i\rangle = \lvert p_i \rangle $ and $ \lbrack i\rvert = \lbrack p_i \rvert $. With this, we can express $p_{i\alpha \dot{\beta}}$ as:
\begin{align}
    p_{i\alpha \dot{\beta}} &= 
    \lvert i\rangle_{\alpha} \lbrack i\rvert_{\dot{\beta}}.
\end{align}

Considering that $\lvert i\rangle_{\alpha}$ transforms under the undotted representation of the Lorentz group $SL(2,\mathbb{C})$ and $\lbrack i\rvert_{\dot{\beta}}$ under the dotted (conjugate) representation, we expect that $\lbrack i\rvert$ will be related to the conjugate of $\lvert i\rangle$.  Furthermore, they should cary the parts of the momentum equally.  Inspection leads us to:
\begin{align}
    \lvert i\rangle_{\alpha}
    &= \sqrt{2\mathcal{E}}
    \left(\begin{array}{c}
    c \\ s
    \end{array}\right)
    \label{eq:|i> def}
    \\
    \lbrack i\rvert_{\dot{\beta}} 
    &= \sqrt{2\mathcal{E}}
    \left(\begin{array}{c}
    c \\ s^*
    \end{array}\right).
    \label{eq:[i| def}
\end{align}

If we raise the Lorentz indices with the epsilon tensor, we have:
\begin{align}
    \langle i\rvert^{\alpha}
    &= \sqrt{2\mathcal{E}}
    \left(\begin{array}{c}
    s \\ -c
    \end{array}\right)
    \\
    \lvert i\rbrack^{\dot{\beta}} 
    &= \sqrt{2\mathcal{E}}
    \left(\begin{array}{c}
    s^* \\ -c
    \end{array}\right),
\end{align}
which is just what we need for:
\begin{align}
    p_i^{\dot{\alpha}\beta} &=
    \lvert i\rbrack^{\dot{\alpha}} \langle i\rvert^{\beta}.
\end{align}

We refer to these as helicity spinors, a term whose significance will become clearer shortly. The Lorentz index on angle spinors is always undotted and lower on right-angle spinors but upper on left-angle spinors, while the Lorentz index on square spinors is always dotted and lower on left-square spinors but upper on right-square spinors.  This notation will allow us to leave the Lorentz index implicit when there is no ambiguity.

To understand how these spinors transform under Lorentz transformations, we might initially guess that the transformation of the spinor is given by $\lvert \bar{1}\rangle_{\alpha} = \Lambda_{\frac{1}{2}\alpha}^{\ \ \beta}\lvert 1\rangle_{\beta}$, where $\bar{p} = \Lambda p$. However, this naive guess is not entirely accurate. Let us clarify this with a specific example.

Consider a rotation by an angle $\bar{\omega}$ around the $z$-axis. In this case, the only change to the momentum is the azimuthal angle, which shifts to $\phi + \bar{\omega}$. However, if we apply this rotation to the spinor, we find:
\begin{align}
    \left(e^{i\bar{\omega}\mathfrak{J}^3}\right)_{\alpha}^{\ \beta} \lvert i\rangle_{\beta} &= 
    \sqrt{2\mathcal{E}}
    \left(\begin{array}{cc}
        c e^{-i\frac{\bar{\omega}}{2}}
        \\
        s e^{i\frac{\bar{\omega}}{2}}
    \end{array}\right)
    \neq
    \sqrt{2\mathcal{E}}
    \left(\begin{array}{cc}
        c 
        \\
        s e^{i\bar{\omega}}
    \end{array}\right).
\end{align}
Instead, we observe:
\begin{align}
    \left(e^{i\bar{\omega}\mathfrak{J}^3}\right)_{\alpha}^{\ \beta} \lvert i\rangle_{\beta} &=
    e^{-i\frac{\bar{\omega}}{2}}
    \lvert \bar{i}\rangle_{\alpha},
\end{align}
where $\bar{p}_i = \Lambda p_i$.  We recognize this as a helicity transformation on the right side.  We see that the angle helicity spinors transform as helicity-$-\frac{1}{2}$ objects under Lorentz transformation.  It turns out that square helicity spinors transform as helicity-$+\frac{1}{2}$ objects.

In order to develop this generally without going through every example, we define these spinors by a standard boost from a standard set of spinors, $\lvert k\rangle$ and $\lbrack k\rvert$.  Therefore,
\begin{align}
    \lvert i\rangle_\alpha 
    &= L_{\frac{1}{2}}(p_i)_{\alpha}^{\ \beta} 
    \lvert k\rangle_{\beta}
    \\
    \lbrack i\rvert_{\dot{\alpha}}
    &= L_{\frac{1}{2}}^*(p_i)_{\dot{\alpha}}^{\ \dot{\beta}}
    \lbrack k\rvert_{\dot{\beta}},
\end{align}
where $L(p)$ is the standard boost discussed in App.~\ref{app:single particle states}.  We can now see that a Lorentz transformation acting on this gives:
\begin{align}
    \Lambda_{\frac{1}{2}\alpha}^{\ \ \beta}\lvert i\rangle_{\beta}
    &= \Lambda_{\frac{1}{2}\alpha}^{\ \ \beta}
    L_{\frac{1}{2}}(p_i)_{\beta}^{\ \gamma}
    L^{-1}_{\frac{1}{2}}(\bar{p}_i)_{\gamma}^{\ \epsilon}\lvert \bar{i}\rangle_{\epsilon}
    \\
    \Lambda_{\frac{1}{2}\dot{\alpha}}^{*\ \ \dot{\beta}}\lbrack i\rvert_{\dot{\beta}}
    &= \Lambda_{\frac{1}{2}\dot{\alpha}}^{*\ \ \dot{\beta}}
    L^*_{\frac{1}{2}}(p_i)_{\dot{\beta}}^{\ \dot{\gamma}}
    L^{*-1}_{\frac{1}{2}}(\bar{p}_i)_{\dot{\gamma}}^{\ \dot{\epsilon}}\lbrack \bar{i}\rvert_{\dot{\epsilon}}.
\end{align}

We have seen the Lorentz transformation on the right in Eq.~(\ref{eq:U[Lambda L(p1)L-1(p2)] Psi^h}).  It is the Lorentz transformation that takes $\bar{p}_i\to k\to p_i\to \bar{p}_i$.  Therefore, it is a transformation that leaves the momentum unchanged.  In the present massless case, it is a rotation around the axis of the momentum.  The generator for this transformation is $\mathcal{J}^{(3)}$ given in Eqs.~(\ref{eq:J^3 around p undotted}) and (\ref{eq:J^3 around p dotted}).  Therefore, we find:
\begin{align}
    \Lambda_{\frac{1}{2}\alpha}^{\ \ \beta}\lvert i\rangle_{\beta}
    &= \left(e^{i\omega_i\mathcal{J}^3}\right)_{\alpha}^{\ \beta}  
    \lvert \bar{i}\rangle_{\beta}
    \label{eq:Lambda |1> = e^iomega J3 |2>}
    \\
    \Lambda_{\frac{1}{2}\dot{\alpha}}^{*\ \ \dot{\beta}}\lbrack i\rvert_{\dot{\beta}}
    &= \left(e^{i\omega_i\mathcal{J}^3}\right)_{\dot{\alpha}}^{\ \dot{\beta}}  
    \lbrack \bar{i}\rvert_{\dot{\beta}},
\end{align}
where $\omega_i=\omega(\Lambda,p_i)$ is the same function of the Lorentz transformation and momentum as in App.~\ref{app:single particle states}.  

Moreover, we can see that $\lvert i\rangle$ and $\lbrack i\rvert$ are eigenstates of $\mathcal{J}^{(3)}$, with
\begin{align}
    \mathcal{J}^{(3)} \lvert i\rangle
    &= -\frac{1}{2}\lvert i\rangle
    \\
    \mathcal{J}^{(3)} \lbrack i\rvert
    &= +\frac{1}{2}\lbrack i\rvert.
\end{align}
Consequently,
\begin{align}
    \Lambda_{\frac{1}{2}\alpha}^{\ \ \beta}\lvert i\rangle_{\beta}
    &= e^{-i\frac{\omega_i}{2}}
    \lvert \bar{i}\rangle_{\alpha}
    \\
    \Lambda_{\frac{1}{2}\dot{\alpha}}^{*\ \ \dot{\beta}}\lbrack i\rvert_{\dot{\beta}}
    &= e^{+i\frac{\omega_i}{2}} 
    \lbrack \bar{i}\rvert_{\dot{\alpha}}.
    \label{eq:Lambda^*[1|=e^iomega/2 [2|}
\end{align}
Lorentz transformation takes these helicity spinors to the spinors with the new momentum, multiplied by a phase and helicity equal to $\mp\frac{1}{2}$ for the angle and square spinors.  These spinors will be very useful for building Lorentz invariants with fields.

We could have approached this process in reverse, starting from the desired transformation properties described in Eqs.~(\ref{eq:Lambda |1> = e^iomega J3 |2>}) through (\ref{eq:Lambda^*[1|=e^iomega/2 [2|}). However, deriving the correct form from the momentum structure led us naturally to the right spinor form. For the spin spinors, however, we will take a different approach. Here, we will begin by specifying the transformation properties we want, informed by the insights gained from the massless case, and then work our way back to the momentum.

As we begin the search for massive spinors, in order to make the notation more economical in the future, we will find it useful boldface the index on massive spin spinors in order to distinguish them from the massless helicity spinors, which will not be boldfaced.  

Looking back at Eqs.~(\ref{eq:e^iomega J Psi^K = U[Lambda]Psi^I}) and (\ref{eq:e^iomega J Psi^K = U[Lambda L(p1)L-1(p2)]Psi^I}), we recognize that the property we seek in our massive spinors is that, under a Lorentz transformation, they should transform according to the spin representation at the new momentum. Specifically, we want to construct a spin spinor, denoted with a spin index, that transforms as follows:
\begin{align}
    \Lambda_{\frac{1}{2}\alpha}^{\ \ \beta}\lvert \mathbf{i}\rangle_{\beta}^{\mathrm{I}}
    &= \left(e^{i\vec{\omega}_i\cdot\vec{\mathcal{J}}}\right)_{\alpha}^{\ \beta}  
    \lvert \bar{\mathbf{i}}\rangle_{\beta}^{\mathrm{I}}
    = \left(e^{i\vec{\omega}_i\cdot\vec{J}}\right)_{\mathrm{K}}^{\ \mathrm{I}}  
    \lvert \bar{\mathbf{i}}\rangle_{\alpha}^{\mathrm{K}}
    \label{eq:Lambda|1>=eJ|2>=eJ|2>}
    \\
    \Lambda_{\frac{1}{2}\dot{\alpha}}^{*\ \ \dot{\beta}}\lbrack \mathbf{i}\rvert_{\dot{\beta}}^{\mathrm{I}}
    &= \left(e^{i\vec{\omega}_i\cdot\vec{\mathcal{J}}}\right)_{\dot{\alpha}}^{\ \dot{\beta}}  
    \lbrack \bar{\mathbf{i}}\rvert_{\dot{\beta}}^{\mathrm{I}}
    = \left(e^{i\vec{\omega}_i\cdot\vec{J}}\right)_{\mathrm{K}}^{\ \mathrm{I}}  
    \lbrack \bar{\mathbf{i}}\rvert_{\dot{\alpha}}^{\mathrm{K}}
    \label{eq:Lambda[1|=eJ[2|=eJ[2|}
    ,
\end{align}
where $\vec{\omega}_i=\vec{\omega}(\Lambda,p_i)$, $\vec{\mathcal{J}}$ are the generators of spin given in Eqs.~(\ref{eq:J^3 around p undotted}) through (\ref{eq:J^- around p dotted}) and $\vec{J}$ are the generators of spin given in Eqs.~(\ref{eq:J^3_K^J on spin}) through (\ref{eq:J^-_K^J on spin}).  In particular, $\vec{\mathcal{J}}$ are the generators that do not change the momentum.  They result in transformations that take $\bar{p}_i\to k\to p_i\to \bar{p}_i$ and correspond to the transformations on the right of
\begin{align}
    \Lambda_{\frac{1}{2}\alpha}^{\ \ \beta}\lvert \mathbf{i}\rangle_{\beta}^{\mathrm{I}}
    &= \Lambda_{\frac{1}{2}\alpha}^{\ \ \beta}
    L_{\frac{1}{2}}(p_i)_{\beta}^{\ \gamma}
    L^{-1}_{\frac{1}{2}}(\bar{p}_i)_{\gamma}^{\ \epsilon}\lvert \bar{\mathbf{i}}\rangle_{\epsilon}^{\mathrm{I}}
    \\
    \Lambda_{\frac{1}{2}\dot{\alpha}}^{*\ \ \dot{\beta}}\lbrack \mathbf{i}\rvert_{\dot{\beta}}^{\mathrm{I}}
    &= \Lambda_{\frac{1}{2}\dot{\alpha}}^{*\ \ \dot{\beta}}
    L^*_{\frac{1}{2}}(p_i)_{\dot{\beta}}^{\ \dot{\gamma}}
    L^{*-1}_{\frac{1}{2}}(\bar{p}_i)_{\dot{\gamma}}^{\ \dot{\epsilon}}\lbrack \bar{\mathbf{i}}\rvert_{\dot{\epsilon}}^{\mathrm{I}},
\end{align}
where we define
\begin{align}
    \lvert \mathbf{i}\rangle_\alpha^{\mathrm{I}}
    &= L_{\frac{1}{2}}(p_i)_{\alpha}^{\ \beta} 
    \lvert \mathbf{k}\rangle_{\beta}^{\mathrm{I}}
    \\
    \lbrack \mathbf{i}\rvert_{\dot{\alpha}}^{\mathrm{I}}
    &= L_{\frac{1}{2}}^*(p_i)_{\dot{\alpha}}^{\ \dot{\beta}}
    \lbrack \mathbf{k}\rvert_{\dot{\beta}}^{\mathrm{I}},
\end{align}
and $k$ is a standard momentum for the paricle, typically the rest momentum.  

In particular, from Eqs.~(\ref{eq:Lambda|1>=eJ|2>=eJ|2>}) and (\ref{eq:Lambda[1|=eJ[2|=eJ[2|}), we want our spinors to satisfy
\begin{align}
    \left(\mathcal{J}^{(j)}\right)_{\alpha}^{\ \beta}
    \lvert \mathbf{i}\rangle_{\beta}^{\mathrm{I}} 
    &= \left(J^{(j)}\right)_{\mathrm{K}}^{\ \ \mathrm{I}}
    \lvert \mathbf{i}\rangle_{\alpha}^{\mathrm{K}}
    \\
    \left(\mathcal{J}^{(j)}\right)_{\dot{\alpha}}^{\ \dot{\beta}}
    \lbrack \mathbf{i}\rvert_{\dot{\beta}}^{\mathrm{I}} 
    &= \left(J^{(j)}\right)_{\mathrm{K}}^{\ \ \mathrm{I}}
    \lbrack \mathbf{i}\rvert_{\dot{\alpha}}^{\mathrm{K}},
\end{align}
and similarly for upper Lorentz indices, where $j$ stands for any of $3, +$ or $-$.  That is, we seek to build non-field objects that transform under the little-group spin.  We want them to live in both Lorentz ``space'' (with a Lorentz index) and in spin ``space'' (with a spin index).  However, since it is the same spin, which is to say, the same little group, the generators of spin in both ``spaces'' should have the same effect on these objects.

Moreover, since the square spinor transforms under the conjugate representation of $SL(2,\mathbb{C})$, it should be the conjugate of the angle spinor.  Since it is the conjugate, we expect the spin index to lower (or raise) under conjugation.  Consequently, we look for a solution that satisfies 
\begin{align}
    \lbrack \mathbf{i}\rvert_{\dot{\alpha}\mathrm{I}} = \left(\lvert \mathbf{i}\rangle_{\alpha}^{\mathrm{I}}\right)^*.
\end{align}
Finally, we would also like our spinors to give the momentum when multiplied with the spin indices contracted, 
\begin{align}
    p_{i\alpha \dot{\beta}} &= 
    \lvert \mathbf{i}\rangle_{\alpha}^{\mathrm{I}} \lbrack \mathbf{i}\rvert_{\dot{\beta}\mathrm{I}}
    \\
    p_i^{\dot{\alpha}\beta} &=
    \lvert \mathbf{i}\rbrack^{\dot{\alpha}}_{\mathrm{I}} \langle \mathbf{i}\rvert^{\beta\mathrm{I}}.
\end{align}
Altogether, these are the defining properties for the spin spinors, to which we look for a solution.

We will write these spinors as a matrix where the Lorentz index gives the row and the spin index gives the column.  In order to find these efficiently, we use the property that the massless limit should give the helicity spinor in one column (where the spin matches the helicity) and zero in the other.  It seems reasonable that $\sqrt{2\mathcal{E}}$ should be replaced with $\sqrt{\mathcal{E}+p}$ in one column and $\sqrt{\mathcal{E}-p}$ in the other.  A quick search for a spin spinor that satisfies these properties yields:
\begin{align}
    \lvert \mathbf{i}\rangle_{\alpha}^{\mathrm{I}}
    &= 
    \left(\begin{array}{cc}
    \sqrt{\mathcal{E}+p}\ c 
    & -\sqrt{\mathcal{E}-p}\ s^*
    \\ 
    \sqrt{\mathcal{E}+p}\ s
    & \sqrt{\mathcal{E}-p}\ c
    \end{array}\right)
    \label{eq:|i>^I def}
    \\
    \lbrack \mathbf{i}\rvert_{\dot{\beta}\mathrm{I}}
    &= 
    \left(\begin{array}{cc}
    \sqrt{\mathcal{E}+p}\ c  
    & -\sqrt{\mathcal{E}-p}\ s
    \\ 
    \sqrt{\mathcal{E}+p}\ s^*
    & \sqrt{\mathcal{E}-p}\ c
    \end{array}\right).
    \label{eq:[i|^I def}
\end{align}
All the other massive spin spinors can be obtained from these by raising and lowering the Lorentz and spin indices using the epsilon tensor, and have been given elsewhere \cite{Arkani-Hamed:2017jhn,Christensen:2018zcq,Christensen:2019mch,Christensen:2022nja,Christensen:2024xzs,Christensen:2024xsg}.  These references give many of their properties and we will not recount them all here.  We will simply focus on the properties relevant to the construction of a Lorentz invariant action.

\subsection{\label{app:spinor products}Lorentz Invariants}

We have already noted that products with the Lorentz indices contracted have the property that the Lorentz transformations cancel [see Eqs.~(\ref{eq:A'_alpha B'^alpha = A_alpha B^alpha}) and (\ref{eq:A'_alpha B'^alpha = A_alpha B^alpha Boosts})].  Therefore, we can immediately see that this applies to products of spinors, with their Lorentz indices contracted.  This includes spinor products with momenta inserted.  For example:
\begin{align}
    \langle i j\rangle 
    &= \langle i\rvert^{\alpha} 
    \lvert j\rangle_{\alpha} 
    = \langle i\rvert^{\alpha}   
    \Lambda_{\frac{1}{2}\ \alpha}^{\ \beta}
    \Lambda_{\frac{1}{2}\beta}^{\ \ \ \eta}
    \lvert j\rangle_{\eta} 
    \\
    \lbrack i j\rbrack 
    &= \lbrack i\rvert_{\dot{\alpha}} 
    \lvert j\rbrack^{\dot{\alpha}} 
    = \lbrack i\rvert_{\dot{\alpha}}   
    \Lambda_{\frac{1}{2}\dot{\beta}}^{*\ \ \dot{\alpha}}
    \Lambda_{\frac{1}{2}\ \dot{\eta}}^{*\dot{\beta}}
    \lvert j\rbrack^{\dot{\eta}} 
    \\
    \langle i \mathbf{j}\rangle^{\mathrm{K}} 
    &= \langle i\rvert^{\alpha} 
    \lvert \mathbf{j}\rangle_{\alpha}^{\mathrm{K}} 
    = \langle i\rvert^{\alpha}   
    \Lambda_{\frac{1}{2}\ \alpha}^{\ \beta}
    \Lambda_{\frac{1}{2}\beta}^{\ \ \ \eta}
    \lvert \mathbf{j}\rangle_{\eta}^{\mathrm{K}}
    \\
    \lbrack \mathbf{i} j\rbrack ^{\mathrm{I}}
    &= \lbrack \mathbf{i}\rvert_{\dot{\alpha}}^{\mathrm{I}} 
    \lvert j\rbrack^{\dot{\alpha}} 
    = \lbrack \mathbf{i}\rvert_{\dot{\alpha}}^{\mathrm{I}}   
    \Lambda_{\frac{1}{2}\dot{\beta}}^{*\ \ \dot{\alpha}}
    \Lambda_{\frac{1}{2}\ \dot{\eta}}^{*\dot{\beta}}
    \lvert j\rbrack^{\dot{\eta}}
    \\
    \langle \mathbf{i} \mathbf{j}\rangle^{\mathrm{IK}} 
    &= \langle \mathbf{i}\rvert^{\alpha\mathrm{I}} 
    \lvert \mathbf{j}\rangle_{\alpha}^{\mathrm{K}} 
    = \langle \mathbf{i}\rvert^{\alpha\mathrm{I}}   
    \Lambda_{\frac{1}{2}\ \alpha}^{\ \beta}
    \Lambda_{\frac{1}{2}\beta}^{\ \ \ \eta}
    \lvert \mathbf{j}\rangle_{\eta}^{\mathrm{K}}
    \\
    \lbrack i \lvert p_l \rvert j\rangle 
    &= \lbrack i\rvert_{\dot{\alpha}} 
    p_{l}^{\dot{\alpha}\beta}
    \lvert j\rangle_{\beta}
    \\
    &= \lbrack i\rvert_{\dot{\alpha}}   
    \Lambda_{\frac{1}{2}\dot{\beta}}^{*\ \ \dot{\alpha}}
    \Lambda_{\frac{1}{2}\ \dot{\eta}}^{*\dot{\beta}}
    p_l^{\dot{\eta}\zeta}
    \Lambda_{\frac{1}{2}\ \zeta}^{\ \kappa}
    \Lambda_{\frac{1}{2}\kappa}^{\ \ \ \iota}
    \lvert j\rangle_{\iota} 
    \nonumber
    \\
    \langle i \lvert p_l \rvert \mathbf{j} \rbrack^{\mathrm{K}} 
    &= \langle i\rvert^{\alpha} 
    p_{l\alpha\dot{\beta}}
    \lvert \mathbf{j}\rbrack^{\dot{\beta}\mathrm{K}} 
    \\
    &= \langle i\rvert^{\alpha}   
    \Lambda_{\frac{1}{2}\ \alpha}^{\ \beta}
    \Lambda_{\frac{1}{2}\beta}^{\ \ \ \eta}
    p_{l\eta\dot{\zeta}}
    \Lambda_{\frac{1}{2}\dot{\kappa}}^{*\ \ \dot{\zeta}}
    \Lambda_{\frac{1}{2}\ \dot{\iota}}^{*\dot{\kappa}}
    \lvert \mathbf{j}\rbrack^{\dot{\iota}\mathrm{K}}
    \nonumber
    \\
    & \vdots &,
    \nonumber
\end{align}
and many more,
where $\Lambda_{\frac{1}{2}}$ could be a rotation, a boost, or a combination.  The Lorentz transformation cancels in all these products.  

In general, the Lorentz transformations will cancel in any spinor product with spinors at the ends and any number of momenta in between, so long as the spinors are of the same type (both angle or both square) if the number of momenta is even or as long as the spnors are of opposite type (one angle and one saquare) if the number of momenta is odd.  

We have also introduced a notation on the left, where we drop the Lorentz indices since they are determined uniquely from the bracket structure.  Right-angle spinors have lower undotted Lorentz indices, left-angle spinors have upper undotted Lorentz indices, right-square spinors have upper dotted Lorentz indices and left-square spinors have lower dotted Lorentz indices.  The Lorentz indices on the momenta are determined by the objects to their left and right, whether spinors or other momenta.  

We can now use the property that the Lorentz transformations acting on the momenta give the transformed momenta, for example $\bar{p}^{\dot{\beta}\kappa}=\Lambda_{\frac{1}{2}\ \dot{\eta}}^{*\dot{\beta}} \Lambda_{\frac{1}{2}\ \zeta}^{\ \kappa}     p_l^{\dot{\eta}\zeta}$.  We can also use the property of the spinors that a Lorentz transformation of them gives the spinors at the new momenta times a little-group transformation [see Eqs.~(\ref{eq:Lambda|1>=eJ|2>=eJ|2>}) and (\ref{eq:Lambda[1|=eJ[2|=eJ[2|})].  That is, for example, we find
\begin{align}
    \langle i\lvert \cdots p_l\cdots \rvert j\rangle &= e^{-\frac{i}{2}\left(\omega_i+\omega_j\right)}
    \langle \bar{i}\lvert \cdots \bar{p}_l\cdots \rvert \bar{j}\rangle
    \label{eq:<i|...pl...|j>= transformation}
    \\
    \lbrack i\lvert \cdots p_l\cdots \rvert j\rbrack &= e^{+\frac{i}{2}\left(\omega_i+\omega_j\right)}
    \lbrack \bar{i}\lvert \cdots \bar{p}_l\cdots \rvert \bar{j}\rbrack
    \\
    \lbrack i\lvert \cdots p_l\cdots \rvert j\rangle &= e^{+\frac{i}{2}\left(\omega_i-\omega_j\right)}
    \lbrack \bar{i}\lvert \cdots \bar{p}_l\cdots \rvert \bar{j}\rangle
    \\
    \langle i\lvert \cdots p_l\cdots \rvert \mathbf{j}\rangle^{\mathrm{K}} &= e^{-\frac{i}{2}\omega_i}
    \left(e^{i\vec{\omega}_j\cdot\vec{J}}\right)_{\mathrm{L}}^{\ \mathrm{K}}
    \langle \bar{i}\lvert \cdots \bar{p}_l\cdots \rvert \bar{\mathbf{j}}\rangle^{\mathrm{L}}
    \\
    \lbrack \mathbf{i}\lvert^{\mathrm{I}} \cdots p_l\cdots \rvert j\rbrack &= e^{+\frac{i}{2}\omega_j}
    \left(e^{i\vec{\omega}_i\cdot\vec{J}}\right)_{\mathrm{L}}^{\ \mathrm{I}}
    \lbrack \bar{\mathbf{i}}\lvert^{\mathrm{L}} \cdots \bar{p}_l\cdots \rvert \bar{j}\rbrack
    \\
    \langle \mathbf{i}\lvert^{\mathrm{I}} \cdots p_l\cdots \rvert \mathbf{j}\rbrack^{\mathrm{K}} 
    &= 
    \left(e^{i\vec{\omega}_i\cdot\vec{J}}\right)_{\mathrm{L}}^{\ \mathrm{I}}
    \left(e^{i\vec{\omega}_j\cdot\vec{J}}\right)_{\mathrm{M}}^{\ \mathrm{K}}
    \langle \bar{\mathbf{i}}\lvert^{\mathrm{L}} \cdots \bar{p}_l\cdots \rvert \bar{\mathbf{j}}\rangle^{\mathrm{M}}
    \label{eq:<i^I|...pl...|j]^K = transformation}
    \\
    &\vdots ,
    \nonumber
\end{align}
where we have separate rotation parameters $\omega_i(\Lambda,p_i)$ and $\omega_j(\Lambda,p_j)$ for each spinor because this parameter depends on both the Lorentz transformation \textit{and} on the momentum of the particle.  In summary, a spinor product in one frame is equal to the spinor product in another frame that is transformed by the little-group representations, and this is exactly what we need to create Lorentz invariant action terms.

We are now in a position to consider action terms that are Lorentz invariant and have the right mass dimension.  We begin by calculating the required mass dimension of the non-field terms.   As we showed in Sec.~\ref{sec:free field}, \textit{all} fields have a mass dimension of $-3$.  Therefore, for an $n-$field action term, the momentum integrals and the momentum-conserving delta function contribute $4n-4$ to the total mass dimension of the term, while the fields contribute $-3n$.  Therefore, in order for the total action term to be dimensionless, requires that the non-field part contributes
\begin{align}
    d &= 4-n.
    \label{eq:d=4-n}
\end{align}
This is equivalent to the well-known dimensionality of the n-point amplitude.  

Next, we know that we must contract all the spin indices of the fields with spin indices on spin spinors with the same momentum and we must have a product of helicity spinors that cancel the helicity transformations of all the massless fields.  [We cannot contract the spin indices directly between fields or cancel the helicity transformations directly between fields because they have different momenta and the phase angle $\omega_i=\omega(\Lambda,p_i)$ is different for each field.]  Since each massive field has $2s_i$ symmetrized spin indices, where $s_i$ is the spin of the $i^{th}$ field, we will need $\sum_i 2s_i$ spin spinors.  We will also need $\sum_j 2\lvert h_j\rvert$ helicity spinors to cancel the transformations of the massless fields.  We have also seen that each spinor has mass dimension of $\frac{1}{2}$.  All together, therefore, the naive mass dimension of the spinors required is
\begin{align}
    \tilde{d} &= \sum_i s_i + \sum_j\lvert h_j\rvert.
    \label{dt = s_i + h_j}
\end{align}
This is without any momentum insertions, which make the dimension of the spinor products higher.

In some cases, this will match the required dimension of the term in Eq.~(\ref{eq:d=4-n}).  However, in cases where it is not already equal, we must obtain the correct dimension in one of two ways.  If there is more than one massless field, then we can divide by spinor products where both spinors are massless helicity spinors.  This works because the helicity transformation is a number and division reverses its helicity transformation, namely $1/e^{\pm \frac{i}2{} \omega_i}=e^{\mp \frac{i}2{} \omega_i}$.  We will come back to this shortly, when we consider terms with only massless fields.  

On the other hand, we can divide by a mass or scale.  If we do this, $\tilde{d}$ is the naive energy growth, or scaling law, for this interaction and we find that the total power of the masses (or scales) should be
\begin{align}
    \tilde{d}_M &= 4-n - \sum_i s_i - \sum_j \lvert h_j\rvert,
\end{align}
or to write it in terms of the number of particles of each spin or helicity up to two, we have
\begin{align}
    \tilde{d}_M &= 4-n - \frac{1}{2}n_{\frac{1}{2}} - n_1 - \frac{3}{2}n_{\frac{3}{2}} - 2n_2,
    \label{eq:dtM = 4-n-n_1/2/2...}
\end{align}
where $n_{\frac{1}{2}}+n_{\frac{3}{2}}$ must be even, since we must have an even number of spinors to make spinor products where the Lorentz transformations cancel.
Therefore, we can enumerate the naive mass (or scale) power required for each action term that does not include division by helicity spinors and have done this in Table~\ref{tab:mass dimensions}.  If there are momentum insertions, $\tilde{d}_M$ will be lower to compensate.
\begin{table}
    \begin{tabular}{|c|c|c|c|c|c|c|c|c|}
        \hline
        & $n$ & $n_0$ & $n_{\frac{1}{2}}$ & $n_{1}$ & $n_{\frac{3}{2}}$ & $n_2$ & $\tilde{d}$ & $\tilde{d}_M$ \\
        \hline\hline
        * & 3 & 3 & 0 & 0 & 0 & 0 & 0 & 1\\
         & 3 & 2 & 0 & 1 & 0 & 0 & 1 & 0 \\
        $\dagger$ & 3 & 2 & 0 & 0 & 0 & 1 & 2 & -1 \\
        *  & 3 & 1 & 2 & 0 & 0 & 0 & 1 & 0 \\
         & 3 & 1 & 1 & 0 & 1 & 0 & 2 & -1 \\
         & 3 & 1 & 0 & 0 & 2 & 0 & 3 & -2 \\
        * & 3 & 1 & 0 & 2 & 0 & 0 & 2 & -1 \\
         & 3 & 1 & 0 & 1 & 0 & 1 & 3 & -2 \\
         & 3 & 1 & 0 & 0 & 0 & 2 & 4 & -3 \\
        * & 3 & 0 & 2 & 1 & 0 & 0 & 2 & -1 \\
        $\dagger$ & 3 & 0 & 2 & 0 & 0 & 1 & 3 & -2 \\
         & 3 & 0 & 1 & 1 & 1 & 0 & 3 & -2 \\
         & 3 & 0 & 1 & 0 & 1 & 1 & 4 & -3 \\
        * & 3 & 0 & 0 & 3 & 0 & 0 & 3 & -2 \\
        $\dagger$ & 3 & 0 & 0 & 2 & 0 & 1 & 4 & -3 \\
         & 3 & 0 & 0 & 1 & 0 & 2 & 5 & -4 \\
         & 3 & 0 & 0 & 0 & 0 & 3 & 6 & -5 \\
        * & 4 & 4 & 0 & 0 & 0 & 0 & 0 & 0 \\
        * & 4 & 2 & 0 & 2 & 0 & 0 & 2 & -2
        \\\hline
    \end{tabular}
    \caption{\label{tab:mass dimensions}Power of mass or scale for minimal interaction that does not involve division by a helicity-spinor product and does not include any momentum insertions.  The first column marks whether the term is present in the CSM (*) or the CSM with gravity ($\dagger$).  The second column gives the total number of fields in the interaction. The next five columns give the number of fields with spin or helicity equal to $0, \frac{1}{2}, 1, \frac{3}{2}$ or $2$.  The eigth column gives the naive energy-growth power for the term and the ninth column gives the total power of the masses or scales in the term.  Three gluons and three gravitons have a lower energy-growth power, $\tilde{d}=1$ and $2$, respectively, since they have division by spinor products, and only include $\tilde{d}_M=0$ and $-1$, respectively, scale powers.  We only include the 4-point interactions present in the CSM.}
\end{table}

Now that we know the general structure of an action term, let us consider an example and show its Lorentz invariance.  We will use the action for a $W$ boson, a charged lepton and a neutrino, which includes a nice mixture of spin and helicity transformations.  We will go through this case slowly to make clear how the cancellation occurs.  We recall from Eq.~(\ref{eq:S_W/Zf}) that half of this action term is given by
\begin{align}
    \mathcal{S} &=
    \frac{e}{M_W s_W}
    \int\frac{d^4p_1d^4p_2d^4p_3}{(2\pi)^{8}} \delta^4(p_1+p_2+p_3)
    \nonumber\\
    &
    \bar{\nu}^+(p_1)l_{\mathrm{I}}(p_2)W_{\mathrm{KL}}(p_3)
    \lbrack\mathbf{32}\rbrack^{\mathrm{KI}} \langle1\mathbf{3}\rangle^{\mathrm{L}}
    .
\end{align}
A quantum Lorentz transformation $U(\Lambda)$ only acts on fields, therefore it will pass through anything that is not a field, including our spinor products.  So, as our first step, we find
\begin{align}
    U(\Lambda)
    &\mathcal{S}
    U^{-1}(\Lambda) =
    \nonumber\\
    &
    \frac{e}{M_W s_W}
    \int\frac{d^4p_1d^4p_2d^4p_3}{(2\pi)^{8}} \delta^4(p_1+p_2+p_3)
    \nonumber\\
    &
    U(\Lambda)\bar{\nu}^+(p_1)U^{-1}(\Lambda)
    U(\Lambda)l_{\mathrm{I}}(p_2)U^{-1}(\Lambda)
    \nonumber\\
    &
    U(\Lambda)W_{\mathrm{KL}}(p_3)U^{-1}(\Lambda)
    \lbrack\mathbf{32}\rbrack^{\mathrm{KI}} \langle1\mathbf{3}\rangle^{\mathrm{L}}
    .
\end{align}
We have already learned how our fields transform, in Eqs.~(\ref{eq:UGU^-1}) through (\ref{eq:UfbU^-1}) and (\ref{eq:UhU^-1}) through (\ref{eq:UWbU^-1}).  Applying these transformations, we have
\begin{align}
    U(\Lambda)
    &\mathcal{S}
    U^{-1}(\Lambda) =
    \nonumber\\
    &
    \frac{e}{M_W s_W}
    \int\frac{d^4p_1d^4p_2d^4p_3}{(2\pi)^{8}} \delta^4(p_1+p_2+p_3)
    \nonumber\\
    &
    e^{i\frac{\omega_1}{2}}\bar{\nu}^+(\bar{p}_1)
    \left(e^{i\vec{\omega}_2\cdot \vec{J}}\right)^{\mathrm{M}}_{\ \ \mathrm{I}}l_{\mathrm{M}}(\bar{p}_2)
    \nonumber\\
    &
    \left(e^{i\vec{\omega}_3\cdot \vec{J}}\right)^{\mathrm{N}}_{\ \mathrm{K}}
    \left(e^{i\vec{\omega}_3\cdot \vec{J}}\right)^{\mathrm{O}}_{\ \mathrm{L}}
    W_{\mathrm{NO}}(\bar{p}_3)
    \lbrack\mathbf{32}\rbrack^{\mathrm{KI}} \langle1\mathbf{3}\rangle^{\mathrm{L}}
    .
\end{align}
In our next step, we can insert cancelling pairs of Lorentz $SL(2,\mathbb{C})$ transformations into our spinor products, as we showed at the beginning of this subsection, resulting in
\begin{align}
    U(\Lambda)
    &\mathcal{S}
    U^{-1}(\Lambda) =
    \nonumber\\
    &
    \frac{e}{M_W s_W}
    \int\frac{d^4p_1d^4p_2d^4p_3}{(2\pi)^{8}} \delta^4(p_1+p_2+p_3)
    \nonumber\\
    &
    e^{i\frac{\omega_1}{2}}\bar{\nu}^+(\bar{p}_1)
    \left(e^{i\vec{\omega}_2\cdot \vec{J}}\right)^{\mathrm{M}}_{\ \ \mathrm{I}}l_{\mathrm{M}}(\bar{p}_2)
    \nonumber\\
    &
    \left(e^{i\vec{\omega}_3\cdot \vec{J}}\right)^{\mathrm{N}}_{\ \mathrm{K}}
    \left(e^{i\vec{\omega}_3\cdot \vec{J}}\right)^{\mathrm{O}}_{\ \mathrm{L}}
    W_{\mathrm{NO}}(\bar{p}_3)
    \nonumber\\
    &\lbrack\mathbf{3}\lvert_{\dot{\alpha}} 
    \left(\Lambda_{\frac{1}{2}}^*\right)_{\dot{\beta}}^{\ \dot{\alpha}}
    \left(\Lambda_{\frac{1}{2}}^*\right)^{\dot{\beta}}_{\ \dot{\eta}}
    \rvert\mathbf{2}\rbrack^{\dot{\eta}\mathrm{KI}}
    \langle1\rvert^{\alpha}   
    \left(\Lambda_{\frac{1}{2}}\right)_{\ \alpha}^{\beta}
    \left(\Lambda_{\frac{1}{2}}\right)_{\beta}^{\ \ \eta}
    \lvert\mathbf{3}\rangle_{\eta}^{\mathrm{L}}
    .
\end{align}
We know that the Lorentz transformations of the spinors give the spinors at the new momentum times little-group transformations, resulting in
\begin{align}
    U(\Lambda)
    &\mathcal{S}
    U^{-1}(\Lambda) =
    \nonumber\\
    &
    \frac{e}{M_W s_W}
    \int\frac{d^4p_1d^4p_2d^4p_3}{(2\pi)^{8}} \delta^4(p_1+p_2+p_3)
    \nonumber\\
    &
    e^{i\frac{\omega_1}{2}}\bar{\nu}^+(\bar{p}_1)
    \left(e^{i\vec{\omega}_2\cdot \vec{J}}\right)^{\mathrm{M}}_{\ \ \mathrm{I}}l_{\mathrm{M}}(\bar{p}_2)
    \nonumber\\
    &
    \left(e^{i\vec{\omega}_3\cdot \vec{J}}\right)^{\mathrm{N}}_{\ \mathrm{K}}
    \left(e^{i\vec{\omega}_3\cdot \vec{J}}\right)^{\mathrm{O}}_{\ \mathrm{L}}
    W_{\mathrm{NO}}(\bar{p}_3)
    \nonumber\\
    &
    \left(e^{i\vec{\omega}_3\cdot \vec{J}}\right)^{\ \mathrm{K}}_{\mathrm{P}}
    \left(e^{i\vec{\omega}_2\cdot \vec{J}}\right)^{\ \mathrm{I}}_{\mathrm{Q}}
    \lbrack\bar{\mathbf{3}}\bar{\mathbf{2}}\rbrack^{\mathrm{PQ}}
    \nonumber\\
    &
    e^{-i\frac{\omega_1}{2}}
    \left(e^{i\vec{\omega}_3\cdot \vec{J}}\right)^{\ \mathrm{L}}_{\mathrm{R}}
    \langle\bar{1}\bar{\mathbf{3}}\rangle^{\mathrm{R}}
    .
\end{align}
At this point, all the little-group transformations cancel among themselves.  The momentum-conserving delta function can be written in terms of the transformed momenta and the integration variables can be transformed to the new momenta, to obtain
\begin{align}
    U(\Lambda)
    &
    \mathcal{S}
    U^{-1}(\Lambda) 
    =
    \nonumber\\
    &
    \frac{e}{M_W s_W}
    \int\frac{d^4\bar{p}_1d^4\bar{p}_2d^4\bar{p}_3}{(2\pi)^{8}} \delta^4(\bar{p}_1+\bar{p}_2+\bar{p}_3)
    \nonumber\\
    &
    \bar{\nu}^+(\bar{p}_1)l_{\mathrm{I}}(\bar{p}_2)W_{\mathrm{KL}}(\bar{p}_3)
    \lbrack\bar{\mathbf{3}}\bar{\mathbf{2}}\rbrack^{\mathrm{KI}} \langle\bar{1}\bar{\mathbf{3}}\rangle^{\mathrm{L}}
    \nonumber\\
    &= \mathcal{S}
    .
\end{align}

Just like traditional field theory is \textit{manifestly} Lorentz invariant if all the Lorentz indices are contracted (and covariant derivatives are used), we also have \textit{manifestly} Lorentz invariant actions here if all the spin indices are contracted among objects with the same momentum and all the helicities (including fields and spinors) add to zero, for each particle.  Of course, the $SL(2,\mathbb{C})$ Lorentz indices must also be contracted, but that is implicit in the fact that we only use spinor products between pairs of spinors.  

Although we have covered the majority of interactions with the rules we have thus far introduced, there is a special case that we need to consider.  If one of the fields is massless and the other two are massive, with the same mass, as in the case of photons, gluons and gravitons interacting with massive fields, it turns out that we need to introduce one more non-field term in order to obtain non-trivial Lorentz-invariant interactions.  

Normally, for each massless field, there are two linearly independent helicity spinors, $\lvert l\rangle$ and $\lvert l\rbrack$, which can be used to construct Lorentz invariant interactions.  However, in the case of interactions with two fields of the same mass, it turns out that these two helicity spinors are proportional to each other \cite{Arkani-Hamed:2017jhn,Christensen:2022nja},
\begin{align}
    x_{ij}\lvert l\rangle &=
    \frac{(p_j-p_i)}{2m}\lvert l\rbrack
    \label{eq:x_{ij} proportional}
    \\
    \tilde{x}_{ij}\lvert l\rbrack &=
    \frac{(p_j-p_i)}{2m}\lvert l\rangle
    \label{eq:xt_{ij} proportional}
    ,
\end{align}
where $m_i=m_j=m$, $m_l=0$, $p_i+p_j+p_l=0$ and $\tilde{x}=1/x$.  We see that $x$ is a proportionality factor.  

If we were to rely solely on spinors for constructing the vertex, we would not be able to distinguish between vertices for positive and negative helicities -- they would be directly related. Since helicity spinors cannot be contracted with themselves ($\langle ii\rangle = \lbrack ii\rbrack = 0$), they must be contracted with the spinors of the other two particles. For concreteness, let's consider a vertex involving a fermion where the third field is massless. In this case, the minimal vertex for the positive-helicity massless particle would involve $\lbrack 3\mathbf{1}\rbrack\lbrack\mathbf{2}3\rbrack$, while the minimal vertex for the negative-helicity particle would involve $\langle3\mathbf{1}\rangle\langle\mathbf{2}3\rangle$.

However, we find that $\lbrack3\mathbf{1}\rbrack=\frac{1}{2m\tilde{x}}\langle3\lvert\left(p_2-p_1\right)\rvert\mathbf{1}\rbrack$. Using momentum conservation, $p_2 = -p_1 - p_3$, and noting that $\langle 3\rvert p_3 = 0$, we obtain $\lbrack3\mathbf{1}\rbrack=-\frac{1}{m\tilde{x}}\langle3\lvert p_1\rvert\mathbf{1}\rbrack$. Additionally, since $p_1\rvert\mathbf{1}\rbrack=-m\lvert\mathbf{1}\rangle$, we have $\lbrack3\mathbf{1}\rbrack=x\langle3\mathbf{1}\rangle$, where $x = 1/\tilde{x}$. Similarly, we find \(\lbrack \mathbf{2}3\rbrack = x\langle \mathbf{2}3\rangle\).

Thus, $\lbrack 3\mathbf{1}\rbrack\lbrack \mathbf{2}3\rbrack=x^2\langle3\mathbf{1}\rangle\langle\mathbf{2}3\rangle$. Since the factors $x$ and $\tilde{x}$ will be present in the interactions regardless, it is preferable to construct a definite vertex. Moreover, the vertex can be constructed using $x$ and $\tilde{x}$ alone, yielding well-defined properties.

A representation of it can be obtained by multiplying on the left with an opposite facing reference helicity spinor, $\langle \xi\rvert$ and $\lbrack \xi\rvert$,
\begin{align}
    x_{ij} &= 
    \frac{\langle \xi\rvert(p_j-p_i)\lvert l\rbrack}{2m\langle \xi l\rangle}
    \\
    \tilde{x}_{ij} &=
    \frac{\lbrack \xi\rvert(p_j-p_i)\lvert l\rangle}{2m\lbrack \xi l\rbrack}
    .
\end{align}
Scattering amplitudes are independent of the choice for $\lvert \xi\rangle$ and $\lvert \xi\rbrack$, and it either takes on a definite form in terms of other particles in the process or drops out completely by the end of the calculation.  

In three-field interactions, in this case, instead of using the helicity spinors, we should use $x$ and $\tilde{x}$ to create Lorentz invariant action terms.  In order to do this, we must determine how these factors transform, for which, we can use our understanding of the transformation properties of spinors and momenta.  We first insert cancelling pairs of $SL(2,\mathbb{C})$ transformations,
\begin{align}
    x_{ij} &= \frac{\langle \xi\rvert
    ^{\alpha}   
    \Lambda_{\frac{1}{2}\ \alpha}^{\ \beta}
    \Lambda_{\frac{1}{2}\beta}^{\ \ \ \eta}
    (p_j-p_i)_{\eta\dot{\zeta}}
    \Lambda_{\frac{1}{2}\dot{\kappa}}^{*\ \ \dot{\zeta}}
    \Lambda_{\frac{1}{2}\ \dot{\iota}}^{*\dot{\kappa}}
    \lvert l\rbrack^{\dot{\iota}}}{2m\langle \xi \rvert^{\lambda}   
    \Lambda_{\frac{1}{2}\ \lambda}^{\ \mu}
    \Lambda_{\frac{1}{2}\mu}^{\ \ \ \nu}
    \lvert l\rangle_{\nu}}
    \\
    \tilde{x}_{ij} &=
    \frac{\lbrack \xi\rvert_{\dot{\alpha}}   
    \Lambda_{\frac{1}{2}\dot{\beta}}^{*\ \ \dot{\alpha}}
    \Lambda_{\frac{1}{2}\ \dot{\eta}}^{*\dot{\beta}}
    (p_j-p_i)^{\dot{\eta}\zeta}
    \Lambda_{\frac{1}{2}\ \zeta}^{\ \kappa}
    \Lambda_{\frac{1}{2}\kappa}^{\ \ \ \iota}\lvert l\rangle_{\iota}}{2m\lbrack \xi\rvert_{\dot{\lambda}}   
    \Lambda_{\frac{1}{2}\dot{\mu}}^{*\ \ \dot{\lambda}}
    \Lambda_{\frac{1}{2}\ \dot{\nu}}^{*\dot{\mu}}
    \lvert l\rbrack^{\dot{\nu}}}
    .
\end{align}
Their Lorentz transformations give the same structure at the new momenta, multiplied by little-group phases,
\begin{align}
    x_{ij} &= 
    \frac{e^{\frac{i}{2}\left(-\omega_{\xi}+\omega_l\right)}\langle \bar{\xi}\rvert(\bar{p}_j-\bar{p}_i)\lvert \bar{l}\rbrack}{2m e^{\frac{i}{2}\left(-\omega_{\xi}-\omega_l\right)}\langle \bar{\xi} \bar{l}\rangle}
    \\
    \tilde{x}_{ij} &=
    \frac{e^{\frac{i}{2}\left(\omega_{\xi}-\omega_l\right)}\lbrack \bar{\xi}\rvert(\bar{p}_j-\bar{p}_i)\lvert \bar{l}\rangle}{2m e^{\frac{i}{2}\left(\omega_{\xi}+\omega_l\right)}\lbrack \bar{\xi} \bar{l}\rbrack}
    .
\end{align}
The $\omega_{\xi}$ phase cancels in each and the $\omega_l$ phase combines to give,
\begin{align}
    x_{ij} &= e^{i\omega_l}\bar{x}_{ij}
    \label{eq:x_ij transform}
    \\
    \tilde{x}_{ij} &= e^{-i\omega_l}\bar{\tilde{x}}_{ij} .
    \label{eq:xt_ij transform}
\end{align}
We find that $x$ and $\tilde{x}$ transform to the same objects at the new momenta times a phase corresponding to helicity $+1$ and $-1$, respectively.  This is exactly what we need to cancel the helicity transformation of the photons, gluons, gravitons and other massless fields in this situation.  With this, we see that the interactions in Sec.~\ref{sec:gluons, photons and gravitons with massive fields} are all Lorentz invariant since the helicity transformations of the massless fields are cancelled with the helicity transformations of $x$ and $\tilde{x}$.  

Although, technically, $x$ and $\tilde{x}$ contain a momentum insertion and division by a mass, they are still considered minimal as we cannot construct a more minimal interaction without them.  Nevertheless, we can see that their energy growth still goes like $\mathcal{E}^1$ and, therefore, the formulas for $\tilde{d}$ and $\tilde{d}_M$ in Eqs.~(\ref{dt = s_i + h_j}) through (\ref{eq:dtM = 4-n-n_1/2/2...}), are still satisfied for interactions of this type.  Consequently, the power counting in Table~\ref{tab:mass dimensions} is valid for these interactions as well.

Interactions where all the fields are massless are special \cite{Arkani-Hamed:2017jhn,Christensen:2018zcq}.  First of all, each interaction can \textit{only} involve angle spinors \textit{or} square spinors, but \textit{not} both.  Also, since the helicity spinors only transform under helicity, their spinor products can appear in both the numerator and denominator.  Furthermore, if none of the fields are gravitons, there is no mass or scale, therefore, the correct mass dimension must be achieved directly with the spinor products alone.  

There are two non-zero possibilities, if the sum of the helicities is $\pm1$.  The non-field terms that would multiply these fields and cancel the helicity transformations of the fields are
\begin{align}
    \mathcal{A}_{\sum_i h_i=-1} =& 
    \langle12\rangle^{1+2h_3}\langle23\rangle^{1+2h_1}\langle31\rangle^{1+2h_2}
    \\
    \mathcal{A}_{\sum_i h_i=+1} =& 
    \lbrack12\rbrack^{1-2h_3}\lbrack23\rbrack^{1-2h_1}\lbrack31\rbrack^{1-2h_2}
    .
\end{align}
A sum or difference of these will give a Hermitian action.
We can see examples of this in Sec.~\ref{sec:massless 3-point}.  Since there is no mass or scale without gravity, we find that $\tilde{d}=1$ and $\tilde{d}_M=0$ for all interactions of this type.

If the graviton is involved, then there is division by one power of the Planck scale but, otherwise, the mass dimension must be achieved with spinor products.  Once again, there are only two non-zero possibilities, which occur when the sum of the helicities is $\pm2$, resulting in the non-field terms
\begin{align}
    \mathcal{A}_{\sum_i h_i=-2} =& 
    \frac{1}{M_P}\langle12\rangle^{2+2h_3}\langle23\rangle^{2+2h_1}\langle31\rangle^{2+2h_2}
    \\
    \mathcal{A}_{\sum_i h_i=+2} =& 
    \frac{1}{M_P}\lbrack12\rbrack^{2-2h_3}\lbrack23\rbrack^{2-2h_1}\lbrack31\rbrack^{2-2h_2}
    .
\end{align}
Once again, these will be combined in a sum in order to achieve Hermiticity.
Examples of these interactions can also be found in Sec.~\ref{sec:massless 3-point}.  With gravity, we find $\tilde{d}=2$ and $\tilde{d}_M=-1$.

The Hermitian conjugates of the spinor products has been covered in \cite{Christensen:2019mch}.  But, we can see from our definitions that
\begin{align}
    \langle ij\rangle^\dagger &= \lbrack ji \rbrack
    \label{eq:<ij>^dagger}
    \\
    \langle \mathbf{i}^{\mathrm{I}}\mathbf{j}^{\mathrm{J}}\rangle^\dagger &= \lbrack \mathbf{j}_{\mathrm{J}}\mathbf{i}_{\mathrm{I}} \rbrack
    \\
    \langle \mathbf{i}^{\mathrm{I}}\mathbf{j}_{\mathrm{J}}\rangle^\dagger &= \lbrack \mathbf{j}^{\mathrm{J}}\mathbf{i}_{\mathrm{I}} \rbrack
    \label{eq:<i^Ij_J>^dagger}
\end{align}
and so on.  The order is reversed, angle and square brackets are interchanged, and spin indices are lowered and raised.

We will also need the Hermitian conjugate of $x$ and $\tilde{x}$.  Using the representation with a reference spinor $\xi$, we have
\begin{align}
    x_{ij}^\dagger &=
    \left(
        -\frac{\langle \xi\lvert \left(p_j-p_i\right) \rvert l\rbrack}{2m\langle \xi l\rangle}
    \right)^\dagger
    \nonumber\\
    &= -\frac{\langle l\lvert \left(p_j-p_i\right) \rvert \xi\rbrack}{2m\lbrack l \xi\rbrack}
    \nonumber\\
    &= -\frac{\lbrack \xi\lvert \left(p_i-p_j\right) \rvert l\rangle}{2m\lbrack \xi l\rbrack}
    \nonumber\\
    &= \tilde{x}_{ji},
    \label{eq:x^dagger = xt}
\end{align}
where on the third row, we used the symmetry of $\langle l\lvert \left(p_j-p_i\right) \rvert \xi\rbrack$ and the antisymmetry of $\lbrack P_{ij} \xi\rbrack$ under reversal.  
We see that Hermitian conjugation interchanges $x$ and $\tilde{x}$ and interchanges the indices.

Given that we expect the CSM to be renormalizable -- based on the fact that the SM is renormalizable and we anticipate that all matrix elements will be equal (though this still requires proof) -- we can use the CSM to hypothesize some conditions for a general renormalizable constructive theory.

The first observation is that none of the interactions in the CSM contain a momentum insertion. By this, we mean that none of the spinor products in the action terms include a momentum between the spinors. For example, we have spinor products of the form $\langle i j\rangle$, but not of the form $\langle i\lvert p_l \rvert j\rbrack$, or $\lbrack i\lvert p_l p_m \rvert j\rbrack$, and so on. Interactions that do not include any additional momentum insertions are referred to as minimal, and we conjecture that renormalizable constructive models only contain minimal interactions.

Inserting additional momenta into an interaction would require dividing by extra powers of a mass or scale, effectively making such terms higher-dimensional interactions. As in traditional field theory, we expect that higher-dimensional interactions lead to an effective theory. While the minimality of interactions seems necessary for renormalizability, it is not sufficient on its own.

Additionally, upon examining Table~\ref{tab:mass dimensions}, we observe that none of the CSM interactions involve a mass or scale with a power less than -2. This suggests another potential necessary condition for renormalizability. However, this condition, along with minimality, is also not sufficient, as some graviton interactions -- which we expect to be non-renormalizable -- also respect this mass/scale constraint.

Regarding four-field interactions, the CSM includes only those involving scalar fields, leading us to suspect that this might be a general feature of renormalizable theories. Furthermore, we do not expect interactions involving five or more fields to be renormalizable.

In summary, while we have identified some potential indicators of renormalizability, the complete set of necessary and sufficient conditions remains an open question. Further exploration and proof are required to fully determine these criteria.

\section{\label{app:locality}Locality}
One of the hallmark features of gauge field theory, which gives rise to Feynman rules, is its manifest locality. This has been considered a great triumph of modern physics. In it, electromagnetism and gravity are not instantaneous forces transmitted over a distance, but are mediated by virtual quantum particles that act locally.  

However, demanding locality along with Lorentz invariance came at a cost. It required the introduction of unphysical degrees of freedom and the gauging of global symmetries to counteract the adverse effects of these unphysical degrees of freedom. As a result, contributions to the amplitude were partitioned into gauge-invariant subsets of diagrams, where significant and crucial cancellations occurred. Importantly, each individual diagram within a gauge-invariant subset was not physical on its own -- only the combined gauge-invariant subset represented a physical process.

This necessity made individual Feynman diagrams less physically insightful and less computationally efficient. Yet, this was the price of maintaining locality. Despite these challenges, the trade-off was worthwhile because Feynman diagrams enabled the calculation of any amplitude, at least in principle, and have driven much of the progress in theoretical particle physics since their inception.

On the other hand, if we aim to avoid introducing unphysical degrees of freedom and their associated gauge symmetries -- and if we want every diagram to be trivially gauge-invariant and physically meaningful -- we must abandon locality. Furthermore, a notable consequence of working within a non-local theory is that it operates as an on-shell theory during intermediate steps of the calculation. However, while the final expressions in such a theory are simpler and physical, it must be admitted that there is a trade off.  Missing even a single on-shell identity can lead to incorrect results. 

In this appendix, we will explore the non-locality of our action.  

\subsection{\label{app:local action example}A Simple Example of a Local Action}

Let's consider locality in a simple example, involving a real scalar field.  Rather than considering every possible term, we will just consider a cubic interaction with an example derivative structure as illustration.  Other cases are simple generalizations of this case.  The position-space action is given by
\begin{align}
    \mathcal{S}_{\phi^3} = 
    \int d^4x \lambda\partial_\mu\phi(x)\partial^\mu\phi(x)\partial^2\phi(x),
\end{align}
where $\lambda$ is dimensionful and this is a higher-dimensional operator.  However, this does not concern us here.  Our objective is to gain an understanding of how a local action appears in momentum space.  This action is local because every part of the action occurs at the same integrated position $x$.

In order to get to momentum space, we Fourier transform, using
\begin{align}
    \phi(x) = \int\frac{d^4p}{(2\pi)^4}\phi(p)e^{ip\cdot x},
\end{align}
where we have used the argument to specify whether this is the original $\phi(x)$ or its Fourier transform $\phi(p)$.  We will always write $\phi$ with its argument and there will be no confusion.
Plugging this in, we have
\begin{align}
    \mathcal{S}_{\phi^3} &=
    \int\frac{d^4p_1d^4p_2d^4p_3}{(2\pi)^{12}} \lambda p_1\cdot p_2 p_3^2\phi(p_1)\phi(p_2)\phi(p_3)
    \nonumber\\
    &\hspace{0.2in}\int d^4x
    e^{i\left(p_1+p_2+p_3\right)\cdot x}
    \nonumber\\
    &= \int\frac{d^4p_1d^4p_2d^4p_3}{(2\pi)^{8}} \lambda p_1\cdot p_2 p_3^2\phi(p_1)\phi(p_2)\phi(p_3)
    \nonumber\\
    &\hspace{0.2in}\delta^4\left(p_1+p_2+p_3\right).
\end{align}
The expression gives the 3-point Feynman vertex we would use in this theory.  We can see that local actions, when written in momentum space have a momentum conserving delta function and all momenta are polynomial and in the numerator of the vertex.  

Although the inverse should be clear, since the actions in the main body of this paper begin in momentum space, we will go through the inverse of this operation for illustration.  The Fourier transform back to position space of the field is given by
\begin{align}
    \phi(p) = \int d^4x \phi(x) e^{-ip\cdot x}.
\end{align}
Plugging this into our momentum-space action, we have
\begin{align}
    \mathcal{S}_{\phi^3} &= 
    \int d^4x_1d^4x_2d^4x_3 \lambda
    \int\frac{d^4p_1d^4p_2d^4p_3}{(2\pi)^{8}} 
    \delta^4\left(p_1+p_2+p_3\right)
    \nonumber\\
    &\hspace{0.2in}
    \left(-\phi(x_1)\partial_{1\mu}e^{-ip_1\cdot x_1}\right)
    \left(-\phi(x_2)\partial_2^{\mu}e^{-ip_2\cdot x_2}\right)
    \nonumber\\
    &\hspace{0.2in}
    \left(\phi(x_3)\partial_3^2e^{-ip_3\cdot x_3}\right).
\end{align}
We can now perform integration by parts for each position-space coordinate separately, moving the derivatives off the exponentials and onto the fields, giving
\begin{align}
    \mathcal{S}_{\phi^3} &= 
    \int d^4x_1d^4x_2d^4x_3 \lambda
    \partial_{1\mu}\phi(x_1)\partial_2^{\mu}\phi(x_2)\partial_3^2\phi(x_3)
    \nonumber\\
    &\hspace{-0.2in}
    \int\frac{d^4p_1d^4p_2d^4p_3}{(2\pi)^{8}} 
    \delta^4\left(p_1+p_2+p_3\right)
    e^{-i\left(p_1\cdot x_1 +p_2\cdot x_2 + p_3\cdot x_3\right)},
\end{align}
where we always assume our fields fall off quickly enough at infinity to make the global boundary term vanish.  
We next use the delta function to set $p_3=-p_1-p_2$, obtaining
\begin{align}
    \mathcal{S}_{\phi^3} &= 
    \int d^4x_1d^4x_2d^4x_3 \lambda
    \partial_{1\mu}\phi(x_1)\partial_2^{\mu}\phi(x_2)\partial_3^2\phi(x_3)
    \nonumber\\
    &\hspace{-0.2in}
    \int\frac{d^4p_1d^4p_2}{(2\pi)^{8}} 
    e^{-i\left(p_1\cdot (x_1-x_3) + p_2\cdot (x_2-x_3)\right)}.
\end{align}
Now, when we do the $p_1$ and $p_2$ integrals, we will obtain two delta functions setting $x_2=x_3$ and $x_3=x_1$,
\begin{align}
    \mathcal{S}_{\phi^3} &= 
    \int d^4x_1d^4x_2d^4x_3 \lambda
    \partial_{1\mu}\phi(x_1)\partial_2^{\mu}\phi(x_2)\partial_3^2\phi(x_3)
    \nonumber\\
    &\hspace{-0.2in}
    \delta^4(x_1-x_3)\delta^4(x_2-x_3).
\end{align}
Finally, performing two of the spatial integrals, we are left with
\begin{align}
    \mathcal{S}_{\phi^3} &= 
    \int d^4x \lambda
    \partial_{\mu}\phi(x)\partial^{\mu}\phi(x)\partial^2\phi(x),
\end{align}
our manifestly local and Lorentz invariant action.  From this example, we can see that when we begin with a manifestly Lorentz invariant action in momentum space that has polynomial momenta in the numerator, its Fourier transform is a manifestly local and Lorentz invariant action in position space.

Before moving to the next subsection, we will briefly remark on the free part of our field-theory action for the CSM.  As we can see in Sec.~\ref{sec:free field}, the momenta only appear polynomially in the numerator.  The free part of the action is local and can be easily Fourier transformed.  It gives a position-space action, whose classical Euler-Lagrange equations are wave equations (with a mass) for each helicity (spin).  

This should make some sense.  For the photon field, this matches what we expect in vacuum.  It also matches Maxwell's source-free equations for $A_{\mu}$ in Lorenz gauge.  For the gravitons, this only makes sense in the weak-field approximation, where the interaction is ignored.  For the gluons, of course, we cannot ignore the interactions since they are strongly coupled and asymptotically free.  For the massive fields, as expected, the wave travels slower than the speed of light.  

A full analysis of the Fourier transform of the field-theory action for the CSM is beyond the scope of this paper.  We will come back to it in a future work.

\subsection{\label{app:non-local denominator momenta}Momenta in the Denominator Make the Action Non-local}

However, on the other hand, we may consider what happens when there is momentum in the denominator.  For example, let's begin with the momentum-space action
\begin{align}
    \mathcal{S}_{\phi^4} &= 
    \int \frac{d^4p_1d^4p_2d^4p_3d^4p_4}{(2\pi)^{12}}\delta^4(p_1+p_2+p_3+p_4)
    \nonumber\\
    &\hspace{0.2in}
    \frac{\phi(p_1)\phi(p_2)\phi(p_3)\phi(p_4)}{(p_1+p_2)^2}.
\end{align}
Let us Fourier transform this to position space.  We have
\begin{align}
    \mathcal{S}_{\phi^4} &=
    \int d^4x_1d^4x_2d^4x_3d^4x_4 \phi(x_1)\phi(x_2)\phi(x_3)\phi(x_4)
    \nonumber\\
    &\hspace{0.2in}
    \int \frac{d^4p_1d^4p_2d^4p_3d^4p_4}{(2\pi)^{12}}\delta^4(p_1+p_2+p_3+p_4)
    \nonumber\\
    &\hspace{0.2in}
    \frac{e^{-i\left(p_1\cdot x_1 +p_2\cdot x_2 + p_3\cdot x_3 + p_4\cdot x_4\right)}}{(p_1+p_2)^2}.
\end{align}
We can easily do the $p_4$ and $p_3$ integrals, giving
\begin{align}
    \mathcal{S}_{\phi^4} &=
    \int d^4x_1d^4x_2d^4x_3 \phi(x_1)\phi(x_2)\phi^2(x_3)
    \nonumber\\
    &\hspace{0.2in}
    \int \frac{d^4p_1d^4p_2}{(2\pi)^{8}}
    \frac{e^{-i\left(p_1\cdot (x_1-x_3) +p_2\cdot (x_2-x_3)\right)}}{(p_1+p_2)^2}.
\end{align}
We next make a change of variables to 
\begin{align}
    k_1 &= p_1+p_2\\
    k_2 &= p_1-p_2,
\end{align}
giving
\begin{align}
    \mathcal{S}_{\phi^4} &=
    \int d^4x_1d^4x_2d^4x_3 \phi(x_1)\phi(x_2)\phi^2(x_3)
    \nonumber\\
    &\hspace{0.2in}
    \int \frac{d^4k_1d^4k_2}{(2\pi)^{8}}
    \frac{e^{-i\left(\frac{1}{2}k_1\cdot (x_1+x_2) + \frac{1}{2}k_2\cdot (x_1-x_2) - k_1\cdot x_3\right)}}{k_1^2}.
\end{align}
Our next step is to do the $k_2$ integral, giving us a delta function in $x_1-x_2$, resulting in
\begin{align}
    \mathcal{S}_{\phi^4} &=
    \int d^4x_1d^4x_3 \phi^2(x_1)\phi^2(x_3)
    \int \frac{d^4k}{(2\pi)^{4}}
    \frac{e^{-i k\cdot (x_1-x_3)}}{k^2}.
\end{align}
The final momentum integral is the well-known Fourier transform of the Feynman propagator $D_F(x_1-x_3)$, which depends on the difference between the two space-time locations.  It is non-local.  
\begin{align}
    \mathcal{S}_{\phi^4} &=
    \int d^4x_1d^4x_3 \phi^2(x_1)\phi^2(x_3)
    D_F(x_1-x_3).
\end{align}
If this were a fundamental vertex in the theory, and not an s-channel amplitude, it would be a non-local vertex.  The same comment applies if the momentum in the denominator were only $p_1^2$ rather than $(p_1+p_2)^2$.  We would still have obtained a non-local vertex.  Momenta in the denominator give non-local interactions.

Suppose we had momentum in both the numerator and the denominator, but they do not cancel.  This time we will consider a 3-point vertex to simplify the math.  Consider the action
\begin{align}
    \mathcal{S} &=
    \int\frac{d^4p_1d^4p_2d^4p_3}{(2\pi)^{8}} \frac{\lambda p_1\cdot p_3}{(p_1+p_2)^2} \phi(p_1)\phi(p_2)\phi(p_3)
    \nonumber\\
    &\hspace{0.2in}\delta^4\left(p_1+p_2+p_3\right).
\end{align}
Inserting the Fourier transform of the fields, we have
\begin{align}
    \mathcal{S} &=
    -\int d^4x_1 d^4x_2 d^4x_3 
    \int\frac{d^4p_1d^4p_2d^4p_3}{(2\pi)^{8}} \frac{\lambda}{(p_1+p_2)^2} 
    \nonumber\\
    &\hspace{0.2in}
    \left[\phi(x_1)\partial_{1\mu}e^{-ip_1\cdot x_1}\right]
    \left[\phi(x_2)\partial_2^{\mu}e^{-ip_2\cdot x_2}\right]
    \phi(x_3)e^{-ip_3\cdot x_3}
    \nonumber\\
    &\hspace{0.2in}\delta^4\left(p_1+p_2+p_3\right).
\end{align}
Integrating $x_1$ and $x_2$ by parts, in order to move the derivatives onto the fields, gives
\begin{align}
    \mathcal{S} &=
    -\lambda\int d^4x_1 d^4x_2 d^4x_3 
    \left[\partial_{1\mu}\phi(x_1)\right] \left[\partial_2^{\mu}\phi(x_2)\right] \phi(x_3)
    \nonumber\\
    &\hspace{0.2in}
    \int\frac{d^4p_1d^4p_2d^4p_3}{(2\pi)^{8}} \frac{1}{(p_1+p_2)^2} 
    \nonumber\\
    &\hspace{0.2in}
    e^{-i\left(p_1\cdot x_1 + p_2\cdot x_2 + p_3\cdot x_3\right)}
    \delta^4\left(p_1+p_2+p_3\right).
\end{align}
Integrating over $p_3$, we have
\begin{align}
    \mathcal{S} &=
    -\lambda\int d^4x_1 d^4x_2 d^4x_3 
    \left[\partial_{1\mu}\phi(x_1)\right] \left[\partial_2^{\mu}\phi(x_2)\right] \phi(x_3)
    \nonumber\\
    &\hspace{0.2in}
    \int\frac{d^4p_1d^4p_2}{(2\pi)^{8}} \frac{1}{(p_1+p_2)^2} 
    e^{-i\left(p_1\cdot (x_1-x_3) + p_2\cdot (x_2-x_3)\right)}.
\end{align}
We next make the change of variables $k_1=p_1+p_2$ and $k_2=p_1-p_2$, leading to
\begin{align}
    \mathcal{S} &=
    -\lambda\int d^4x_1 d^4x_2 d^4x_3 
    \left[\partial_{1\mu}\phi(x_1)\right] \left[\partial_2^{\mu}\phi(x_2)\right] \phi(x_3)
    \nonumber\\
    &\hspace{0.2in}
    \int\frac{d^4k_1d^4k_2}{(2\pi)^{8}} \frac{1}{k_1^2} 
    e^{-\frac{i}{2}k_1\cdot\left(x_1+x_2-2x_3\right)-\frac{i}{2}k_2\cdot\left(x_1-x_2\right)}.
\end{align}
Performing the $k_2$ integral, we find $\delta^4(x_1-x_2)$, resulting in
\begin{align}
    \mathcal{S} &=
    -\lambda\int d^4x_1 d^4x_3 
    \left[\partial_{1\mu}\phi(x_1)\right] \left[\partial_1^{\mu}\phi(x_1)\right] \phi(x_3)
    \nonumber\\
    &\hspace{0.2in}
    \int\frac{d^4k_1}{(2\pi)^{4}} \frac{e^{-i k_1\cdot\left(x_1-x_3\right)}}{k_1^2} 
    .
\end{align}
Once again, we have the Fourier transform of the Feynman propagator, giving us the non-local 3-point vertex
\begin{align}
    \mathcal{S} =&
    -\lambda\int d^4x_1 d^4x_3 
    \nonumber\\
    &
    \left[\partial_{1\mu}\phi(x_1)\right] \left[\partial_1^{\mu}\phi(x_1)\right] \phi(x_3)
    D_F(x_1-x_3)
    .
\end{align}
We see the general property that, momenta in the denominator lead to non-local interactions, even in combination with momenta in the numerator.

\subsection{\label{app:non-local spinor products}Spinor Products Make the Action Non-local}

We can already observe that spinor products in the denominator indicate non-locality. This non-locality arises in the case of all-massless field interactions, such as those described in Sec.~\ref{sec:massless 3-point}. The helicity-spinor products that appear in the denominator can be brought to the numerator by multiplying by their Hermitian conjugates. For example, $1/\langle ij\rangle$ can be rewritten as $\lbrack ji\rbrack/(2p_i\cdot p_j)$. However, this transformation leaves momentum in the denominator, which leads to a non-local interaction, as we showed in the previous subsection.

Therefore, let us consider the contributions to the action where spinor products appear in the numerator. Our goal here is not to find the full Fourier transform of the constructive action -- that is beyond the scope of this paper. Instead, we aim to provide evidence that the action is indeed non-local when spinor products are present, even in the absence of explicit momenta in the denominator.

Generally speaking, if elements other than the $e^{ip \cdot x}$ factor cannot be extracted from the integral, the result will not yield a delta function that equates positions in the spacetime integrals. In App.~\ref{app:local action example}, we demonstrated that polynomial momenta in the numerator could be converted into derivatives on position, allowing them to be removed from the integral. This property enabled the action to be written as a local interaction in position space.
 
However, in App.~\ref{app:non-local denominator momenta}, we showed that this was not possible when momenta appeared in the denominator, leading to non-local actions. In this subsection, we will show that while some components can indeed be turned into derivatives and extracted from the integrals, others cannot, and these will lead to non-local interactions.

If the spinor is massless, then we see that it can be written as [see Eqs.~(\ref{eq:|i> def}) and (\ref{eq:[i| def})],
\begin{align}
    \lvert i\rangle_{\alpha}
    &= \sqrt{\frac{2}{\mathcal{E}_i}}
    \left(\begin{array}{c}
    p_i^3 \\ p_i^1 + i p_i^2
    \end{array}\right)
    \\
    \lbrack i\rvert_{\dot{\beta}} 
    &= \sqrt{\frac{2}{\mathcal{E}_i}}
    \left(\begin{array}{c}
    p_i^3 \\ p_i^1 - i p_i^2
    \end{array}\right).
\end{align}
If the spinor is massive, on the other hand, we can write [see Eqs.~(\ref{eq:|i>^I def}) and (\ref{eq:[i|^I def})],
\begin{align}
    \lvert \mathbf{i}\rangle_{\alpha}^{\mathrm{I}}
    &= 
    \left(\begin{array}{cc}
    \frac{\sqrt{\mathcal{E}_i+p_i}}{p_i}\ p_i^3 
    & -\frac{\sqrt{\mathcal{E}_i-p_i}}{p_i}\left( p_i^1 - i p_i^2\right)
    \\ 
    \frac{\sqrt{\mathcal{E}_i+p_i}}{p_i}\left( p_i^1 + i p_i^2\right)
    & \frac{\sqrt{\mathcal{E}_i-p_i}}{p_i}\ p_i^3
    \end{array}\right)
    \\
    \lbrack \mathbf{i}\rvert_{\dot{\beta}\mathrm{I}}
    &= 
    \left(\begin{array}{cc}
    \frac{\sqrt{\mathcal{E}_i+p_i}}{p_i}\ p_i^3 
    & -\frac{\sqrt{\mathcal{E}_i-p_i}}{p_i}\left( p_i^1 + i p_i^2\right)
    \\ 
    \frac{\sqrt{\mathcal{E}_i+p_i}}{p_i}\left( p_i^1 - i p_i^2\right)
    & \frac{\sqrt{\mathcal{E}_i-p_i}}{p_i}\ p_i^3
    \end{array}\right),
\end{align}
where $p_i=\lvert\vec{p}_i\rvert$.    
For massive fields, to gain further insight, we could also pull out a $1/\left(p_i\sqrt{\mathcal{E}_i+p_i}\right)$, resulting in
\begin{align}
    \lvert \mathbf{i}\rangle_{\alpha}^{\mathrm{I}}
    =& 
    \frac{1}{p_i\sqrt{\mathcal{E}_i+p_i}}
    \nonumber\\
    &
    \left(\begin{array}{cc}
    p_i^3 \left(\mathcal{E}_i+p_i\right) 
    & -m_i\left( p_i^1 - i p_i^2\right)
    \\ 
    \left( p_i^1 + i p_i^2\right)\left(\mathcal{E}_i+p_i\right)
    & m_i\ p_i^3
    \end{array}\right)
    \\
    \lbrack \mathbf{i}\rvert_{\dot{\beta}\mathrm{I}}
    =& 
    \frac{1}{p_i\sqrt{\mathcal{E}_i+p_i}}
    \nonumber\\
    &
    \left(\begin{array}{cc}
    p_i^3 \left(\mathcal{E}_i+p_i\right)
    & -m_i\left( p_i^1 + i p_i^2\right)
    \\ 
    \left( p_i^1 - i p_i^2\right) \left(\mathcal{E}_i+p_i\right)
    & m_i\ p_i^3
    \end{array}\right).
\end{align}
After removing $1/\sqrt{\mathcal{E}_i}$ in the massless case and $1/\left(p_i\sqrt{\mathcal{E}_i+p_i}\right)$ in the massive case, we see that everything else in these spinors is polynomial in the momenta.  Therefore, we can turn everything that isn't in the denominator into derivatives and pull them out of the momentum integrals.  At this point, we will be left with momentum integrals that are schematically of the form
\begin{align}
    \mathcal{S} =& \int d^4x_1\cdots d^4x_n \cdots \langle\partial_i\partial_k\rangle_{\mathrm{K}}\cdots \phi_i(x_i)\cdots \phi_k(x_k)^{\mathrm{K}}\cdots
    \nonumber\\
    &
    \int \frac{d^4p_1\cdots d^4p_n}{(2\pi)^{4n-4}}\delta^4(p_1+\cdots+p_n)
    \nonumber\\
    &\frac{e^{-i\left(p_1\cdot x_1+\cdots p_n\cdot x_n\right)}}{\cdots \left(\sqrt{\mathcal{E}_i}\right)^{n_i} \cdots \left(p_k\sqrt{\mathcal{E}_k+p_k}\right)^{n_k}\cdots},
\end{align}
where $\langle \partial_i\partial_k\rangle_{\mathrm{K}}$ represents, schematically, the parts of the spinor products that are in the numerator and have been turned into derivatives and pulled out of the momentum integrals.  The denominators, including factors of $1/\sqrt{\mathcal{E}_i}$ and $1/\left(p_i\sqrt{\mathcal{E}_i+p_i}\right)$ for each massless and massive field, respectively, are left inside the momentum integrals.  The remaining momentum integrals, which have powers of energy and momentum in the denominator, will result in non-local contributions to the spacetime action.  We will fully pursue the non-local spacetime action in future research.

\section{\label{app:constructive rules}Constructive Rules}
In this appendix, we will review the process of going from the action to the interaction vertices.  We have no intention of being exhaustive.  We only consider a few examples to illustrate the process.

\subsection{\label{app:ggg vertex helicity}Triple-Gluon Vertex}
We begin by considering the interaction of three gluons.  The action terms for massless fields have the opposite helicity on the fields as in the spinor products.  This is required to have a Lorentz invariant action.  We would like to understand how the action terms lead to the correct helicities for the amplitudes. 

To be concrete, let's work with the pure gluon action,
\begin{align}
    \mathcal{S}_{\textsl{g}\textsl{g}} &=
    \int\frac{d^4p_1d^4p_2}{(2\pi)^4} \delta^4(p_1+p_2) p_1^2
    \textsl{g}_a^+(p_1) \textsl{g}_a^-(p_2)
    \nonumber\\
    \mathcal{S}_{\textsl{g}\textsl{g}\textsl{g}} &=
    \frac{g_s}{\sqrt{2}}\int\frac{d^4p_1d^4p_2d^4p_3}{(2\pi)^8} \delta^4(p_1+p_2+p_3) 
    \nonumber\\
    &
    \hspace{-0.1in}\textsl{g}_a^+(p_1) f_{abc}
        \left(\frac{\langle12\rangle^3}{\langle23\rangle\langle31\rangle}\textsl{g}_b^+(p_2)
        -\frac{\lbrack23\rbrack^3}{\lbrack12\rbrack\lbrack31\rbrack}\textsl{g}_b^-(p_2)\right)
    \textsl{g}_c^-(p_3),
\end{align}
coming from Eqs.~(\ref{eq:S_2m=0}) and (\ref{eq:S massless 3-point}).
Let's suppose we are finding the 3-point amplitude for two positive helicity and one negative helicity gluon with momenta $k_1, k_2$ and $k_3$, respectively.  A quick look at the action may lead us to believe that we would get $\langle12\rangle^3/(\langle23\rangle\langle31\rangle)$, which would be incorrect.  But, in fact, as we will show in this appendix, we get the correct $\lbrack12\rbrack^3/(\lbrack23\rbrack\lbrack31\rbrack)$.

We would like to calculate 
\begin{align}
    i\tilde{\mathcal{M}}^{++-} &=
    \mathcal{N}\int d\textsl{g}^+ d\textsl{g}^-
    \textsl{g}_d^+(k_1)\textsl{g}_e^+(k_2)\textsl{g}_f^-(k_3)
    e^{i\mathcal{S}_{\textsl{g}\textsl{g}}+i\mathcal{S}_{\textsl{g}\textsl{g}\textsl{g}}},
\end{align}
where
\begin{align}
    \mathcal{N}^{-1} = \int d \textsl{g}^+ d\textsl{g}^-
    e^{i\mathcal{S}_{\textsl{g}\textsl{g}}}.
\end{align}

In order to find this amplitude, we will introduce a source $\mathcal{J}^{\pm}$ for the gluon, and write
\begin{align}
    \mathcal{S}_{\textsl{g}\textsl{g}}[\mathcal{J}] =& 
    \int\frac{d^4p_1d^4p_2}{(2\pi)^4} \delta^4(p_1+p_2) \Bigg[
    p_1^2
    \textsl{g}_a^+(p_1) \textsl{g}_a^-(p_2)
    \nonumber\\
    &
    +\mathcal{J}_a^+(p_1)\textsl{g}_a^-(p_2)
    +\textsl{g}_a^+(p_1)\mathcal{J}_a^-(p_2)
    \Bigg],
\end{align}
so that $\mathcal{S}_{\textsl{gg}}=\mathcal{S}_{\textsl{gg}}[0]$.
Now, the amplitude can be written as
\begin{align}
    i\tilde{\mathcal{M}}^{++-} &=
    \lim_{\mathcal{J}\to0}
    \left(-i\frac{\delta}{\delta \mathcal{J}_d^-(-k_1)}\right)
    \left(-i\frac{\delta}{\delta \mathcal{J}_e^-(-k_2)}\right)
    \nonumber\\
    &
    \left(-i\frac{\delta}{\delta \mathcal{J}_f^+(-k_3)}\right)
    \mathcal{N}\int d\textsl{g}^+ d\textsl{g}^-
    e^{i\mathcal{S}_{\textsl{g}\textsl{g}}[\mathcal{J}]+i\mathcal{S}_{\textsl{g}\textsl{g}\textsl{g}}}.
\end{align}

Next, we can rewrite $\mathcal{S}_{\textsl{g}\textsl{g}\textsl{g}}$ as
\begin{align}
    \mathcal{S}_{\textsl{g}\textsl{g}\textsl{g}}\left[\frac{\delta}{\delta \mathcal{J}}\right] =&
    \frac{g_s}{\sqrt{2}}\int\frac{d^4p_1d^4p_2d^4p_3}{(2\pi)^8} \delta^4(p_1+p_2+p_3) 
    \nonumber\\
    &
    \left(-i\frac{\delta}{\delta \mathcal{J}_a^-(-p_1)}\right)
    f_{abc}
    \nonumber\\
    &\Bigg[
        \frac{\langle12\rangle^3}{\langle23\rangle\langle31\rangle}
        \left(-i\frac{\delta}{\delta \mathcal{J}_b^-(-p_2)}\right)
        \nonumber\\
        &-\frac{\lbrack23\rbrack^3}{\lbrack12\rbrack\lbrack31\rbrack}
        \left(-i\frac{\delta}{\delta \mathcal{J}_b^+(-p_2)}\right)
    \Bigg]
    \nonumber\\
    &
    \left(-i\frac{\delta}{\delta \mathcal{J}_c^+(-p_3)}\right)
    .
\end{align}
This can also be pulled out of the gluon integral, giving us
\begin{align}
    i\tilde{\mathcal{M}}^{++-} &=
    \lim_{\mathcal{J}\to0}
    \left(-i\frac{\delta}{\delta \mathcal{J}_d^-(-k_1)}\right)
    \left(-i\frac{\delta}{\delta \mathcal{J}_e^-(-k_2)}\right)
    \nonumber\\
    &
    \left(-i\frac{\delta}{\delta \mathcal{J}_f^+(-k_3)}\right)
    e^{i\mathcal{S}_{\textsl{g}\textsl{g}\textsl{g}}\left[\delta/\delta\mathcal{J}\right]}
    \mathcal{N}\int d\textsl{g}^+ d\textsl{g}^-
    e^{i\mathcal{S}_{\textsl{g}\textsl{g}}[\mathcal{J}]}.
\end{align}

We next go back to working on $\mathcal{S}_{\textsl{g}\textsl{g}}[\mathcal{J}]$.  We complete the square, obtaining
\begin{align}
    \mathcal{S}_{\textsl{g}\textsl{g}}[\mathcal{J}] &= 
    \int\frac{d^4p_1d^4p_2}{(2\pi)^4} \delta^4(p_1+p_2) \Bigg[
    \nonumber\\
    &
    p_1^2
    \left(\textsl{g}_a^+(p_1)+\frac{1}{p_1^2}\mathcal{J}_a^+(p_1)\right)
    \left(\textsl{g}_a^-(p_2)+\frac{1}{p_1^2}\mathcal{J}_a^-(p_2)\right)
    \nonumber\\
    &
    -\frac{1}{p_1^2}\mathcal{J}_a^+(p_1)\mathcal{J}_a^-(p_2)
    \Bigg].
\end{align}
For convenience, we will split this into two pieces,
\begin{align}
    \mathcal{S}_{\textsl{g}\textsl{g}}[\mathcal{J}] &= 
    \bar{\mathcal{S}}_{\textsl{g}\textsl{g}}[\mathcal{J}]
    +\mathcal{S}_{\mathcal{J}\mathcal{J}},
\end{align}
where
\begin{align}
    \mathcal{S}_{\mathcal{J}\mathcal{J}} &= 
    -\int\frac{d^4p_1d^4p_2}{(2\pi)^4} \delta^4(p_1+p_2) 
    \frac{1}{p_1^2}\mathcal{J}_a^+(p_1)\mathcal{J}_a^-(p_2).
\end{align}
Before we continue with the amplitude, let's take a minute to find the propagator.  We get it with two derivatives acting on $e^{i\mathcal{S}_{\mathcal{JJ}}}$, giving
\begin{align}
    \lim_{\mathcal{J}\to0}
    \left(-i\frac{\delta}{\delta \mathcal{J}_b^+(-q_1)}\right)
    \left(-i\frac{\delta}{\delta \mathcal{J}_b^-(-q_2)}\right)
    e^{i\mathcal{S}_{\mathcal{JJ}}}
    =
    \nonumber\\
    \frac{(-i)^3}{q_1^2}(2\pi)^4\delta^4(q_1+q_2).
\end{align}
After removing the momentum conserving delta function, we have $i/q^2$.

Focusing back on the amplitude, we can pull $\mathcal{S}_{\mathcal{J}\mathcal{J}}$ out of the gluon amplitude integral, leaving us with
\begin{align}
    i\tilde{\mathcal{M}}^{++-} &=
    \lim_{\mathcal{J}\to0}
    \left(-i\frac{\delta}{\delta \mathcal{J}_d^-(-k_1)}\right)
    \left(-i\frac{\delta}{\delta \mathcal{J}_e^-(-k_2)}\right)
    \nonumber\\
    &
    \left(-i\frac{\delta}{\delta \mathcal{J}_f^+(-k_3)}\right)
    e^{i\mathcal{S}_{\textsl{g}\textsl{g}\textsl{g}}\left[\delta/\delta\mathcal{J}\right]}
    e^{i\mathcal{S}_{\mathcal{JJ}}}
    \nonumber\\
    &
    \mathcal{N}\int d\textsl{g}^+ d\textsl{g}^-
    e^{i\bar{\mathcal{S}}_{\textsl{g}\textsl{g}}[\mathcal{J}]}.
\end{align}

We next make a change of variables from $\textsl{g}\to \textsl{g}+\mathcal{J}/p^2$.  This change eliminates $\mathcal{J}$ from the integral, and we find
\begin{align}
    \mathcal{N}\int d\textsl{g}^+ d\textsl{g}^-
    e^{i\bar{\mathcal{S}}_{\textsl{g}\textsl{g}}[\mathcal{J}]}
    &= \mathcal{N}\int d\textsl{g}^+ d\textsl{g}^-
    e^{i\mathcal{S}_{\textsl{g}\textsl{g}}}
    = 1.
\end{align}

With this, we are left with
\begin{align}
    i\tilde{\mathcal{M}}^{++-} &=
    \lim_{\mathcal{J}\to0}
    \left(-i\frac{\delta}{\delta \mathcal{J}_d^-(-k_1)}\right)
    \left(-i\frac{\delta}{\delta \mathcal{J}_e^-(-k_2)}\right)
    \nonumber\\
    &
    \left(-i\frac{\delta}{\delta \mathcal{J}_f^+(-k_3)}\right)
    e^{i\mathcal{S}_{\textsl{g}\textsl{g}\textsl{g}}\left[\delta/\delta\mathcal{J}\right]}
    e^{i\mathcal{S}_{\mathcal{JJ}}}.
\end{align}

At the moment, we are interested in the connected tree-level amplitude.  We begin by expanding the interaction exponential and only keeping the first order term, 
\begin{align}
    \tilde{\mathcal{M}}^{++-} &=
    \lim_{\mathcal{J}\to0}
    \left(-i\frac{\delta}{\delta \mathcal{J}_d^-(-k_1)}\right)
    \left(-i\frac{\delta}{\delta \mathcal{J}_e^-(-k_2)}\right)
    \nonumber\\
    &
    \left(-i\frac{\delta}{\delta \mathcal{J}_f^+(-k_3)}\right)
    \mathcal{S}_{\textsl{g}\textsl{g}\textsl{g}}\left[\delta/\delta\mathcal{J}\right]
    e^{i\mathcal{S}_{\mathcal{JJ}}}.
\end{align}
To keep the equations more compact, we will begin with the external derivatives.  We begin with the derivative with respect to $\mathcal{J}_f^+(-k_3)$.  This derivative acting on $\mathcal{S}_{\mathcal{JJ}}$ is
\begin{align}
    \left(-i\frac{\delta}{\delta \mathcal{J}_f^+(-k_3)}\right)\mathcal{S}_{\mathcal{JJ}} &=
    i\int\frac{d^4p_1d^4p_2}{(2\pi)^4}\delta^4(p_1+p_2)\frac{1}{p_1^2}
    \nonumber\\
    &\mathcal{J}_f^-(p_2)(2\pi)^4\delta^4(p_1+k_3)
    \nonumber\\
    &= 
    \frac{i}{k_3^2}\mathcal{J}_f^-(k_3),
\end{align}
resulting in
\begin{align}
    \tilde{\mathcal{M}}^{++-} &=
    \frac{-1}{k_3^2}\lim_{\mathcal{J}\to0}
    \mathcal{S}_{\textsl{g}\textsl{g}\textsl{g}}\left[\delta/\delta\mathcal{J}\right]
    \left(-i\frac{\delta}{\delta \mathcal{J}_d^-(-k_1)}\right)
    \nonumber\\
    &
    \left(-i\frac{\delta}{\delta \mathcal{J}_e^-(-k_2)}\right)
    \mathcal{J}_f^-(k_3)
    e^{i\mathcal{S}_{\mathcal{JJ}}}.
\end{align}
The connected amplitude will come from all the external derivatives acting on the exponential, therefore,
\begin{align}
    \tilde{\mathcal{M}}^{++-} &=
    \frac{-1}{k_1^2k_2^2k_3^2}\lim_{\mathcal{J}\to0}
    \mathcal{S}_{\textsl{g}\textsl{g}\textsl{g}}\left[\delta/\delta\mathcal{J}\right]
    \nonumber\\
    &
    \mathcal{J}_d^+(k_1)
    \mathcal{J}_e^+(k_2)
    \mathcal{J}_f^-(k_3)
    e^{i\mathcal{S}_{\mathcal{JJ}}}.
\end{align}
At the end, we have to take $\mathcal{J}\to0$, so the derivatives inside the interaction should only act on the $\mathcal{J}$ that are outside the exponential at this point.  We will begin with the $\mathcal{J}(p_1)$ derivative, giving
\begin{align}
    \tilde{\mathcal{M}}^{++-} &=
    \frac{-i g_s}{\sqrt{2}k_1^2k_2^2k_3^2}
    f_{fbc}
    \nonumber\\
    &
    \lim_{\mathcal{J}\to0}
    \int\frac{d^4p_2d^4p_3}{(2\pi)^4} \delta^4(-k_3+p_2+p_3) 
    \frac{\lbrack p_2 p_3 \rbrack^3}{\lbrack k_3 p_2 \rbrack\lbrack p_3 k_3 \rbrack}
    \nonumber\\
    &
    \left(-i\frac{\delta}{\delta \mathcal{J}_b^+(-p_2)}\right)
    \left(-i\frac{\delta}{\delta \mathcal{J}_c^+(-p_3)}\right)
    \nonumber\\
    &
    \mathcal{J}_d^+(k_1)
    \mathcal{J}_e^+(k_2)
    e^{i\mathcal{S}_{\mathcal{JJ}}},
\end{align}
where we have made the momenta in the spinor products explicit to eliminate confusing label redundancy, and the sign comes from the sign in front of the square-bracket term in the action.
Doing the other two derivatives leaves us with
\begin{align}
    \tilde{\mathcal{M}}^{++-} &=
    \frac{-i^3\sqrt{2}g_s}{k_1^2k_2^2k_3^2}
    \frac{\lbrack 12 \rbrack^3}{\lbrack 31 \rbrack\lbrack 23 \rbrack}
    f_{fde}
    (2\pi)^4\delta^4(k_1+k_2+k_3) 
    ,
\end{align}
where we have taken into account the fact that there are two ways the derivatives can be taken by including the factor of $2$ (they give the same result).
Finally, we amputate the external propagators (removing three factors of $i/k^2$) and drop the momentum conserving delta function [$(2\pi)^4\delta^4(k_1+k_2+k_3)$] leaving 
\begin{align}
    i\mathcal{M}^{++-} &=
    -i\sqrt{2}g_s
    \frac{\lbrack 12 \rbrack^3}{\lbrack 23 \rbrack \lbrack 31 \rbrack}
    f_{def}
    .
\end{align}
This gives us the vertex that we should expect, based on the action. 

The opposite helicity combination gives
\begin{align}
    \tilde{\mathcal{M}}^{--+} &=
    \frac{-1}{k_1^2k_2^2k_3^2}\lim_{\mathcal{J}\to0}
    \mathcal{S}_{\textsl{g}\textsl{g}\textsl{g}}\left[\delta/\delta\mathcal{J}\right]
    \nonumber\\
    &
    \mathcal{J}_d^-(k_1)
    \mathcal{J}_e^-(k_2)
    \mathcal{J}_f^+(k_3)
    e^{i\mathcal{S}_{\mathcal{JJ}}}.
\end{align}
Performing the derivatives gives
\begin{align}
    \tilde{\mathcal{M}}^{--+} &=
    \frac{i^3\sqrt{2}g_s}{k_1^2k_2^2k_3^2}
    \frac{\langle 12 \rangle^3}{\langle 23 \rangle\langle 31 \rangle}
    f_{def}
    (2\pi)^4\delta^4(k_1+k_2+k_3) 
    ,
\end{align}
resulting in the vertex
\begin{align}
    i\mathcal{M}^{--+} &=
    i\sqrt{2}g_s
    \frac{\langle 12 \rangle^3}{\langle 31 \rangle\langle 23 \rangle}
    f_{def}
    ,
\end{align}
as expected.

\subsection{\label{app:W-nu vertex chirality}$\textbf{W}$ Lepton Vertex}
We will also consider an example with both massless and massive fields by considering a neutrino interaction.
We will be more concise in this subsection, using the features of the previous subsection.  Suppose we would like the 3-point amplitude at tree level for a neutrino, electron and a $W$ boson,
\begin{align}
    i\tilde{\mathcal{M}}^{-\mathrm{IJK}} &=
    \mathcal{N}
    \int d\nu^- d\bar{\nu}^+ de d\bar{e} dW d\bar{W}
    \nonumber\\
    &
    \nu^-(k_1) \bar{e}^{\mathrm{I}}(k_2) \bar{W}^{\mathrm{JK}}(k_3)
    e^{i\mathcal{S}_{2}+i\mathcal{S}_{int}},
\end{align}
where
\begin{align}
    \mathcal{N}^{-1} &= 
    \int d\nu^- d\bar{\nu}^+ de d\bar{e} dW d\bar{W} e^{i\mathcal{S}_{2}} 
\end{align}
and
\begin{align}
    \mathcal{S}_{2} &= 
    \int\frac{d^4p_1d^4p_2}{(2\pi)^4} \delta^4(p_1+p_2) 
    \Big[
    \left(p_1^2-m_e^2\right)\bar{e}_{\mathrm{I}}(p_1)e^{\mathrm{I}}(p_2)
     \nonumber\\
     &
    +
    p_1^2\bar{\nu}^{+}(p_1) \nu^{-}(p_2)
    +\left(p_1^2-M_W^2\right)\bar{W}_{\mathrm{IJ}}(p_1)W^{\mathrm{IJ}}(p_2)
    \Big]
    \\
    \mathcal{S}_{int} &=
    \frac{e}{M_W s_W}
    \int\frac{d^4p_1d^4p_2d^4p_3}{(2\pi)^{8}} \delta^4(p_1+p_2+p_3)\Big[
    \nonumber\\
    &
    \bar{\nu}^+(p_1)e_{\mathrm{I}}(p_2)W_{\mathrm{KL}}(p_3)
    \lbrack\mathbf{32}\rbrack^{\mathrm{KI}} \langle1\mathbf{3}\rangle^{\mathrm{L}}
    \nonumber\\
    &\hspace{0.5in}
    -
    \bar{e}_{\mathrm{I}}(p_1)\nu^-(p_2)\bar{W}_{\mathrm{KL}}(p_3)
    \langle\mathbf{13}\rangle^{\mathrm{IL}} \lbrack\mathbf{3}2\rbrack^{\mathrm{K}}
    \Big],
\end{align}
come from Eqs.~(\ref{eq:S_2m=0}), (\ref{eq:S p^2-m^2}) and (\ref{eq:S_W/Zf}).

We first introduce sources for the fields,
\begin{align}
    \mathcal{S}_{2}\left[\mathcal{J}\right] &= 
    \int\frac{d^4p_1d^4p_2}{(2\pi)^4} \delta^4(p_1+p_2) 
    \Big[
    \left(p_1^2-m_e^2\right)\bar{e}_{\mathrm{I}}(p_1)e^{\mathrm{I}}(p_2)
     \nonumber\\
     &
    + \bar{e}_{\mathrm{I}}(p_1)\mathcal{J}_e^{\mathrm{I}}(p_2)
    + \bar{\mathcal{J}}_{e\mathrm{I}}(p_1)e^{\mathrm{I}}(p_2)
    +
    p_1^2\bar{\nu}^{+}(p_1) \nu^{-}(p_2)
    \nonumber\\
    &
    +\bar{\nu}^+(p_1)\mathcal{J}_{\nu}^-(p_2)
    +\bar{\mathcal{J}}_{\nu}^+(p_1)\nu^-(p_2)
    \nonumber\\
    &
    +\left(p_1^2-M_W^2\right)\bar{W}_{\mathrm{IJ}}(p_1)W^{\mathrm{IJ}}(p_2)
    +\bar{W}_{\mathrm{IJ}}(p_1)\mathcal{J}_W^{\mathrm{IJ}}(p_2)
    \nonumber\\
    &
    +\bar{\mathcal{J}}_{W\mathrm{IJ}}(p_1)W^{\mathrm{IJ}}(p_2)
    \Big].
\end{align}
We complete the square, giving
\begin{align}
    \mathcal{S}_{2}\left[\mathcal{J}\right] &= 
    \int\frac{d^4p_1d^4p_2}{(2\pi)^4} \delta^4(p_1+p_2) 
    \Big[
    \left(p_1^2-M_W^2\right)
    \nonumber\\
    &
    \left(\bar{W}_{\mathrm{IJ}}(p_1)+\bar{\mathcal{J}}_{W\mathrm{IJ}}(p_1)\right) 
    \left(W^{\mathrm{IJ}}(p_2)+\mathcal{J}_W^{\mathrm{IJ}}(p_2)\right)
    \nonumber\\
    &
    +\left(p_1^2-m_e^2\right)
    \left(\bar{e}_{\mathrm{I}}(p_1)+\bar{\mathcal{J}}_{e\mathrm{I}}(p_1)\right) 
    \left(e^{\mathrm{I}}(p_2)+\mathcal{J}_e^{\mathrm{I}}(p_2)\right)
    \nonumber\\
    &
    +p_1^2
    \left(\bar{\nu}^{+}(p_1)+\bar{\mathcal{J}}_{\nu}^+(p_1)\right)
    \left(\nu^{-}(p_2)+\mathcal{J}_{\nu}^-(p_2)\right)
     \nonumber\\
     &
    -\frac{1}{\left(p_1^2-M_W^2\right)}
    \bar{\mathcal{J}}_{W\mathrm{IJ}}(p_1)
    \mathcal{J}_W^{\mathrm{IJ}}(p_2)
    \nonumber\\
    &-\frac{1}{\left(p_1^2-m_e^2\right)}
    \bar{\mathcal{J}}_{e\mathrm{I}}(p_1)
    \mathcal{J}_e^{\mathrm{I}}(p_2)
    -\frac{1}{p_1^2}
    \bar{\mathcal{J}}_{\nu}^+(p_1)
    \mathcal{J}_{\nu}^-(p_2)
    \Big].
\end{align}
Therefore, the interaction action becomes
\begin{align}
    \mathcal{S}_{int}\left[\frac{\delta}{\delta \mathcal{J}}\right] &=
    \frac{(-i)^3e}{M_W s_W}
    \int\frac{d^4p_1d^4p_2d^4p_3}{(2\pi)^{8}} \delta^4(p_1+p_2+p_3)\Bigg[
    \nonumber\\
    &
    \frac{\delta}{\delta\mathcal{J}_{\nu}^-(-p_1)}
    \frac{\delta}{\delta\bar{\mathcal{J}}_e^{\mathrm{I}}(-p_2)}
    \frac{\delta}{\delta\bar{\mathcal{J}}_W^{\mathrm{KL}}(-p_3)}
    \lbrack\mathbf{32}\rbrack^{\mathrm{KI}} \langle1\mathbf{3}\rangle^{\mathrm{L}}
    \nonumber\\
    &\hspace{-0.25in}
    -
    \frac{\delta}{\delta\mathcal{J}_e^{\mathrm{I}}(-p_1)}
    \frac{\delta}{\delta\bar{\mathcal{J}}_{\nu}^+(-p_2)}
    \frac{\delta}{\delta\mathcal{J}_W^{\mathrm{KL}}(-p_3)}
    \langle\mathbf{13}\rangle^{\mathrm{IL}} \lbrack\mathbf{3}2\rbrack^{\mathrm{K}}
    \Bigg].
\end{align}

Putting this all together, we have
\begin{align}
    \tilde{\mathcal{M}}^{-\mathrm{IJK}} &=
    (-i)^3\lim_{\mathcal{J}\to0}
    \mathcal{S}_{int}\left[\frac{\delta}{\delta \mathcal{J}}\right]
    \frac{\delta}{\delta\bar{\mathcal{J}}_{\nu}^+(-k_1)}
    \frac{\delta}{\delta\mathcal{J}_{e\mathrm{I}}(-k_2)}
    \nonumber\\
    &
    \frac{\delta}
    {\delta\mathcal{J}_{W\mathrm{JK}}(-k_3)}
    e^{i\mathcal{S}_{\mathcal{JJ}}},
\end{align}
at tree-level,
where
\begin{align}
    \mathcal{S}_{\mathcal{JJ}} &=
    \int\frac{d^4p_1d^4p_2}{(2\pi)^4} \delta^4(p_1+p_2) 
    \Big[
     \nonumber\\
     &
    -\frac{1}{\left(p_1^2-M_W^2\right)}
    \bar{\mathcal{J}}_{W\mathrm{IJ}}(p_1)
    \mathcal{J}_W^{\mathrm{IJ}}(p_2)
    \nonumber\\
    &-\frac{1}{\left(p_1^2-m_e^2\right)}
    \bar{\mathcal{J}}_{e\mathrm{I}}(p_1)
    \mathcal{J}_e^{\mathrm{I}}(p_2)
    -\frac{1}{p_1^2}
    \bar{\mathcal{J}}_{\nu}^+(p_1)
    \mathcal{J}_{\nu}^-(p_2)
    \Big].
\end{align}
As before, we are interested in the connected amplitude, so we use the external derivatives to pull out three factors of $\mathcal{J}$, leading to
\begin{align}
    \tilde{\mathcal{M}}^{-\mathrm{IJK}} &=
    \frac{-(-1)^2}{k_1^2(k_2^2-m_e^2)(k_3^2-M_W^2)}
    \nonumber\\
    &\lim_{\mathcal{J}\to0}
    \mathcal{S}_{int}\left[\frac{\delta}{\delta \mathcal{J}}\right]
    \mathcal{J}_\nu^-(k_1)
    \bar{\mathcal{J}}_e^{\mathrm{I}}(k_2)
    \bar{\mathcal{J}}_W^{\mathrm{JK}}(k_3)
    e^{i\mathcal{S}_{\mathcal{JJ}}},
\end{align}
where there are two extra signs.  The first is due to the $\mathcal{J}_e$ derivative passing through the $\bar{\mathcal{J}}_e$, where $\mathcal{J}_e$ is Grassman.  The second is due to the need to raise and lower one spin index $\mathrm{I}$.  We need to raise and lower the spin indices on $\mathcal{J}_W$, however, there are two, so the signs cancel each other.
Performing the derivatives inside the interaction term, we have
\begin{align}
    \tilde{\mathcal{M}}^{-\mathrm{IJK}} =&
    \frac{e}{M_W s_W}
    \frac{-(-1)(-i)^3}{k_1^2(k_2^2-m_e^2)(k_3^2-M_W^2)}
    \lbrack\mathbf{32}\rbrack^{\mathrm{JI}}
    \langle1\mathbf{3}\rangle^{\mathrm{K}}
    \nonumber\\
    &
    (2\pi)^4\delta^4(k_1+k_2+k_3),
\end{align}
where we get another minus sign from the $\bar{\mathcal{J}}_e$ derivative passing through the $\mathcal{J}_{\nu}$.
After amputating and dropping the momentum conserving delta function, we have the vertex
\begin{align}
    i\mathcal{M}^{-\mathrm{IJK}} =&
    \frac{-i e}{M_W s_W}
    \lbrack\mathbf{32}\rbrack^{\mathrm{JI}}
    \langle1\mathbf{3}\rangle^{\mathrm{K}},
\end{align}
which is the correct vertex for these particles.

If we had, instead, calculated the amplitude with a positive-helicity antineutrino, we would have
\begin{align}
    \tilde{\mathcal{M}}^{+\mathrm{IJK}} &=
    \frac{-(-1)}{k_1^2(k_2^2-m_e^2)(k_3^2-M_W^2)}
    \nonumber\\
    &\lim_{\mathcal{J}\to0}
    \mathcal{S}_{int}\left[\frac{\delta}{\delta \mathcal{J}}\right]
    \bar{\mathcal{J}}_\nu^+(k_1)
    \mathcal{J}_e^{\mathrm{I}}(k_2)
    \mathcal{J}_W^{\mathrm{JK}}(k_3)
    e^{i\mathcal{S}_{\mathcal{JJ}}},
\end{align}
where this time, we get one extra sign from moving the $\mathcal{J}_{\nu}$ derivative past the $\bar{\mathcal{J}}_{\nu}$.
Performing the derivatives and setting $\mathcal{J}\to0$, gives
\begin{align}
    \tilde{\mathcal{M}}^{+\mathrm{IJK}} =&
    \frac{e}{M_W s_W}
    \frac{i^3}{k_1^2(k_2^2-m_e^2)(k_3^2-M_W^2)}
    \langle\mathbf{23}\rangle^{\mathrm{IL}} \lbrack\mathbf{3}1\rbrack^{\mathrm{K}}
    \nonumber\\
    &
    (2\pi)^4\delta^4(k_1+k_2+k_3),
\end{align}
where there is an extra sign from the relative sign between terms in the action, but no need to move a fermionic $\mathcal{J}$ derivative past a fermionic $\mathcal{J}$.  This gives the vertex
\begin{align}
    i\mathcal{M}^{+\mathrm{IJK}} =&
    \frac{i e}{M_W s_W}
    \langle\mathbf{23}\rangle^{\mathrm{IL}} \lbrack\mathbf{3}1\rbrack^{\mathrm{K}},
\end{align}
as we would expect.

\end{document}